# Development of SyReC Based Expandable Reversible Logic Circuits

*A*

***Dissertation***

*Submitted in*

*Partial fulfillment*

*for the award of the Degree of*

***MASTER OF TECHNOLOGY***

***in Department of Computer Science Engineering***

***(With specialization in COMPUTER SCIENCE & ENGINEERING)***

**Supervisor**

Dr. S.C. Jain
(Professor)

**Submitted By**

Vandana Maheshwari
Enrolment No.:
11E2UCCSF4XP613

**DEPARTMENT OF COMPUTER SCIENCEENGINEERING**
**UNIVERSITY COLLEGE OF ENGINEERING**
RAJASTHAN TECHNICAL UNIVERSITY
KOTA (RAJASTHAN)

**(March, 2014)**

# CANDIDATE'S DECLARATION

I hereby declare that the work, which is being presented in the Dissertation, entitled **"Development of SyReC Based Expandable Reversible Logic Circuits"** in partial fulfillment for the award of Degree of **"Master of Technology"** in Dept. of Computer Science Engineering with Specialization in Computer Science, **and submitted to the Department of Computer Science Engineering,** University College of Engineering, Kota, Rajasthan Technical University is a record of my own investigations carried under the Guidance of **Dr. S.C. Jain**, Department of Computer Science Engineering, University College of Engineering, Kota**.**

I have not submitted the matter presented in this Dissertation anywhere for the award of any other Degree.

**Vandana Maheshwari**
Computer Science & Engineering
Enrolment No.: 11E2UCCSF4XP613
University College of Engineering,
Kota (Rajasthan)

Under Guidance of

**Dr. S. C. Jain**
Professor,
Department of Computer Science & Engineering
University College of Engineering,
Kota (Rajasthan)

# CERTIFICATE

This is to certify that this Dissertation entitled **"Development of SyReC Based Expandable Reversible Logic Circuits"** has been successfully carried out by **Vandana Maheshwari** (Enrolment No.: 11E2UCCSF4XP613), under my supervision and guidance, in partial fulfillment of the requirement for the award of **Master of Technology** Degree in **Computer Science & Engineering** from **University College of Engineering**, Rajasthan Technical University, Kota for the year 2011-2013.

**Dr. S. C. Jain**
Professor,
Department of Computer Science & Engineering
University College of Engineering,
Kota (Rajasthan)

# ACKNOWLEDGEMENTS

It is matter of great pleasure for me to submit this report on dissertation entitled **“Development of SyReC Based Expandable Reversible Logic Circuits”**, as a part of curriculum for award of “Master in Technology” with specialization in “Computer Science & Engineering” degree of Rajasthan Technical University, Kota.

I am thankful to my dissertation guide **Dr. S.C. Jain**, Professor in department computer science for his constant encouragement, able guidance and for giving me a platform to build by career by giving me a chance to learn different fields of this technology. I am also thankful to **Mr. C.P. Gupta**, Associate Professor& Head of Computer Science Department for this valuable support.

I would like to acknowledge my thanks to entire faculty and supporting staff of Computer Engineering Department in general and particularly for their help, directly or indirectly during my Dissertation work.

I express my deep sense of reverence to my parents and family members for their unconditional support, patience and encouragement.

**Date** **Vandana Maheshwari**

# CONTENTS

# LIST OF FIGURES

# LIST OF TABLES

# ABSTRACT

Reversible computing is gaining high interest from researchers due to its various promises. One of the prominent advantages perceived from reversible logic is that of reduced power dissipation with many reversible gates at hand, designing a reversible circuit (combinational) has received due attention and achievement. A proposed language for description of reversible circuit, namely SyReC, is also in place. What remain are the software tools which would help in reversible circuit synthesis through simulation.

Beginning with the smallest reversible circuit realizations the SyReC statements and expressions, we employ a hierarchal approach to develop a complete reversible circuit, entirely from its SyReC code. We implement this as a software tool. The tool allows a user to expand a reversible circuit of choice in terms of bit width of its inputs. The background approach of expansion of a reversible circuit has also been proposed as a part of this dissertation. Also, a user can use the tool to observe the effect of expansion on incurred costs, in terms of increase in number of lines, number of gates and quantum cost. The importance of observing the change in costs with respect to scale of expansion is important not only from analysis point of view, but also because the cost depends on the approach used for expansion.

This dissertation also proposes a reversible circuit design for elevator controller (combinational) and the related costs. The aim is to emphasize use of the proposed approach is designing customized circuits.

Chapter 1

# INTRODUCTION

Computing technology is advancing at a high pace, achieving ever-higher transistor densities and increasing the computational utility achieved in a given quantity of time, space, material, energy and cost. The availability of affordable computing enables new applications in all fields thus driving up demand for more computing power.

This feedback loop of increasing demand and improving technology faces the challenges of limitations like power dissipation. As predicted by Gordon Moore in 1960 [1], popularly known as Moore's low, the transistor counts in a chip will double every one and half year on average. ITRS has also drawn a road-map of required feature size in future at atomic level in 2050 as shown in figure 1.1. Such shrinking in feature size would result in a number of implementation and operational difficulties like heat dissipation, requirement of very thin laser beam, clock distribution etc.

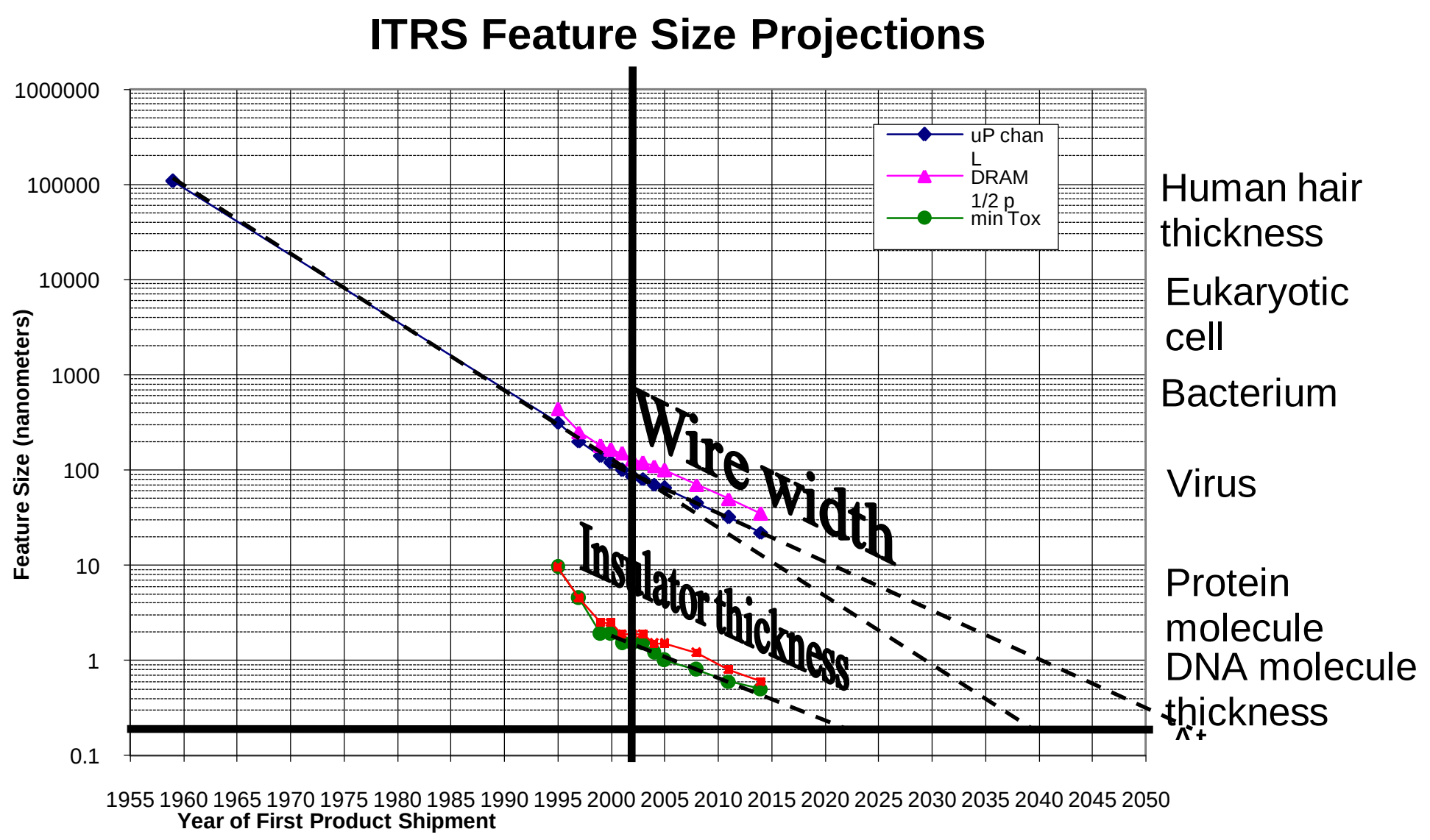

**Figure 1.1: ITRS Feature size Projection**

Reversible computing is emerging as a promising alternative to the computational CMOS technology, since it can reduce or even eliminated power dissipation. Furthermore, reversible logic builds the basis for quantum computation – a completely new way of processing which enables to solve certain problems exponentially faster compared to conventional methods. In this chapter section 1.1 describes limitations of current technologies, section 1.2 introduces Reversible Computation as an alternative which gives solution to overcome the limitations, section 1.3 describes the objectives of the dissertation work and finally section 1.4 states the thesis organization.

## 1.1 Limitations of Conventional Computing System

Researchers expect that "traditional" technologies like CMOS will reach their limits in the near future. The problems faced in the development of these technologies due to following problems:

### 1.1.1 Physical problems

Most of the efforts of development in traditional technologies have been towards miniaturization of integrated circuits, but it has the related growing problem of power dissipation, which is a crucial issue in today's hardware design process. While due to new fabrication processes, energy loss has significantly been reduced over the last decades, physical limits still exist. Landauer [2] proved that using conventional (irreversible) logic, gate operations always lead to energy dissipation regardless of the underlying technology. More precisely, exactly $kT.Ln2$ Joule of energy is dissipated for each "lost" bit of information during an irreversible operation where $k$ is the Boltzmann constant and $T$ is the temperature. While this amount of power currently does not sound significant, it may become crucial when considering that (1) today millions of operations are performed in some seconds (i.e. increasing processor frequency multiplies this amount)and (2) more and more operations are performed with smaller and smaller transistor sizes (i.e. in a smaller area) [3, 4].

### 1.1.2 Computational problem

A large number of computation intensive problems like NP-complete type problems demand high computational speed, but have not been solved by classical computers. Complex problem like 8-queen problem which needs lot of backtracking cannot be solved by irreversible logic gates because we cannot achieve input from the output. Security is also the essential feature required by many applications which make use of cryptanalysis methods but heat generation in conventional system directly affects the security. Memory-intensive problems like travelling salesman problem cannot be solved in conventional computing environment.

### 1.1.3 Economic problem

The above stated limitations of heat dissipation, memory etc lead to high cost of synthesis using conventional technologies, which in turn increase the cost at which computing power, is available for users.

## 1.2 Reversible Computation

In 60s and 70s theoretical physicists considered the problem of circuit synthesis as "whether it is possible to compute without generating heat". A possible direction of thought was towards reducing computation steps (or circuit depth). In 1960, Landauer [2] showed that the energy used for computations is not correlated with the number of computation steps, but instead with the amount of information that is discarded. Deleting information in a computing device necessitates dissipation of a small amount of heat. In1970s, Bennett [5] showed that zero energy dissipation is only possible, if information-lossless computation is performed. He showed that classical computation can be done reversibly with no energy dissipated per computational step through a reversible model of the Turing machine. He thus demonstrated that any problem that can be simulated on the original irreversible machine can also be simulated with the same efficiency on the reversible model. This does not hold for conventional circuits but reversible circuits, i.e. circuits where all operations are performed in an invertible manner.

Thus, the above problem of heat dissipation can be eliminated by using reversible logic, which can be performed through reversible gates. Reversible logic is a logic design style in which there is a one to one mapping between the input and the output vectors. Reversible gates [4] are circuits (gates) that have one-to-one mapping between vectors of inputs and outputs; thus the vector of input states can be always reconstructed from the vector of output states. This prevents the loss of information which is the root cause of power dissipation in irreversible logic circuits. Reversible circuits need a corresponding computation paradigm, referred to as Reversible computation. It describes computational models that are both forward and backward deterministic. A computation is *reversible* if it can be 'undone' in the sense that the output contains sufficient information to reconstruct the input, i.e., no input information is erased.

There were two related issues, logical reversibility and physical reversibility, which were intimately connected. Logical reversibility refers to the ability to reconstruct the input from the output of a computation, or gate function. A process is said to be *physically reversible* if it results in no increase in physical entropy. The reversible logic circuits must be constructed under two main constraints. They are (1) Fan-out is not permitted. (2) Loops or feedbacks are not permitted. These constraints are yet under discussion whether to be followed strictly or not, like for sequential circuits the constraints can be relaxed. Hence, reversible circuits are seen as future alternative conventional circuit technologies with certain low-power applications. Reversible computing also has the applications in emerging nanotechnologies such as quantum dot cellular automata, optical computing, quantum computing, mobile computing and low power computing, etc.

With the introduction of the concept of reversible logic, synthesis of reversible circuits has become an interesting and growing field of research. Methodologies used for conventional circuit's synthesis like transformation based, cycle based, search based, ESOP based, BDD based have well been adopted for reversible circuit synthesis [6, 7, 8, 9, 10, 11].

Similarly, synthesis approaches using language like VHDL for reversible circuits has been proposed by [12]. SyReC is a programming language to specify reversible circuits. It has also been claimed that SyReC can be used for automatic synthesis. We aim to explore this further.

## 1.3 Objectives

There is dearth of simulation packages for reversible circuit synthesis. Physical realization of a reversible circuit depends on feasibility of a quantum computer. Thus making research in reversible circuits to be completely dependent on simulators and software tools. We aim to construct a tool which can synthesize a reversible circuit when provided a SyReC specification. Also, we emphasize on how to achieve expansion of a reversible circuit, once the circuit has been realized, without the need of rewriting or modifying SyReC specification.

We also implement the approach by incorporating it in a software tool. The tool takes a SyReC specification as input, realizes it into a reversible circuit, and expands it as per user inputs. We also predict the costs related to such expansion. To emphasize the generality of the proposed approach we present a design for reversible circuit of elevator controller and synthesize it using a tool.

## 1.4 Organization of Thesis

Chapter 2 titled "Literature survey" covering domain of reversible logic, reversible circuit synthesis and SyReC. Chapter 3 titled "Circuit realizations from SyReC" describes circuit realizations for various SyReC statements and expressions. Chapter 4 titled "Expandable circuit design" describes the general approach of reversible circuit synthesis for SyReC and expansion of reversible circuit along with typical examples and a special example of elevator controller. Chapter 5 titled "Implementation and results" describes the implementation of our approach as a software tool "RCHDL Realizer" and corresponding results. Chapters 6 titled "Conclusion and future scope" concludes our dissertation and discuss future scope of the work.

**Chapter 2**

# LITERATURE SURVEY

As discussed earlier, demand scenario is posing a number of challenges as current technologies to meet out computing need. Reversible computing emerging as a potential candidate to replace conventional logic and quantum technology may be a target implementation. We categorize our survey in the following categories.

- Reversible logic gates
- Circuit representation formats
- Reversible circuit design
- Tools
- Hardware Description Language SyReC

Section 2.1 presents the progress in development of reversible gates with time which includes basic gates to complex gates developed for specific applications. Section 2.2 will explain the different representation formats available for reversible circuits. Section 2.3 presents the reversible circuits developed for different applications and their design methodology. Section 2.4 will presents the development of tools available in this area. Section 2.5 introduces the basics of hardware description language SyReC.

## 2.1 Reversible Logic Gates

Reversible logic is a logic design style in which there is a one to one mapping between the input and the output vectors. Reversible are circuits (gates) that have one-to-one mapping between vectors of inputs and outputs; thus the vector of input states can be always reconstructed from the vector of output states. This prevents the loss of information which is the root cause of power dissipation in irreversible logic circuits [2]. Interest in reversible logic gates is continuously growing and number of gates has been developed. We categorize out survey of gates in 2 categories.

### 2.1.1 Basic Gates

A reversible gate realizes a reversible function. For a gate $g$, the gate $g-1$ implements the inverse transformation. Some of the basic gates are given below.

#### 2.1.1.1 Not Gate

Negation is an important operation on Boolean function and also important for any Computational system whether is **CLASSICAL, MULTIVALUED OR REVERSIBLE**. Schematic representation of reversible Not Gate is show in figure 2.1

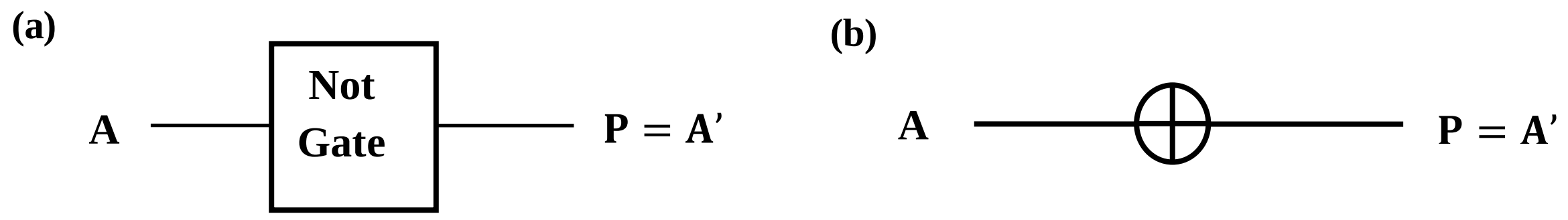

**Figure 2.1: Reversible Not Gate (a) Block Diagram (b) Schematic Representation**

#### 2.1.1.2 Feynman Gate

Feynman gate is given by Richard P Feynman in 1982; it is basically 2x2 reversible gates. Feynman gate can perform negation operation but in controlled way and also known as Controlled NOT gate. If two lines are A and B, the first line A is known as CONTROL line and second line B is known as TARGET line. Operation on target line is negation and only performed when control line is set to 1 otherwise no operation on target line is observed. It is widely used for fan-out purposes. Block diagram and schematic representation of 2x2 Feynman gate is shown in figure 2.2.

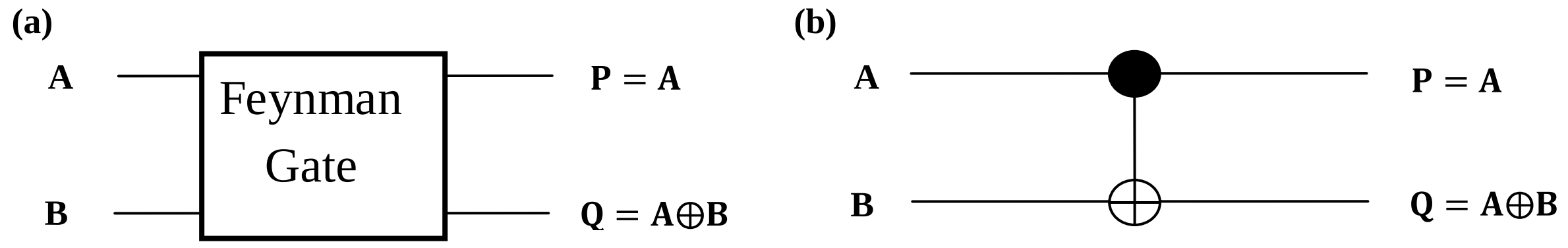

**Figure 2.2: Feynman Gate (a) Block Diagram (b) Schematic Representation**

#### 2.1.1.3 3*3 Toffoli Gate

In 1982 Toffoli gives a new gate called Toffoli gate [6], it is a 3x3; this gate is generalized further up to n lines. The TOFFOLI gate is also called the controlled-controlled-NOT gate since it can be understood as flipping the third input bit if, and only if, the first two input bits are both 1. In other words, the values of the first two input bits control whether the third input bit is flipped. The block diagram and schematic representation of 3x3 Toffoli gate is shown in figure 2.3. Toffoli gate plays an important role in the reversible logic synthesis. It is also used in the design of any Boolean function and hence it can be considered as a **Universal Reversible Gate**.

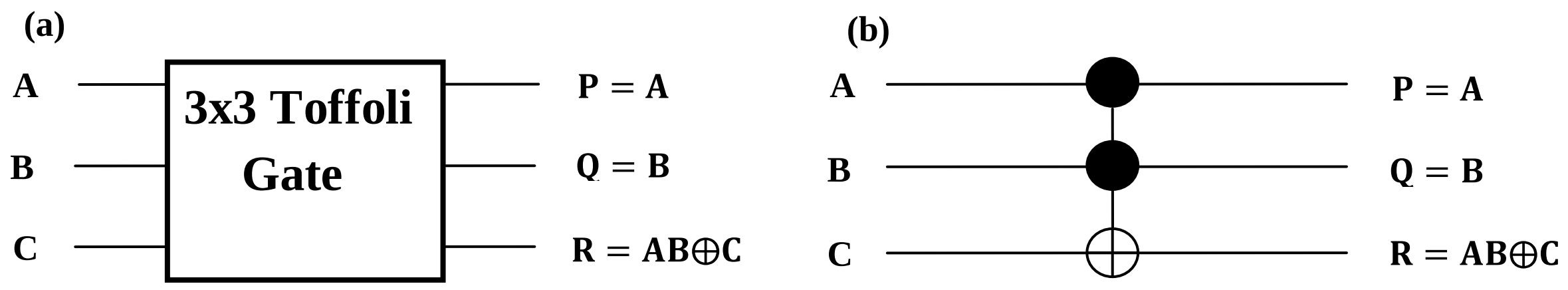

**Figure 2.3: Toffoli Gate (a) Block Diagram (b) Schematic Representation**

#### 2.1.1.4 3x3 Fredkin Gate

In 1982 Edword Fredkin and Tommaso Toffoli [7] proposed a new gate called 3x3 Fredkin gates which are further generalized up to n lines. The block diagram and schematic representation of 3x3 Fredkin gate is given in figure 2.4. The FREDKIN gate can also be seen as a controlled-SWAP gate in that it swaps the values of the second and third bits, if, and only if, the first bit is set to 1.

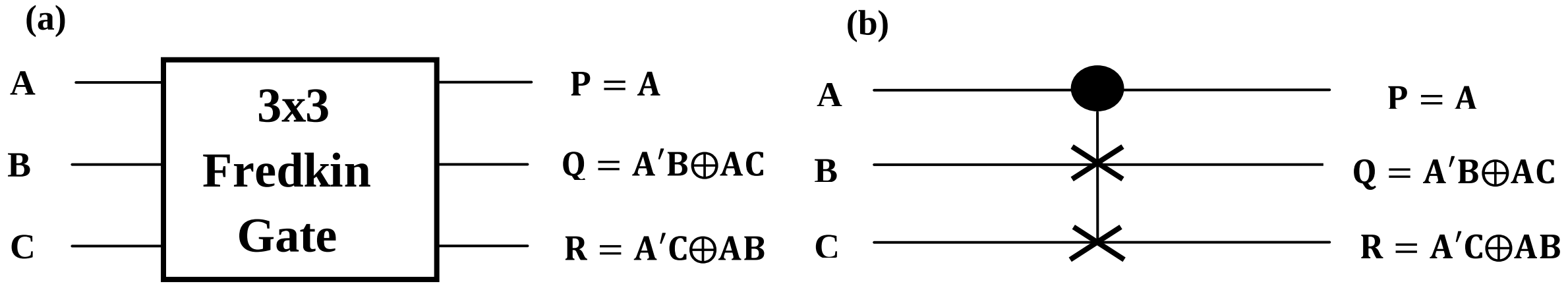

**Figure 2.4: Fredkin Gate (a) Block Diagram (b) Schematic Representation**

#### 2.1.1.5 3x3 Peres Gate

Peres Gate is proposed by A. Peres in 1985 [8], it is also 3x3 Reversible Gate, a Peres gate P (a1, a2, a3) has one control line a1 and two target lines a2 and a3. Peres Gate is the combination of Feynman Gate (a1, a2) and Toffoli Gate (a1, a2, a3), and so it can simultaneously generate two output functions (from Q and R). Figure 2.5 shows block diagram and schematic representation of 3x3 Peres gate.

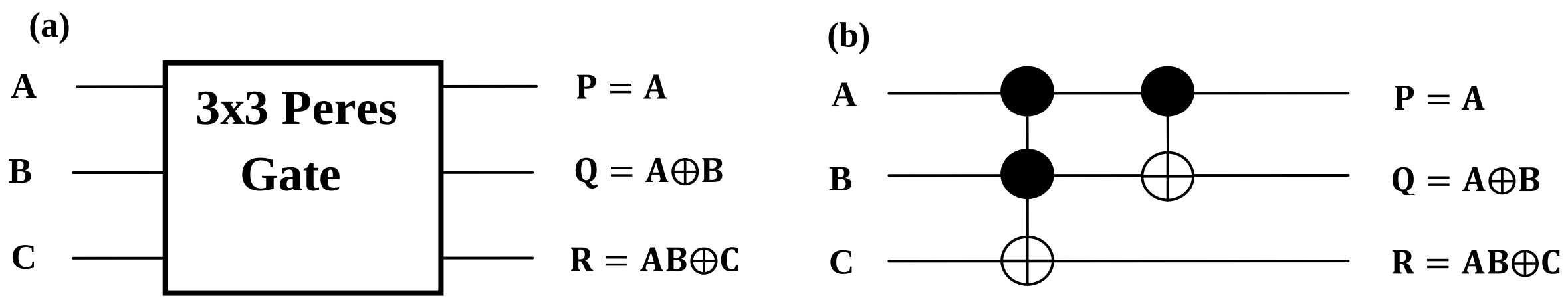

**Figure 2.5: Peres Gate (a) Block Diagram (b) Schematic Representation**

#### 2.1.1.6 Swap Gate

Swapping of lines is important in many systems, Swap gate is reversible gate which SWAP simply exchanges the bit values it is handed. Swap gate is basically a Fredkin gate with m=0. Swap gate is basically 2x2 gates S (a1, a2) which swaps values of a1 and a2 [6]. Figure 2.6 shows schematic representation of Swap gate.

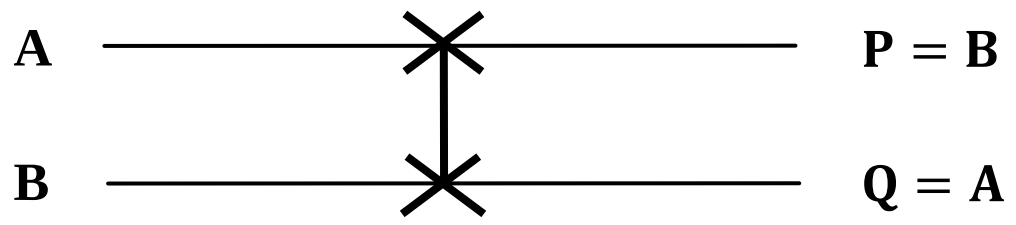

**Figure 2.6: Schematic Representation of Swap Gate**

### 2.1.2 Complex Gates

1. **Multiple Controlled Toffoli Gate**: In 1980 Toffoli gives a gate which is generalized up to n lines [6]. A multiple bit Toffoli gate ($x_1$; $x_2$; ....; $x_{m+1}$) passes the first m lines, control lines, unchanged. This gate flips the (m+1)-th line, target line, if and only if each positive (negative) control line carries the 1 (0) value. For $m = 0; 1; 2$ the gates are named NOT (N), CNOT (C), and Toffoli (T), respectively.

2. **Multiple Controlled Fredkin Gate:** In 1982 Edward Fredkin and Toffoli gives a gate which is generalized up to n lines [7]. A multiple bit Fredkin gate ($x_1$; $x_2$; ....; $x_{m+2}$) has two target lines $x_{m+1}$; $x_{m+2}$ and m control lines $x_1$; $x_2$; ....; $x_m$. The gate interchanges the values of the targets if the conjunction of all m positive (negative) controls evaluates to 1 (0). For m = 0; 1 the gates are called SWAP (S) and Fredkin (F), respectively.

3. **MAJ gate and UMA Gate:** In 2005 Steven A Cuccaro and team presents quantum ripple carry addition circuit, In which they proposed new gates called *Majority in place (MAJ)* and *Un-Majority and add (UMA)* [9]. They compute the majority of three bits in place and provide the carry bit for addition. Cascading it with an *Un-majority and Add* (UMA) gate forms a full adder. MAJ and UMA gates are basically made-up of two CNOT gates and one TOFFOLI gate.

4. **Double Peres Gate:** H.R. Bhagyalakshmi, M.K. Venkatesha proposed a new gate, they observe that in many circuits there is use two consecutive peres gates [10], They propose a new gate which can perform functionality of two consecutive peres gates.

## 2.2 Circuit Representation Models

Reversible functions have different properties than traditional functions, so some new methods have been proposed from 1982-2011 for representation of reversible functions. Reversible circuits can be described in many ways and each format of representation can be used in different synthesis approach.

### 2.2.1. Truth Tables

The simplest method to describe a reversible function of size n is a truth table with n columns and $2^n$ rows [7]. In this method input and output vector of a reversible circuit (reversible functions) are shown in row. The given functions often need thereby to be reversible. Since this is not the case for many practical functions, a pre-processing step called embedding often is performed first. This creates a reversible description of the given function which afterwards can be used to realize the desired circuit. Truth table can be used in transformation based synthesis approach.

A reversible truth table contains input vector which include both Primary and Constant inputs and output vector contains both Garbage and Primary outputs, while an irreversible truth table contains only Primary inputs and corresponding Primary outputs in each row.

#### 2.2.2. Binary Decision Diagrams

In 1986, a reversible function can be represented by a *Binary Decision Diagram* (BDD) [11]. A BDD is a directed acyclic graph where the Shannon decomposition is applied on each non-terminal node. Bryant proposed Reduced Ordered BDDs (ROBDDs), which offer canonical representations of Boolean functions. An ROBDD can be constructed from a BDD by ordering variables, merging equivalent sub-graphs and removing nodes with identical children. Several more specialized BDD variants have emerged for reversible and quantum circuits. In general, a BDD of a function may need an exponential number of nodes. However, BDD variants can represent many practical functions with only polynomial numbers of nodes.

#### 2.2.3. Cycle Form

Cyclic form (1996) is one of the shortest formats for representation of reversible circuits, it represents the cyclic chain of inputs and outputs and useful in cycle based synthesis approach for reversible circuits, so viewing a reversible function as a permutation, one can represent it as a product of disjoint cycles [12].

#### 2.2.4. Reed-Muller expansion (PPRM)

Search based synthesis approach uses the PPRM representation (1996) in which any Boolean function can be represented using Boolean variables and XOR operators, a reversible Boolean function can also be represented with XOR Sum of Product [13]. PPRM expansion uses only uncomplemented variables and can be derived from the EXOR-Sum-of-Products (ESOP) description by replacing $a'$ with $a \oplus 1$ for a complemented variable $a$. The PPRM expansion of a function is unique and is defined as follows.

$$f(x_1;\ x_2; \ldots \ldots;\ x_n) = a_0 \oplus a_1x_1 \oplus \ \ldots \ldots \oplus a_nx_n \oplus a_{12}x_1x_2 \oplus \ldots \ldots \oplus a_{n,n-1}\, x_{n-1}\, x_n \oplus \ldots \ldots \oplus a_{12\ldots\ldots n}\, x_1\, x_2 \ \ldots \ldots\ x_n \qquad (2.1)$$

A compact way to represent PPRM expansions is the vector of coefficients $a_0$, $a_1$, ... , $a_{12...n}$, called the *RM spectrum* of the function. Consider an $n$-variable function and record its values (from the truth table) in a $2^n$-element bit vector $F$. Then, the RM spectrum ($R$) of $F$ over the two-element field[8] GF(2) is defined as $R = M^nF$ where

$$M^0 = [1]; \qquad M^n = \begin{bmatrix} M^{n-1} & 0 \\ M^{n-1} & M^{n-1} \end{bmatrix} \qquad (2.2)$$

### 2.2.5. Matrix representations

A Boolean reversible function (permutation) $f$ can be described by a 0-1 matrix with a single 1 in each column and in each row (a *permutation matrix*), where the non-zero element in row $i$ appears in column $f(i)$ [14].

## 2.3 Reversible Circuit Design

In software and hardware applications of reversible information processing, sequences of reversible operations can be viewed as reversible circuits. A combinational *Reversible Circuit* is an acyclic combinational logic circuit in which all gates are reversible, and are interconnected without explicit fan out and loops.

### 2.3.1 CAD flow for Reversible Circuit Design

In this section we outline key steps in generation and optimization of reversible circuits, as illustrated in Figure 2.7 [15].

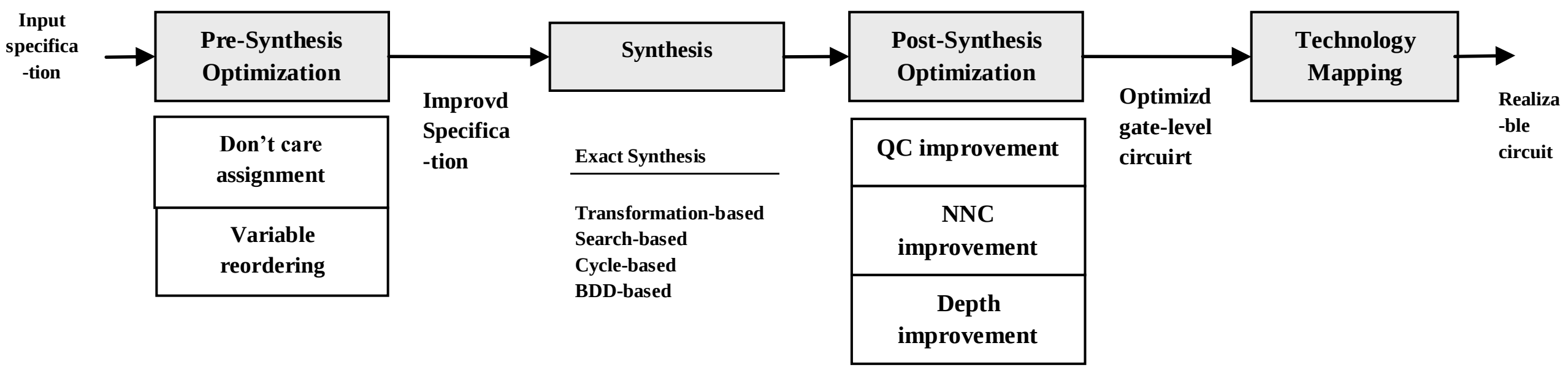

**Figure 2.7: General flow used in reversible circuit synthesis**

#### 2.3.1.1 Pre-Synthesis Optimization

To implement an irreversible specification using reversible gates, ancillae should be added to the original specification where the number of added lines, their values, and the ordering of output lines affect the cost of synthesized circuits. This process can be either performed prior to synthesis or in a unified approach during synthesis [15].

#### 2.3.1.2 Synthesis

Synthesis seeks reversible circuits that satisfy a reversible specification. It can be performed optimally or heuristically. Many synthesis methods are given in literature like transformation based, search based, cycle based, BDD based, programming based etc.

1. **Transformation-based method:** In 2003 D. M. Miller, D. Maslov, and G. W. Dueck [16] proposed the Transformation Based Method which was synthesis of reversible circuit in terms of n x n Toffoli gates. Iteratively select a gate so as to make a function’s truth table or RM spectrum more similar to the identity function. These methods are mainly efficient for permutations where output codeword follow a regular (repeating) pattern.

2. **Cycle-based method:** In 2003, M. Saeedi, MS Zamani, M Sedighi [17, 18] proposed a Cycle Based Method. This method decomposes a given permutation into a set of disjoint (often small) cycles and synthesizes individual cycles separately. Compared to other algorithms, these methods are mainly efficient for permutations without regular patterns and reversible functions that leave many input combinations unchanged.

3. **Search-based method:** In 2006, P. Gupta, A Agarwal, NK Jha [19] proposed a Search Based Method which traverses a search tree to find a reasonably good circuit. These methods mainly use the PPRM expansion to represent a reversible function. The efficiency of these methods is highly dependent on the number of circuit lines and the number of gates in the final circuit.

4. **ESOP-based method:** In 2007, K. Fazel, M. Thornton, and J. Rice [20] proposed an ESOP Based Method. The algorithm was capable of generating a cascade of reversible gates for logic functions with large numbers of qubits. The algorithm was fast as it uses a simple cost metric heuristic during a recursive divide-and-conquer function to determine NOT and Toffoli gate placement.

5. **BDD-based method:** In 2009 Robert Wille, Rolf Drechsler [21] proposed a BDD Based Method. The basic idea was to create a Binary Decision Diagram for the function to be synthesized and afterwards substituting each node by a cascade of Toffoli or elementary quantum gates, respectively. This approach can synthesize circuits for functions with more than hundred variables in just a few CPU seconds. Use binary decision diagrams to improve sharing between controls of reversible gates. These techniques scale better than others. However, they require a large number of ancilla qubits — a valuable resource in fledgling quantum computers.

6. **Programming Language based Synthesis:** All these synthesis approaches are available that relies on Boolean function representations, thus do not allow the design of complex reversible system. Consequently, higher levels of abstractions have been considered leading to the development of hardware description languages. In 2010, Robert Wille, Sebastian Offermann, Rolf Drechsler [22] proposed the programming language SyReC that allows to specify and afterwards to automatically synthesize reversible circuits. SyReC is based on the reversible software language Janus [23], which has been enriched by further concepts, new operations, and some restrictions. It provides fundamental constructs to define control and data operations, while still preserving reversibility. Detailed discussion of research work about presented in sec. 2.5.

#### 2.3.1.3 Post-synthesis Optimization

The results obtained by heuristic synthesis methods are often sub-optimal. Further improvements can be achieved by local optimization.

- **Improving gate count and quantum cost [24]:** To improve the quantum cost of a circuit, several techniques attempt to improve individual sub-circuits one at a time. Sub-circuit optimization may be performed based on offline synthesis of a set of functions using pre-computed tables, online synthesis of candidates, or circuit transformations that involve additional ancillae.

- **Reducing circuit depth [25]:** To realize a low-depth implementation of a given function, consecutive elementary gates with disjoint sets of control and target lines should be used to provide the possibility of parallel gate execution. Circuit depth may also be improved by restructuring controls and targets of different gates in a synthesized circuit.

- **Improving locality [26]:** For the implementation of a given computation on quantum architecture with restricted qubit interactions, one may use SWAP gates to move gate qubits towards each other as much as required. The interaction cost of a given computation can be hand-optimized for particular applications. A generic approach can also be used to either reduce the number of SWAP gates or find the minimal number of SWAP gates for a circuit.

## 2.4 Tools

- **Rev Kit:** Rev Kit is open source toolkit [27]. It provides functionality like parsers, export function, cost calculation etc but also elaborated methods for synthesis, optimization, and verification of reversible (and quantum) circuits. It accepts reversible circuit in PLA form.

- **RC Viewer:** It is circuit viewer tool, accepts circuit in *.real form. Its improved version is introduced with the name of RC Viewer+.

These tools are not user friendly for developing reversible circuits. Some works have been done by our department which is as follows:

- **RCDEV (2012):** It is developed by Nitin Purohit and Dr. S.C. Jain. Its features are:
  - Creation of Schematic with standard libraries
  - Editing of generated circuit
  - Save and retrieve developed circuit
  - Inter conversions of different design entry format
  - Joining different circuit given in different specifications
  - Partitioning and Reversing a circuit
  - Reversibility verifier
  - Easy to learn and user friendly

- **RCTEST (2013):** It is developed by Anugrah Jain and Dr. S.C. Jain. Its features are:
  - Reversible circuit schematic display
  - Parity-Preservation of the reversible circuit
  - Generation of a reversible circuit from ESOP synthesis
  - Simulation of ESOP based online testable reversible circuits
  - Implementation of Toffoli based online testable approach
  - Simulation of the proposed extension for a reversible circuit
  - Convert a parity-preserving reversible circuit into an online testable reversible circuit

- Simulate the testing of all single-bit faults occurred in a testable reversible circuit

- **RPGASim(2013):** It is developed by Pankaj Kumar Israni and Dr. S.C. Jain. Its features are:
  - Entire project is implemented in java platform
  - To develop simulation tool using RPGA structure that can simulate any reversible circuit
  - Symmetry analyzer for any reversible circuit
  - The tool can generate RPGA structure of any given input
  - Generate response for any symmetry circuit
  - Can run in step by step mode

## 2.5 Hardware Description Language SyReC

The scalability of all these approaches like transformation-based method, search-based method, BDD-based method etc. is limited, i.e. the methods are applicable for relatively small functions only and they rely on Boolean description, thus, do not allow the design of complex reversible systems. In 2010, Robert Wille, Sebastian Offermann, Rolf Drechsler [22, 28] proposed the programming language SyReC that allows to specify and afterwards to automatically synthesize reversible circuits. It is based on the reversible software language Janus [23], which has been enriched by further concepts (e.g. declaring circuit signals of different bit-widths), new operations (e.g. bit-access and shifts), and some restrictions (e.g. the prohibition of dynamic loops). It provides fundamental constructs to define control and data operations, while still preserving reversibility. The syntax of a SyReC specification is outlined in Figure 2.8 and briefly described in the following:

### 2.5.1 Module and Signal Declarations

Each SyReC program (denoted by <program>) consists of one or more modules (denoted by <module>) [28]. A module is introduced with the keyword **module** and includes an identifier (represented by a string), a list of parameters representing global signals (denoted by <parameter-list>), local signal declarations (denoted by <signal-list>), and a sequence of statements (denoted by <statement-list>). The top-module of a program is defined by the special identifier **main**. If no module with this name exists, the last module declared is used instead by convention.

SyReC uses a signal representing a non-negative integer as its sole data type. The bit width of signals can optionally be defined by round brackets after the signal name. If no bit-width is specified, a default value is assumed. For each signal, an access modifier has to be given. For a parameter signal (used in a module declaration) this can be **in**, **out**, and **inout**. Local signals can either work as internal signals (denoted by **wire**) or in case of sequential circuits as state signals2 (denoted by **state**). The access modifier influences properties in the synthesized circuits as given in Table 2.1.

**Table 2.1: Signal access modifiers and implied circuit properties**

| **Modifier** | **Constant Value** | **Garbage** | **State** | **Initial Value** |
|---|---|---|---|---|
| in | - | yes | No | given by primary input |
| out | 0 | no | No | 0 |
| inout | - | no | No | given by primary input |
| wire | 0 | yes | No | 0 |
| state | - | no | Yes | given by pseudo-primary input |

Signals can be grouped to multi-dimensional arrays of constant length using square brackets after the signal name and before the optional bit-width declaration.

**Program and Modules**

‹program› ::= ‹module› {‹module›}

‹module› ::= ‘**module**’ ‹identifier› ‘**(**’ [‹parameter-list›] ‘**)**’ {‹signal-list›} ‹statement-list›

‹parameter-list› ::= ‹parameter› {‘**,**’ ‹parameter›}

‹parameter› ::= (‘**in**’ | ‘**out**’ | ‘**inout**’) ‹signal-declaration›

‹signal-list› ::= (‘**wire**’ | ‘**state**’) ‹signal-declaration› {‘**,**’ ‹signal-declaration›}

‹signal-declaration› ::= ‹identifier› {‘**[**’‹int›‘**]**’} [‘**(**’‹int›‘**)**’]

**Statements**

‹statement-list› ::= ‹statement› {‘**;**’ ‹statement›}

‹statement› ::= ‹call-statement› | ‹for-statement› | ‹if-statement› | ‹unary-statement› | ‹assign-statement› | ‹swap-statement› | ‹skip-statement›

‹call-statement› ::= (‘**call**’ | ‘**uncall**’) ‹identifier› ‘**(**’ (‹identifier› {‘**,**’ ‹identifier›}) ‘**)**’

‹for-statement› ::= ‘**for**’ [[‘**$**’ ‹identifier› ‘**=**’] ‹number› ‘**to**’] ‹number› [‘**step**’ [‘**-**’] ‹number›] ‹statement-list› ‘**rof**’

‹if-statement› ::= ‘**if**’ ‹expression› ‘**then**’ ‹statement-list› ‘**else**’ ‹statement-list› ‘**fi**’ ‹expression›

‹unary-statement› ::= (‘~’ | ‘++’ | ‘--’) ‘**=**’ ‹signal›

‹assign-statement› ::= ‹signal› (‘^’ | ‘+’ | ‘-’) ‘**=**’ ‹expression›

‹swap-statement› ::= ‹signal› ‘<=>’ ‹signal›

‹skip-statement› ::= ‘**skip**’

‹signal› ::= ‹identifier› {‘**[**’ ‹expression› ‘**]**’} [‘**.**’ ‹number› [‘**:**’ ‹number›]]

**Expressions**

‹expression› ::= ‹number› | ‹signal› | ‹binary-expression› | ‹unary-expression› | ‹shift-expression›

‹binary-expression› ::= ‘**(**’ ‹expression› (‘+’ | ‘-’ | ‘^’ | ‘*’ | ‘/’ | ‘**&&**’ | ‘||’ | ‘**&**’ | ‘|’ | ‘<’ | ‘>’ | ‘**=**’ | ‘**!=**’ | ‘<=’ | ‘>=’) ‹expression› ‘**)**’

‹unary-expression› ::= (‘**!**’ | ‘~’) ‹expression›

‹shift-expression› ::= ‘**(**’ ‹expression› (‘<<’ | ‘>>’) ‹number› ‘**)**’

**Data-types**

‹letter› ::= (‘**A**’ | . . . | ‘**Z**’ | ‘**a**’ | . . . | ‘**z**’)

‹digit› ::= (‘**0**’ | . . . | ‘**9**’)

‹identifier› ::= (‘ _’ | ‹letter›) {(‘ ’ | ‹letter› | ‹digit›)}

‹int› ::= ‹digit› {‹digit›}

‹number› ::= ‹int› | ‘#’ ‹identifier› | ‘**$**’ ‹identifier› | (‘**(**’ ‹number› (‘+’ | ‘-’ | ‘*’ | ‘/’) ‹number› ‘**)**’)

**Figure 2.8: Syntax of the hardware language SyReC**

**Examples**

```
// module declaration with two inputs and one output:
module adder( in a, in b, out c )

// the same declaration with 16-bit signals:
module adder( in a(16), in b(16), out c(16) )

// an adder summing up 4 inputs given as an array:
module adder( in inputs[4](16), out c(16) )
```

### 2.5.2 Statements

Statements include call and uncall of other modules, loops, conditional statements, and various data operations (i.e. unary operations, reversible assignment operations, swap statements). The empty statement can explicitly be modeled using the **skip** keyword. Statements are separated by semicolons. Signals within statements are denoted by <signal> allowing access to the whole signal (e.g. x), a certain bit (e.g. x.4), or a range of bits (e.g. x.2:4). The bit-width of a signal can also be accessed (e.g. #x).

To call another module, simply the keyword **call** (**uncall**) together with the identifier of the module to be called along with the parameters have to be applied.

**Example**

```
// calling a module identified by adder
wire a, b, c
call adder( a, b, c )
```

In loops, the number of iterations (and therewith the number of duplications of the respective code block) needs to be defined. This number has to be available prior to the compilation, i.e., dynamic loops are not allowed. Therefore, e.g. fix integer values, the bit-width of a signal, or internal **$**-variables can be applied. Furthermore, the current value of internal counter variables

can be accessed during the iterations. Using the optional keyword **step**, also the iteration itself can be modified. A loop is terminated by **rof**.

**Examples**

```
for 1 to 10 do
// statements
rof

// iterating over the bit-width of a variable
wire x
for $i = 0 to #x do
/* statements (possibly using $i)
the i-th bit of x can be accessed by x. $i
a range of bits can be accessed e .g. by x .0: $i */
rof

for $counter = 1 to 10 step 2 do
// statements
// the loops iterates 5 times ( i. e., $counter is set to 1, 3, 5, 7, and 9 only)
rof
```

Conditional statements need an expression to be evaluated followed by the respective then-block and else-block. Each of these blocks is a sequence of statements. In order to ensure reversibility, a conditional statement is terminated by **fi** and an adjusted expression.

**Example**

```
if ( x = 5 ) then
a + = b // statements that are executed if x = 5
else
a - = b // statements that are executed if x! = 5
fi ( x = 5 )
```

Finally, statements can express various data operations. Operations are thereby distinguished between reversible assignment operations (denoted by <assign-statement>), unary operations (denoted by <unary-statement>), the swap operation (denoted by <swap-statement>), and the not necessarily reversible binary operations (denoted by <binary expression>) as well as the shift operations (denoted by <shift-expression>). Table 2.2 and 2.3 lists all statements and expressions, respectively, along with their semantics. Variable accesses are referred to as x and y, expressions as e and f, and natural numbers as n.

**Table 2.2: Statements in SyReC**

| Operation | Semantic |
|---|---|
| ˜ = x | Bit-wise negation of x |
| + + = x | Increment of x |
| − − = x | Decrement of x |
| x ^ = e | Bit-wise XOR assignment of e to x, i.e. $x \leftarrow x \oplus e$ |
| x += e | Increase by value of e to x, i.e. $x \leftarrow x + e$ |
| x −= e | Decrease by value of e to x, i.e. $x \leftarrow x + e$ |
| x <=> $y$ | Swapping value of x with value of y |

**Table 2.3: Expressions in SyReC**

| Operation | Semantic |
|---|---|
| e + f | Addition of e and f |
| e − f | Subtraction of e and f |
| e ∗ f | Multiplication of e and f |
| e / f | Division of e and f |
| e ˆ f | Bit-wise XOR of e and f |
| e & $f$ | Bit-wise AND of e and f |
| $e \mid f$ | Bit-wise OR of e and f |
| ˜e | Bit-wise negation of e |
| e && $f$ | Logical AND of e and f |
| $e \mid\mid f$ | Logical OR of e and f |
| ! e | Logical NOT of e |
| e < $f$ | True, if and only if e is less than f |
| e > $f$ | True, if and only if e is greater than f |
| e = f | True, if and only if e equals f |
| e != f | True, if and only if e not equals f |
| e <= $f$ | True, if and only if e is less or equal to f |
| e >= $f$ | True, if and only if e is greater or equal to f |
| e << $n$ | Logical left shift of e by n |
| e >> $n$ | Logical right shift of e by n |

Reversible assignment operations assign values to a signal on the left-hand side. Therefore, the respective signal must not appear in the expression on the right-hand side. Furthermore, only a restricted set of assignment operations exists, namely increase($+=$), decrease($-=$), bit-wise XOR($\hat{}=$). These operations preserve the reversibility (i.e., it is possible to compute these operations in both directions). The same holds for the unary operations, namely bit-wise negation($\tilde{}$), increase by one($++$), and decrease by one($--$), as well as for the swap operation($<=>$).

In contrast, binary operations, i.e., arithmetic($+,-,*$), bit-wise($\&, |, \hat{}$), logical($\&\&, ||$), relational($<, >, =, != , <=, >=$), and shifting($<<, >>$) operations, may not be reversible. Thus, they can only be used in right-hand expressions (denoted by <expression>) which preserve, i.e., do not modify, the values of the respective inputs. In doing so, all computations remain reversible since the input values can be applied to reverse any operation. For example, to specify the multiplication $a * b$ in SyReC, a new free signal c must be introduced which is used to store the product. That result in the expression $\mathrm{c}\hat{} = \mathrm{a} * \mathrm{b}$. In comparison to common (irreversible) programming languages, statements such as $\mathrm{a} = \mathrm{b} + (5 * \mathrm{a})$ are not allowed.

### 2.5.3 Examples

Using SyReC, complex reversible circuits can be specified. Some examples are provided in the following.

**Program Counter**

```
module program_counter (in reset(1), in inc(1), in jmp(2), inout  pc(2))
wire zero(2)

 if ( reset = 1 ) then
        pc <=> zero
 else
        if ( inc = 1 ) then
                pc += 1
         else
                pc <=> jmp
        fi ( inc = 1 )
fi ( reset = 1 )
```

**Arithmetic Logic Unit**

```
module alu ( in op(2), in a, in b, out c )
 if ( op = 0 ) then
    c ^ = ( a + b )
 else
    if ( op = 1 ) then
          c ^ = ( a − b )
   else
          if ( op = 2 ) then
                  c ^ = ( a ∗ b )
          else
                  c ^ = a
          fi ( op = 2 )
    fi ( op = 1 )
fi ( op = 0 )
```

Driven by its promising applications, reversible logic received significant attention. As a result, an impressive progress had been made in the development of synthesis approaches, implementation of sequential elements, and hardware description languages. In 2011, Robert Wille, Mathias Soeken, Daniel Große, Eleonora Schonborn and Rolf Drechsler [30], these recent achievements were employed in order to design a RISC CPU in reversible logic that can execute software programs written in an assembler language. The respective combinational and sequential components are designed using state-of-the-art design techniques. However, existing HDL synthesizers lead to circuits with a significant number of additional lines. So in 2012, Robert Wille, Mathias Soeken, Eleonora Schonborn, Rolf Drechsler[31], focused on the reduction of additional circuit lines which were caused by buffering intermediate results. They proposed an approach that reuses these lines as soon as the intermediate results were not required anymore. Experiments confirm that this approach decreases the number of circuit lines by up to two orders of magnitude and 60% on average. So far, existing methods realize control logic with a significant amount of redundant circuit structures. So in 2012, Sebastian Offerman, Robert

Wille, and Rolf Drechsler [32] presented that avoids large parts of these redundancies by buffering the results of recurring computations in one additional circuit line. Accordingly, the proposed approach enables to realize control logic with significantly less circuit lines, while the increase of the circuit cost.

## 2.6 Survey Extraction

Reversible computing is emerging as an alternative technology, followed by requirements of development tools. Synthesis of reversible circuits should be considered from all aspects. Use of an HDL, dedicated for reversible synthesis is necessary and has been satisfied to some extent through SyReC. Yet the concepts of sequential reversible circuit remain to be introduced. Instead of inventing a fully equipped HDL for reversible circuit from scratch, we could enrich the existing tool set of SyReC. Specifically it would be desirable if a circuit, once defined as a SyReC specification could be realized into a reversible circuit and be expanded a reversible circuit and generate a respective SyReC code. This can definitely be achieved through a software tool.

**Chapter 3**

# CIRCUIT REALIZATIONS FROM SYREC

In this chapter we discuss how various constructs of SyReC language can be realized into reversible circuits. These would go further to form a background library from which circuit components for a complete circuit specified in SyReC can be picked. We begin by distinguishing in reversible assignment operations, binary operations, unary operations and swap operation, along with the respective reversible circuit realizations and finally describe the realization of control operations as reversible cascades.

## 3.1 Reversible Assignment Operations

Reversible assignment operations include those which are reversible even if they assign a new value to the variable on the left-hand side of a statement. In the following, we use the notation depicted in Figure 3.1 to denote such an operation in a circuit structure. Solid lines represent the variable(s) on the right-hand side of the operation, i.e. the variable(s) whose values are preserved.

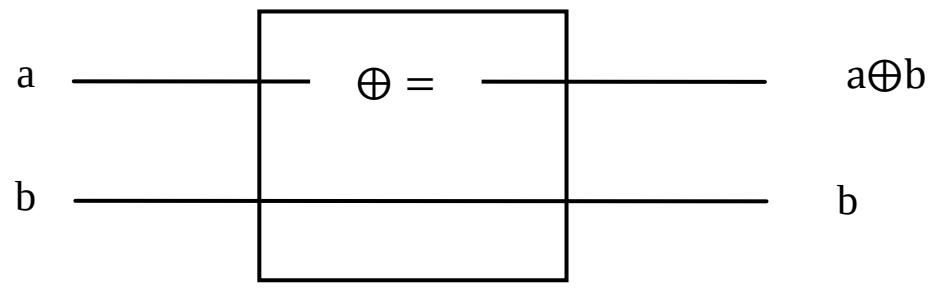

**Figure 3.1: General realization of reversible assignment operation: a ⊕= b**

Given below are the specific reversible assignment operations for varying bit-size of operands:

### 3.1.1 Increase Operation

Operation denoted as a+= b means increasing value of a by b, i.e. $a \leftarrow a + b$. The truth table for 1-bit operands is given in table 3.1 and reversible circuit in figure 3.2.

**Table 3.1: Truth table of a+= b (1-bit)**

| Input | | Output | |
|---|---|---|---|
| **a** | **b** | **a+= b** | **b** |
| 0 | 0 | 1 | 0 |
| 0 | 1 | 0 | 1 |
| 1 | 0 | 0 | 0 |
| 1 | 1 | 1 | 1 |

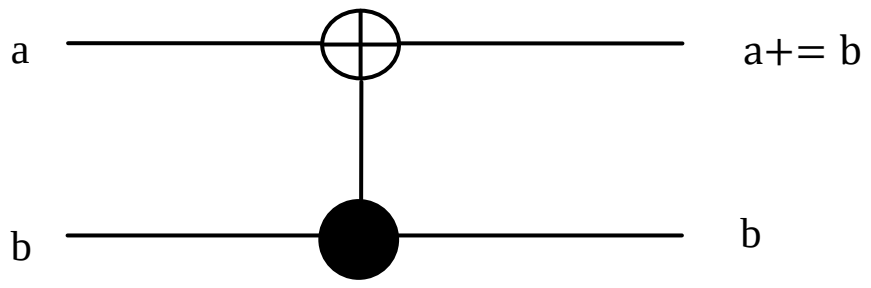

**Figure 3.2: Realization of a+= b (1-bit)**

When operands are 2-bit wide the truth table is given in table 3.2 and the reversible circuit in figure 3.3.Note that the lower most circuit line in figure 3.3 indicates how the circuit can be expanded when number of bits in the operands increased.

**Table 3.2: Truth table of a+= b (2-bit)**

| Inputs | | | | Outputs | | | |
|---|---|---|---|---|---|---|---|
| **b1** | **b0** | **a1** | **a0** | **b1** | **b0** | **a1'** | **a0'** |
| 0 | 0 | 0 | 0 | 0 | 0 | 0 | 0 |
| 0 | 0 | 0 | 1 | 0 | 0 | 0 | 1 |
| 0 | 0 | 1 | 0 | 0 | 0 | 1 | 0 |
| 0 | 0 | 1 | 1 | 0 | 0 | 1 | 1 |
| 0 | 1 | 0 | 0 | 0 | 1 | 0 | 1 |
| 0 | 1 | 0 | 1 | 0 | 1 | 1 | 0 |
| 0 | 1 | 1 | 0 | 0 | 1 | 1 | 1 |
| 0 | 1 | 1 | 1 | 0 | 1 | 0 | 0 |
| 1 | 0 | 0 | 0 | 1 | 0 | 1 | 0 |
| 1 | 0 | 0 | 1 | 1 | 0 | 1 | 1 |
| 1 | 0 | 1 | 0 | 1 | 0 | 0 | 0 |
| 1 | 0 | 1 | 1 | 1 | 0 | 0 | 1 |
| 1 | 1 | 0 | 0 | 1 | 1 | 1 | 1 |
| 1 | 1 | 0 | 1 | 1 | 1 | 0 | 0 |
| 1 | 1 | 1 | 0 | 1 | 1 | 0 | 1 |
| 1 | 1 | 1 | 1 | 1 | 1 | 1 | 0 |

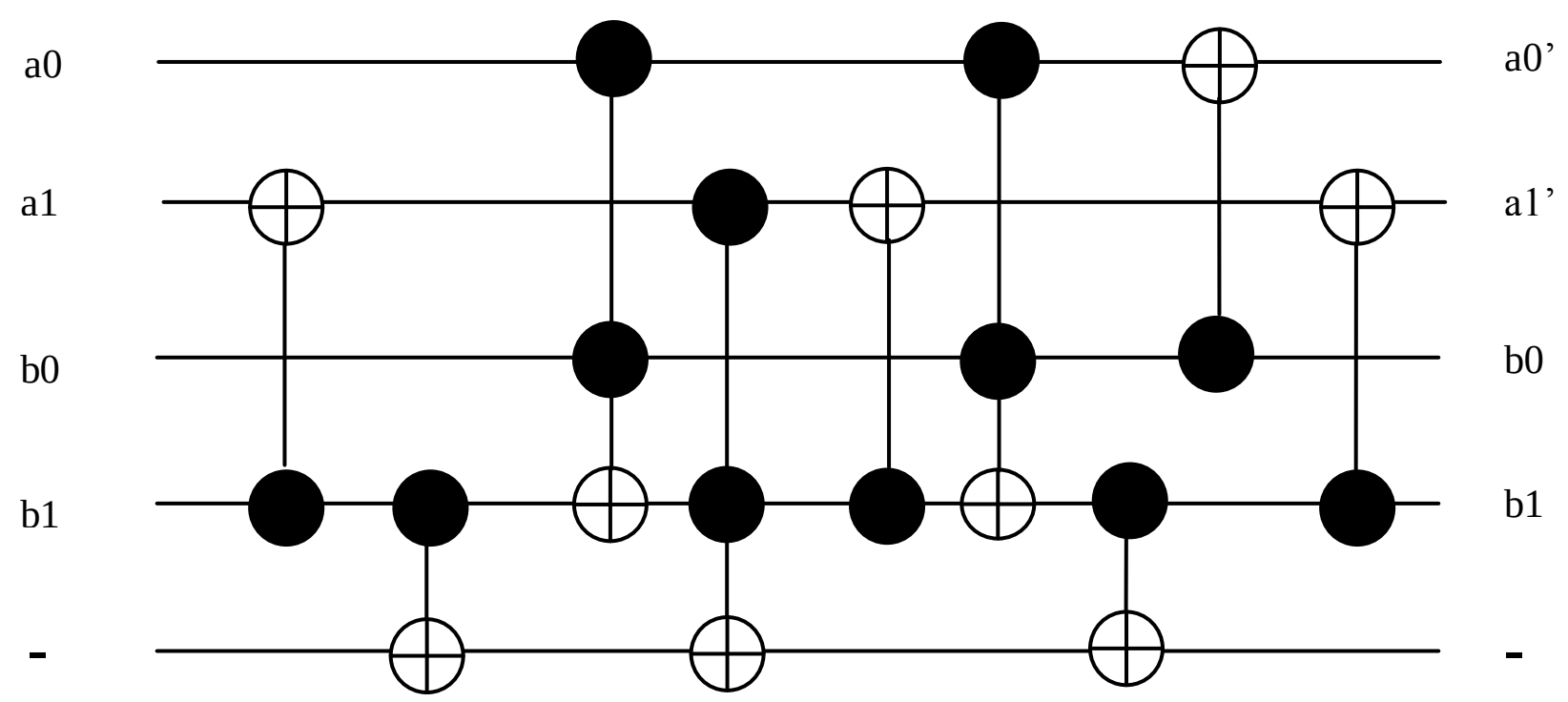

**Figure 3.3: Realization of a+= b (2-bit)**

### 3.1.2 Decrease Operation

Operation denoted as a − = b means decreasing value of a by b, i.e. a ← a − b. The truth table and the reversible circuit of decrease operation for 1-bit operand are same as increase operation for 1-bit operand. When operands are 2-bit wide the truth table is given in table 3.3 and the reversible circuit in figure 3.4.Note that the lower most circuit line in figure 3.4 indicates how the circuit can be expanded when number of bits in the operands increased.

**Table 3.3: Truth table of a−= b (2-bit)**

| **Inputs** | | | | **Outputs** | | | |
|---|---|---|---|---|---|---|---|
| **b1** | **b0** | **a1** | **a0** | **b1** | **b0** | **a1'** | **a0'** |
| 0 | 0 | 0 | 0 | 0 | 0 | 0 | 0 |
| 0 | 0 | 0 | 1 | 0 | 0 | 0 | 1 |
| 0 | 0 | 1 | 0 | 0 | 0 | 1 | 0 |
| 0 | 0 | 1 | 1 | 0 | 0 | 1 | 1 |
| 0 | 1 | 0 | 0 | 0 | 1 | 1 | 1 |
| 0 | 1 | 0 | 1 | 0 | 1 | 0 | 0 |
| 0 | 1 | 1 | 0 | 0 | 1 | 0 | 1 |
| 0 | 1 | 1 | 1 | 0 | 1 | 1 | 0 |
| 1 | 0 | 0 | 0 | 1 | 0 | 1 | 0 |
| 1 | 0 | 0 | 1 | 1 | 0 | 1 | 1 |
| 1 | 0 | 1 | 0 | 1 | 0 | 0 | 0 |
| 1 | 0 | 1 | 1 | 1 | 0 | 0 | 1 |
| 1 | 1 | 0 | 0 | 1 | 1 | 0 | 1 |
| 1 | 1 | 0 | 1 | 1 | 1 | 1 | 0 |
| 1 | 1 | 1 | 0 | 1 | 1 | 1 | 1 |
| 1 | 1 | 1 | 1 | 1 | 1 | 0 | 0 |

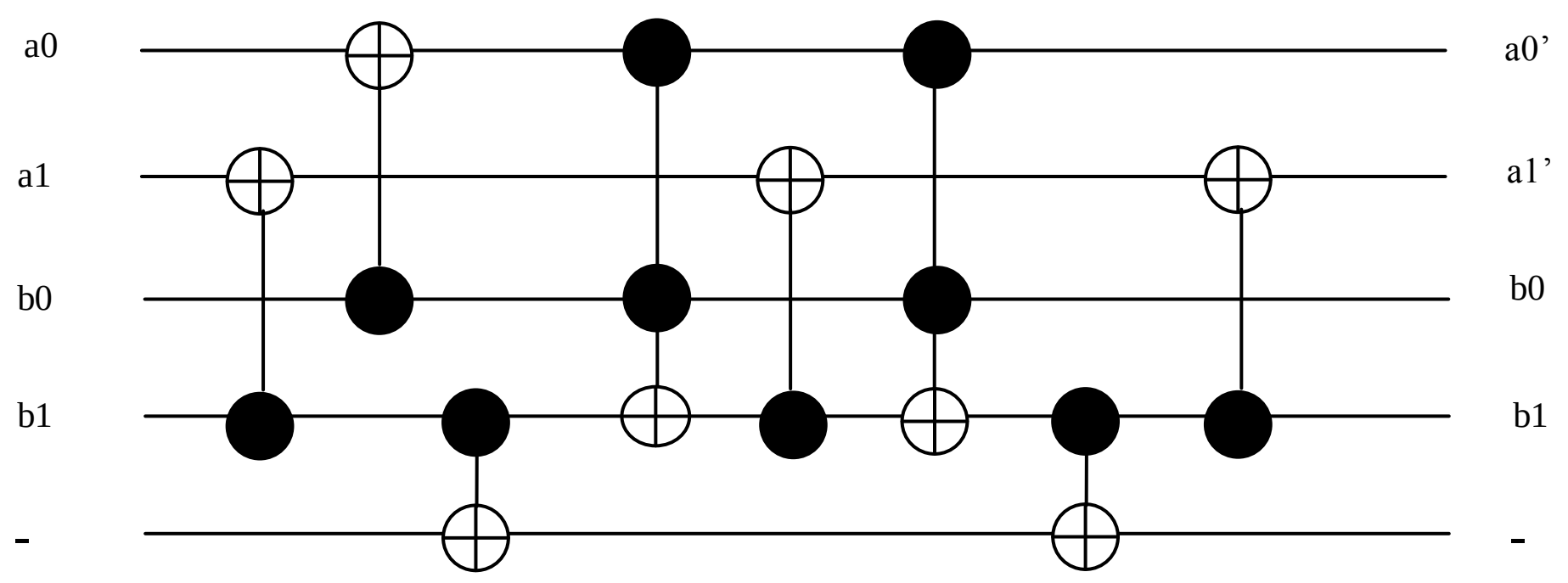

**Figure 3.4: Realization of a−= b (2-bit)**

### 3.1.3 XOR Operation

Operation denoted as a ^ = b means bitwise XOR assignment of b to a, i.e. a ← a^b. The truth table and the reversible circuit of bitwise XOR operation for 1-bit operand are same as increase operation for 1-bit operand. When operands are 2-bit wide the truth table is given in table 3.4 and the reversible circuit in figure 3.5.

**Table 3.4: Truth table of a^ = b (2-bit)**

| **Inputs** | | | | **Outputs** | | | |
|---|---|---|---|---|---|---|---|
| **b1** | **b0** | **a1** | **a0** | **b1** | **b0** | **a1'** | **a0'** |
| 0 | 0 | 0 | 0 | 0 | 0 | 0 | 0 |
| 0 | 0 | 0 | 1 | 0 | 0 | 1 | 0 |
| 0 | 0 | 1 | 0 | 0 | 0 | 0 | 1 |
| 0 | 0 | 1 | 1 | 0 | 0 | 1 | 1 |
| 0 | 1 | 0 | 0 | 0 | 1 | 1 | 0 |
| 0 | 1 | 0 | 1 | 0 | 1 | 0 | 0 |
| 0 | 1 | 1 | 0 | 0 | 1 | 1 | 1 |
| 0 | 1 | 1 | 1 | 0 | 1 | 0 | 1 |
| 1 | 0 | 0 | 0 | 1 | 0 | 0 | 1 |
| 1 | 0 | 0 | 1 | 1 | 0 | 1 | 1 |
| 1 | 0 | 1 | 0 | 1 | 0 | 0 | 0 |
| 1 | 0 | 1 | 1 | 1 | 0 | 1 | 0 |
| 1 | 1 | 0 | 0 | 1 | 1 | 1 | 1 |
| 1 | 1 | 0 | 1 | 1 | 1 | 0 | 1 |
| 1 | 1 | 1 | 0 | 1 | 1 | 1 | 0 |
| 1 | 1 | 1 | 1 | 1 | 1 | 0 | 0 |

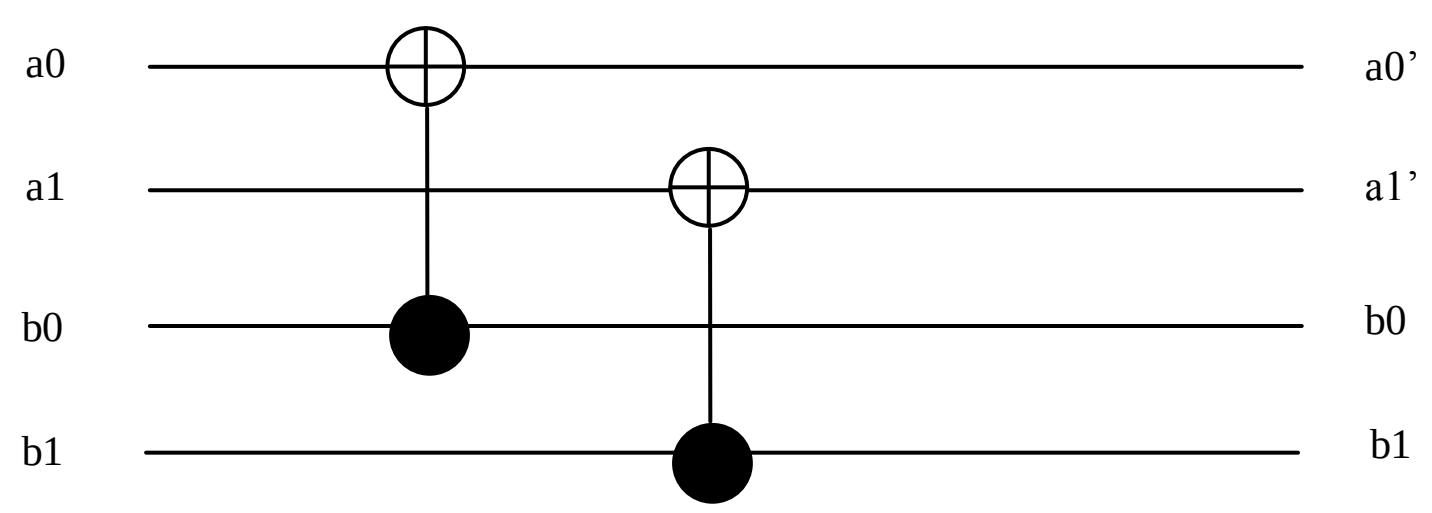

**Figure 3.5: Realization of a^ = b (2-bit)**

## 3.2 Binary Operations

Binary operations include operations that are not necessarily reversible so that its inputs have to be preserved to allow a (reversible) computation in both directions. To denote such operations, the notation depicted in Figure 3.6 is used. Again here, solid lines represent the variable(s) whose values are preserved (i.e. in this case the input variables).

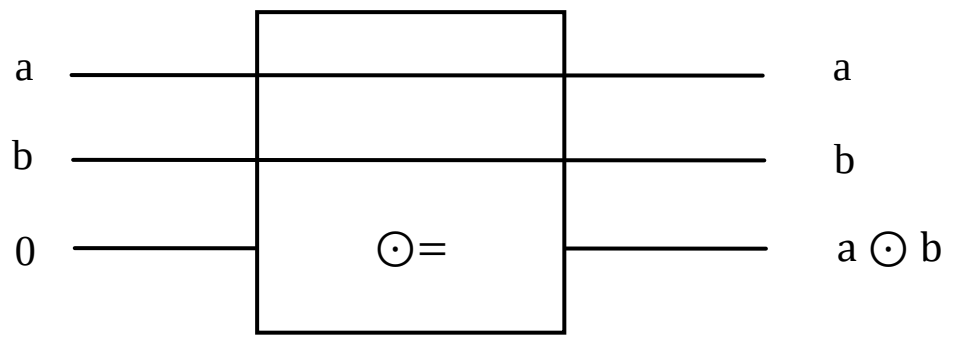

**Figure 3.6: General realization of binary assignment operation: a⊙ = b**

Now we list various binary operations and include circuits for 1-bit and 2-bit operands in following subsections.

### 3.2.1 Bitwise Operations

**i) Bitwise AND operation:** It is denoted in SyReC as a&*b*, implying Bitwise AND of a and b. The truth table for 1-bit operands is given in table 3.5 and reversible circuit in figure 3.7.

**Table 3.5: Truth table of a&b (1-bit)**

| Input | | | Output | | |
|---|---|---|---|---|---|
| **a** | **b** | **0** | **a** | **b** | **a&b** |
| 0 | 0 | 0 | 0 | 0 | 0 |
| 0 | 1 | 0 | 0 | 1 | 0 |
| 1 | 0 | 0 | 1 | 0 | 0 |
| 1 | 1 | 0 | 1 | 1 | 1 |

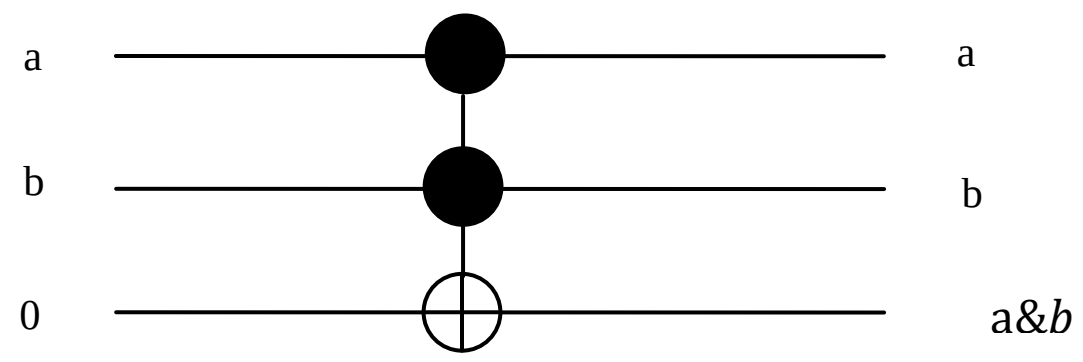

**Figure 3.7: Realization of a&b (1-bit)**

When operands are 2-bit wide the circuit for a&$b$ is realized as shown in figure 3.8.

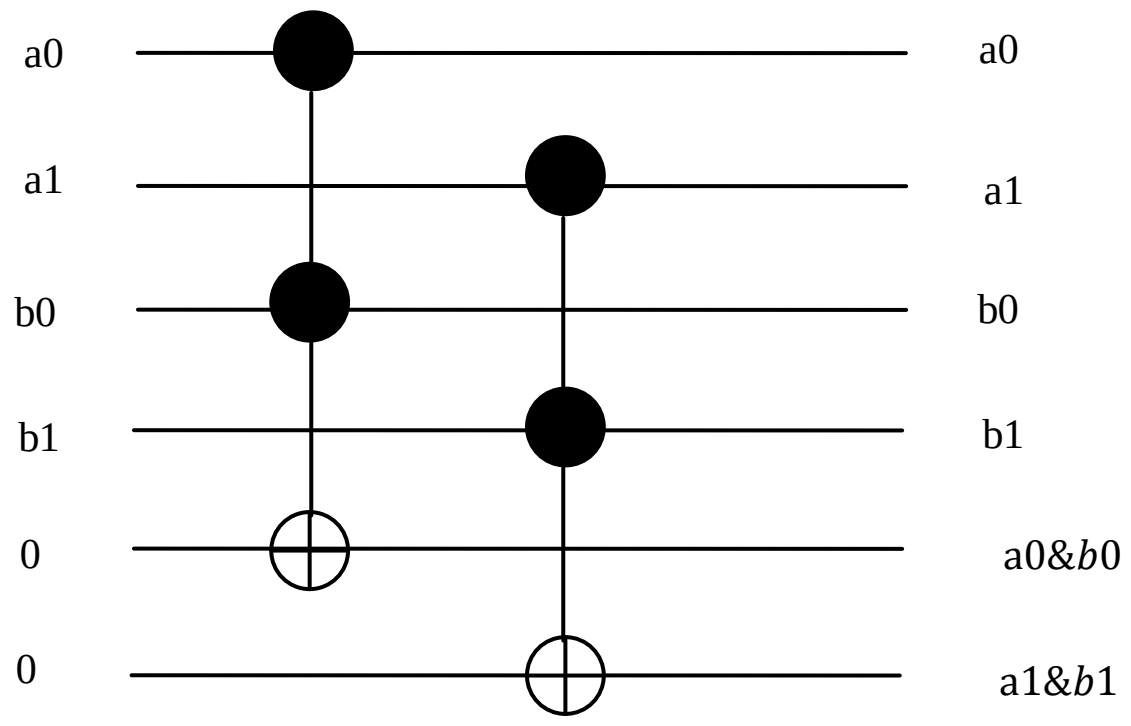

**Figure 3.8: Realization of a&b (2-bit)**

**ii) Bitwise OR operation:** It is denoted in SyReC as a|$b$, implying Bitwise OR of a and b. The truth table for 1-bit operands is given in table 3.6 and reversible circuit in figure 3.9.

**Table 3.6: Truth table of a|b (1-bit)**

| **Input** | | | **Output** | | |
|---|---|---|---|---|---|
| **A** | **b** | **0** | **a** | **b** | **a\|b** |
| 0 | 0 | 0 | 0 | 0 | 0 |
| 0 | 1 | 0 | 0 | 1 | 1 |
| 1 | 0 | 0 | 1 | 0 | 1 |
| 1 | 1 | 0 | 1 | 1 | 1 |

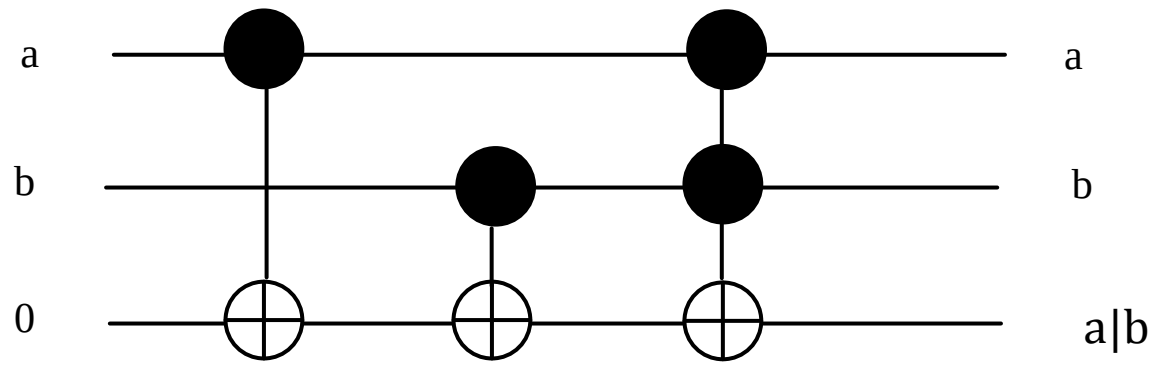

**Figure 3.9: Realization of a|b (1-bit)**

When operands are 2-bit wide the circuit for a|$b$ is realized as shown in figure 3.10.

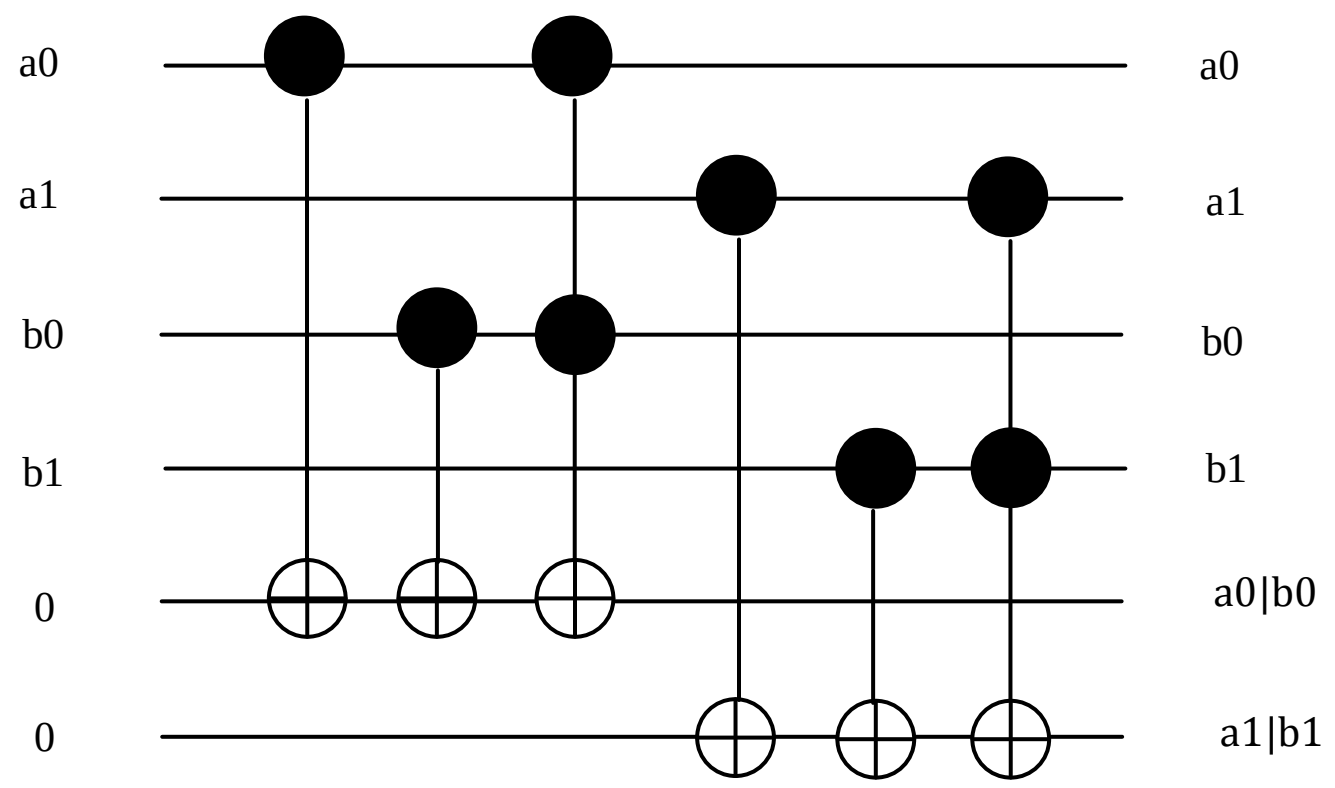

**Figure 3.10: Realization of a|b (2-bit)**

**iii) Bitwise XOR operation:** It is denoted in SyReC as a^$b$, implying Bitwise OR of a and b. The truth table for 1-bit operands is given in table 3.7 and reversible circuit in figure 3.11.

**Table 3.7: Truth table of a^b (1-bit)**

| **Input** | | | **Output** | | |
|---|---|---|---|---|---|
| **A** | **b** | **0** | **a** | **b** | **a^b** |
| 0 | 0 | 0 | 0 | 0 | 0 |
| 0 | 1 | 0 | 0 | 1 | 1 |
| 1 | 0 | 0 | 1 | 0 | 1 |
| 1 | 1 | 0 | 1 | 1 | 0 |

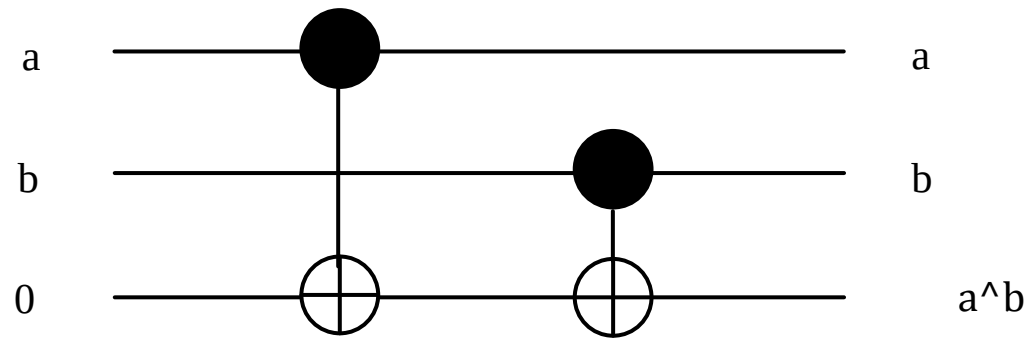

**Figure 3.11: Realization of a^b (1-bit)**

When operands are 2-bit wide the circuit for a^$b$ is realized as shown in figure 3.12.

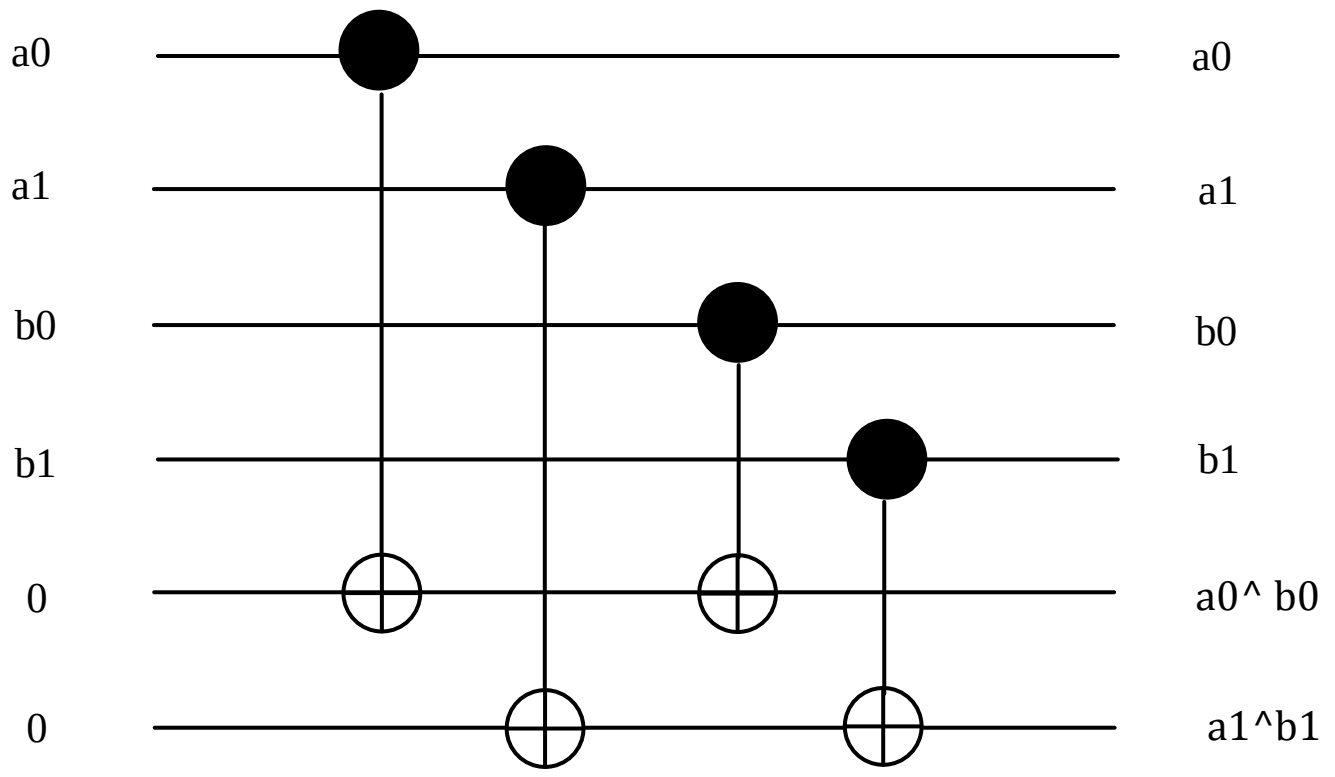

**Figure 3.12: Realization of a^b (2-bit)**

**iv) Bitwise Negation:** It is denoted in SyReC as~a, implying Bitwise negation of a. The truth table for 1-bit operand is given in table 3.8 and reversible circuit is given in figure 3.13.

**Table 3.8: Truth table of ~a (1-bit)**

| Input | Output |
|---|---|
| **a** | **a'** |
| 0 | 1 |
| 1 | 0 |

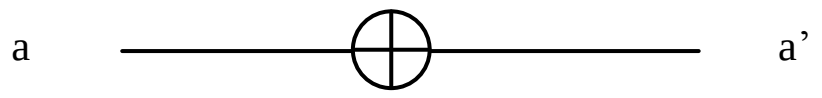

**Figure 3.13: Realization of ~a (1-bit)**

When operands are 2-bit wide the circuit for ~a is realized as shown in figure 3.14.

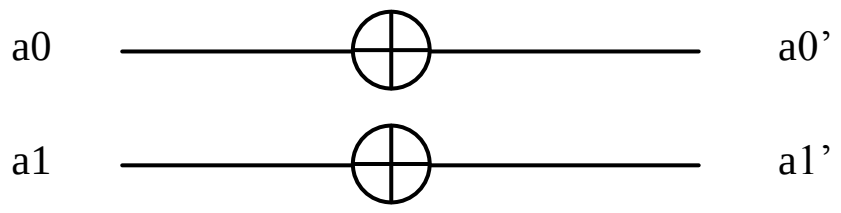

**Figure 3.14: Realization of ~a (2-bit)**

### 3.2.2 Comparison Operations

While using relative operators between 2-operands the returned value is Boolean. A comparison like less than, greater than, equals and combinations are possible in SyReC. In this section we present the SyReC statements for relative operators and corresponding reversible circuit for 1-bit operands.

**i) Less Than:** Denoted as a < b, implying the output is True, if and only if a less than b.The truth table is given in table 3.9 and reversible circuit in figure 3.15.

**Table 3.9: Truth table of a < b**

| **Input** | | | **Output** | | |
|---|---|---|---|---|---|
| **a** | **b** | **0** | **a** | **b** | **a < b** |
| 0 | 0 | 0 | 0 | 0 | 0 |
| 0 | 1 | 0 | 0 | 1 | 1 |
| 1 | 0 | 0 | 1 | 0 | 0 |
| 1 | 1 | 0 | 1 | 1 | 0 |

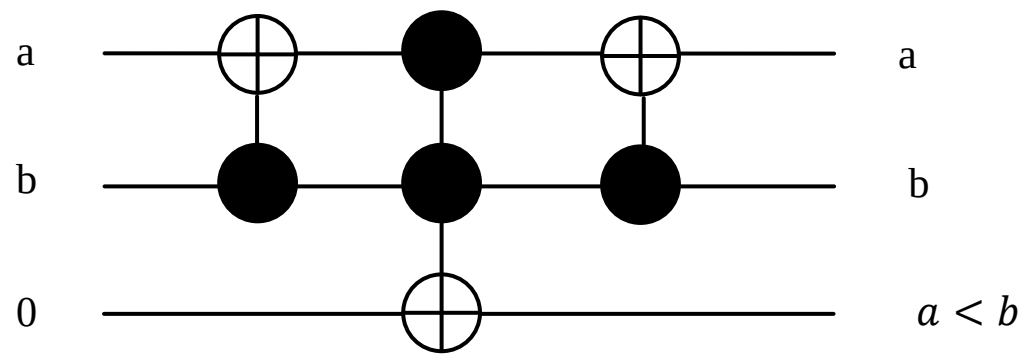

**Figure 3.15: Realization of a < b**

**ii) Greater than:** Denoted asa > b, implying the output is True, if and only if a greater than b. The truth table is given in table 3.10 and reversible circuit in figure 3.16.

**Table 3.10: Truth table of a > b**

| **Input** | | | **Output** | | |
|---|---|---|---|---|---|
| **a** | **b** | **0** | **a** | **b** | **a > b** |
| 0 | 0 | 0 | 0 | 0 | 0 |
| 0 | 1 | 0 | 0 | 1 | 0 |
| 1 | 0 | 0 | 1 | 0 | 1 |
| 1 | 1 | 0 | 1 | 1 | 0 |

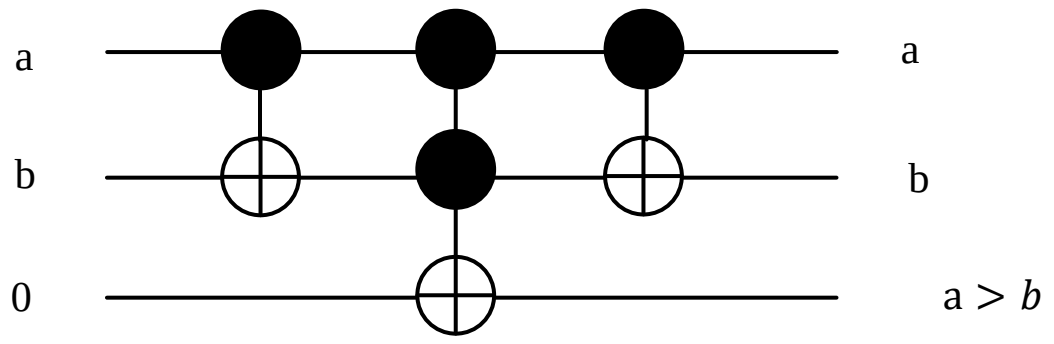

**Figure 3.16: Realization of a > *b***

**iii) Less or equal to:** Denoted asa ≤ b, implying the output is True, if and only if a is less or equal to b. The truth table is given in table 3.11 and reversible circuit in figure 3.17.

**Table 3.11: Truth table of a ≤ b**

| Input | | | Output | | |
|---|---|---|---|---|---|
| **a** | **b** | **0** | **a** | **b** | **a ≤ b** |
| 0 | 0 | 0 | 0 | 0 | 1 |
| 0 | 1 | 0 | 0 | 1 | 1 |
| 1 | 0 | 0 | 1 | 0 | 0 |
| 1 | 1 | 0 | 1 | 1 | 1 |

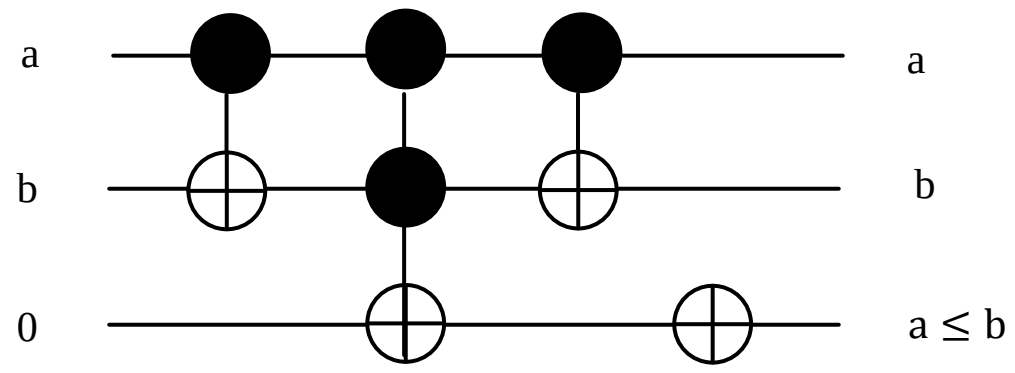

**Figure 3.17: Realization of a ≤ b**

**iv) Greater or equal to:** Denoted as a ≥ b, implying the output is True, if and only if a is greater or equal to b. The truth table is given in table 3.12 and reversible circuit in figure 3.18.

**Table 3.12: Truth table of a ≥ b**

| Input | | | Output | | |
|---|---|---|---|---|---|
| **a** | **b** | **0** | **a** | **b** | **a ≥ b** |
| 0 | 0 | 0 | 0 | 0 | 1 |
| 0 | 1 | 0 | 0 | 1 | 0 |
| 1 | 0 | 0 | 1 | 0 | 1 |
| 1 | 1 | 0 | 1 | 1 | 1 |

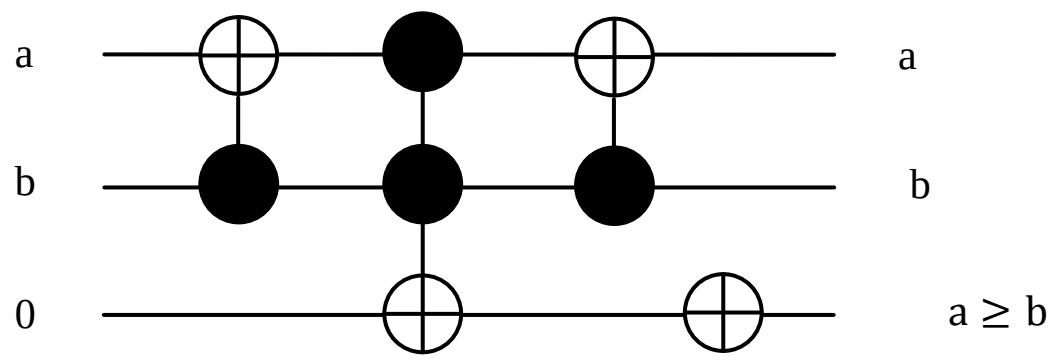

**Figure 3.18: Realization of a ≥ b**

**v) Equal to:** Denoted as a = b, implying the output is True, if and only if a is equal to b. The truth table is given in table 3.13 and reversible circuit in figure 3.19.

**Table 3.13: Truth table of a = b**

| Input | | | Output | | |
|---|---|---|---|---|---|
| **a** | **b** | **0** | **a** | **b** | **a = b** |
| 0 | 0 | 0 | 0 | 0 | 1 |
| 0 | 1 | 0 | 0 | 1 | 0 |
| 1 | 0 | 0 | 1 | 0 | 0 |
| 1 | 1 | 0 | 1 | 1 | 1 |

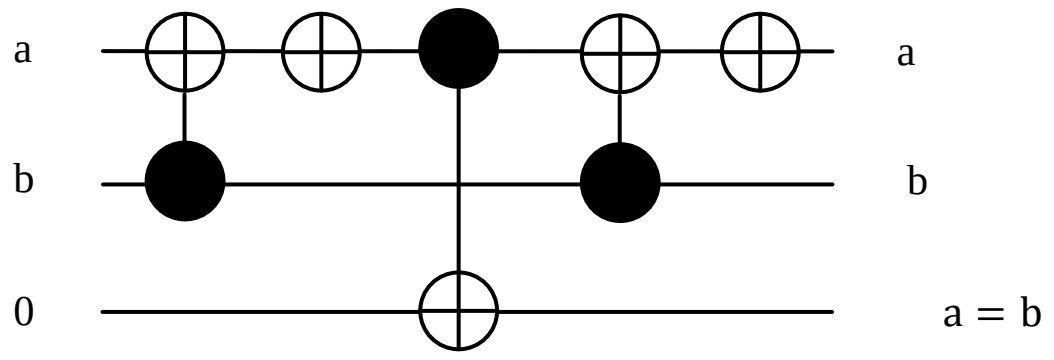

**Figure 3.19: Realization of a = b**

**vi) Not Equal to:** Denoted as a! = b, implying the output is True, if and only if a is not equal to b. The truth table is given in table 3.14 and reversible circuit in figure 3.20.

**Table 3.14: Truth table of a! = b**

| Input | | | Output | | |
|---|---|---|---|---|---|
| **a** | **b** | **0** | **a** | **b** | **a! = b** |
| 0 | 0 | 0 | 0 | 0 | 0 |
| 0 | 1 | 0 | 0 | 1 | 1 |
| 1 | 0 | 0 | 1 | 0 | 1 |
| 1 | 1 | 0 | 1 | 1 | 0 |

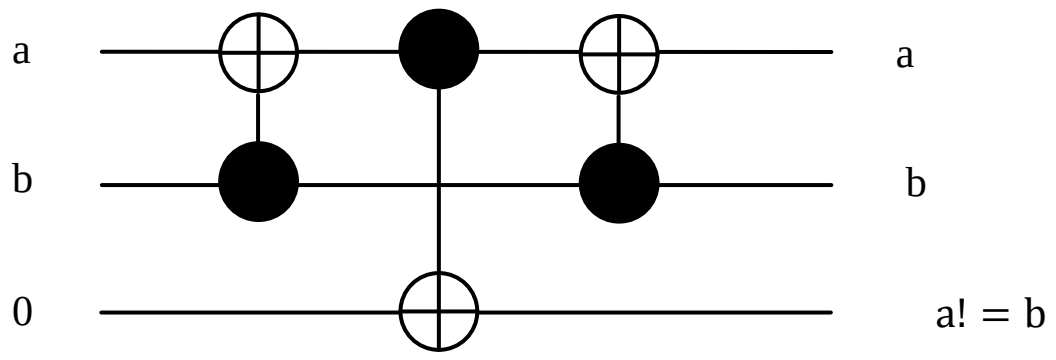

**Figure 3.20: Realization of a! = b**

### 3.2.3 Logical Expressions

Conditional statements in SyReC are formed using the relative operators as described above and also these outputs can further be combined using logical operators.

**(i) Logical AND of a and b denoted as a&&b:** The truth table for 1-bit operands is given in table 3.15 and reversible circuit in figure 3.21.

**Table 3.15: Truth table of a&&b**

| **Input** | | | **Output** | | |
|---|---|---|---|---|---|
| **a** | **b** | **0** | **a** | **b** | **a&&b** |
| 0 | 0 | 0 | 0 | 0 | 0 |
| 0 | 1 | 0 | 0 | 1 | 0 |
| 1 | 0 | 0 | 1 | 0 | 0 |
| 1 | 1 | 0 | 1 | 1 | 1 |

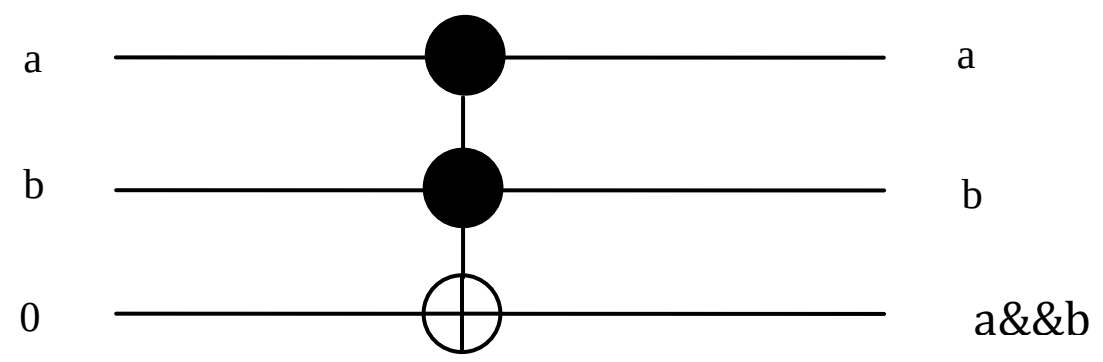

**Figure 3.21: Realization of a&&b**

**ii) Logical OR of a and b denoted as a||*b*:** The truth table for 1-bit operands is given in table 3.16 and reversible circuit in figure 3.22.

**Table 3.16: Truth table of a||b**

| **Input** | | | **Output** | | |
|---|---|---|---|---|---|
| **A** | **b** | **0** | **a** | **b** | **a\|\|b** |
| 0 | 0 | 0 | 0 | 0 | 0 |
| 0 | 1 | 0 | 0 | 1 | 1 |
| 1 | 0 | 0 | 1 | 0 | 1 |
| 1 | 1 | 0 | 1 | 1 | 1 |

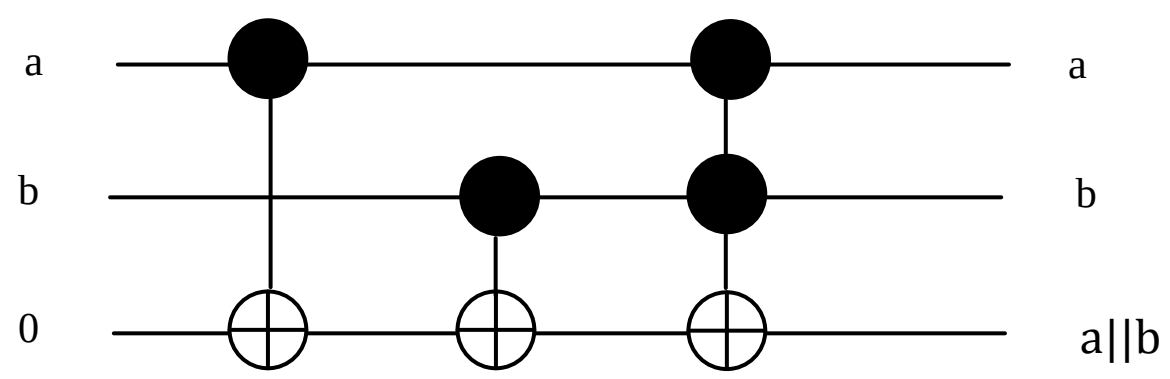

**Figure 3.22: Realization of a||b (1-bit)**

**iii) Logical NOT of a denoted as ! a:** The truth table for 1-bit operands is given in table 3.17 and reversible circuit in figure 3.23.

**Table 3.17: Truth table of ! a**

| Input | Output |
|---|---|
| **a** | **a'** |
| 0 | 1 |
| 1 | 0 |

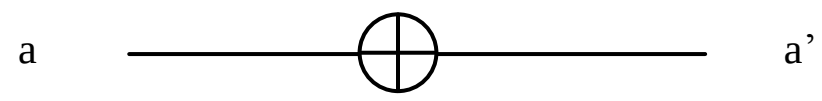

**Figure 3.23: Realization of ! a**

### 3.2.4 Shift Operations

SyReC allows only two kinds of shift operation over operands. Each of these takes two arguments – the operand and a number 'n' denoting number of bits to be shifted.

**i) Logical left shift of a by n written as a << n:** Figure 3.24 shows reversible circuit for the logical left shift operation of a by 1 and result stored in c, here a and c is 3 bit-wide. We need some constant inputs too for reversibility, and the number of constant input lines depends on bit width of operands, here three. Accordingly, three extra output lines will be obtained, here treated as garbage.

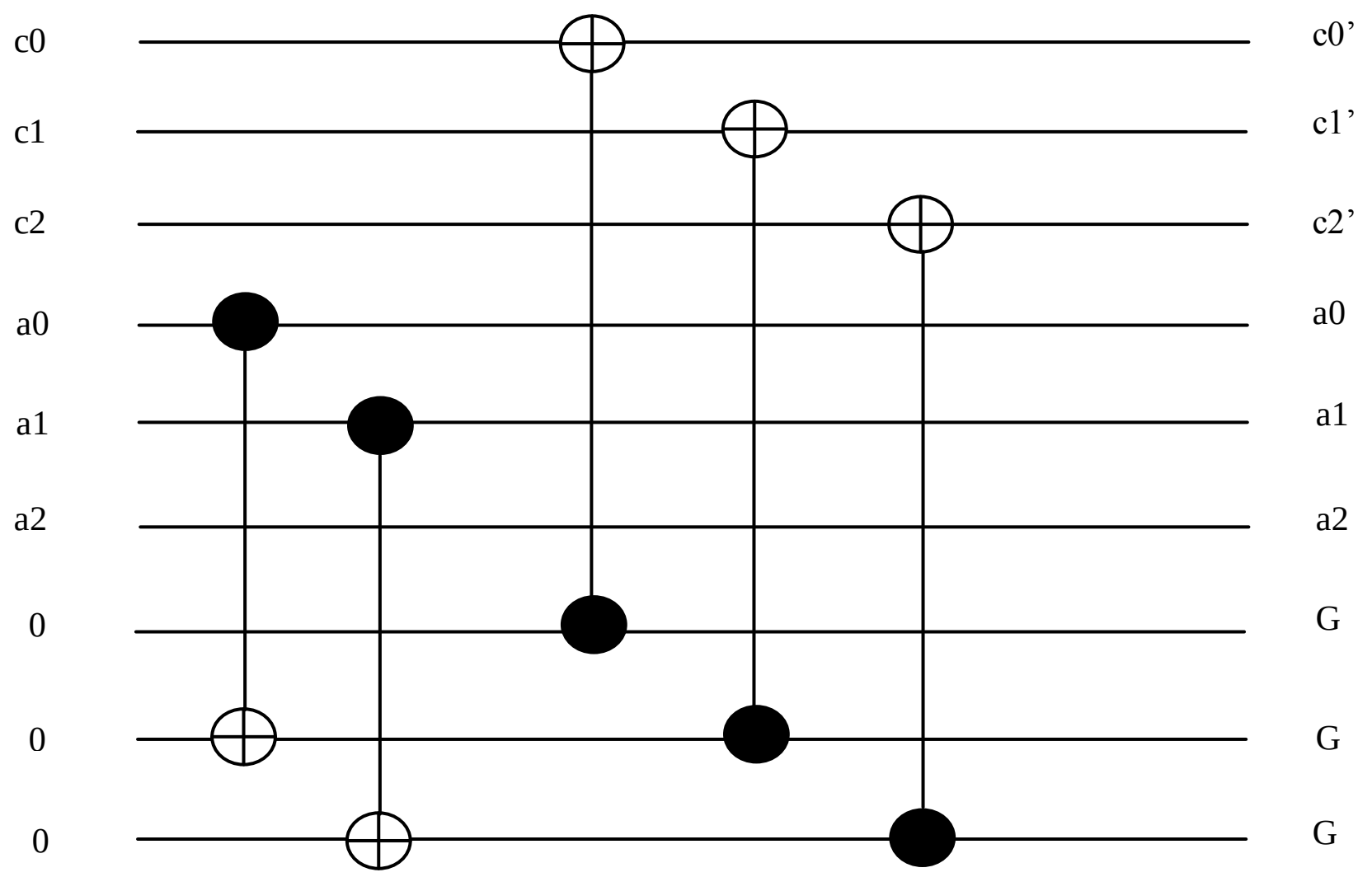

**Figure 3.24: Realization of a << 1**

**ii)** ***a* >> n Logical right shift of a by n:** Figure 3.25 shows reversible circuit for the logical right shift operation of a by 1 and result stored in c, here a and c is 3 bit-wide. We need some constant inputs too for reversibility, and the number of constant input lines depends on bit width of operands, here three. Accordingly, three extra output lines will be obtained, here treated as garbage.

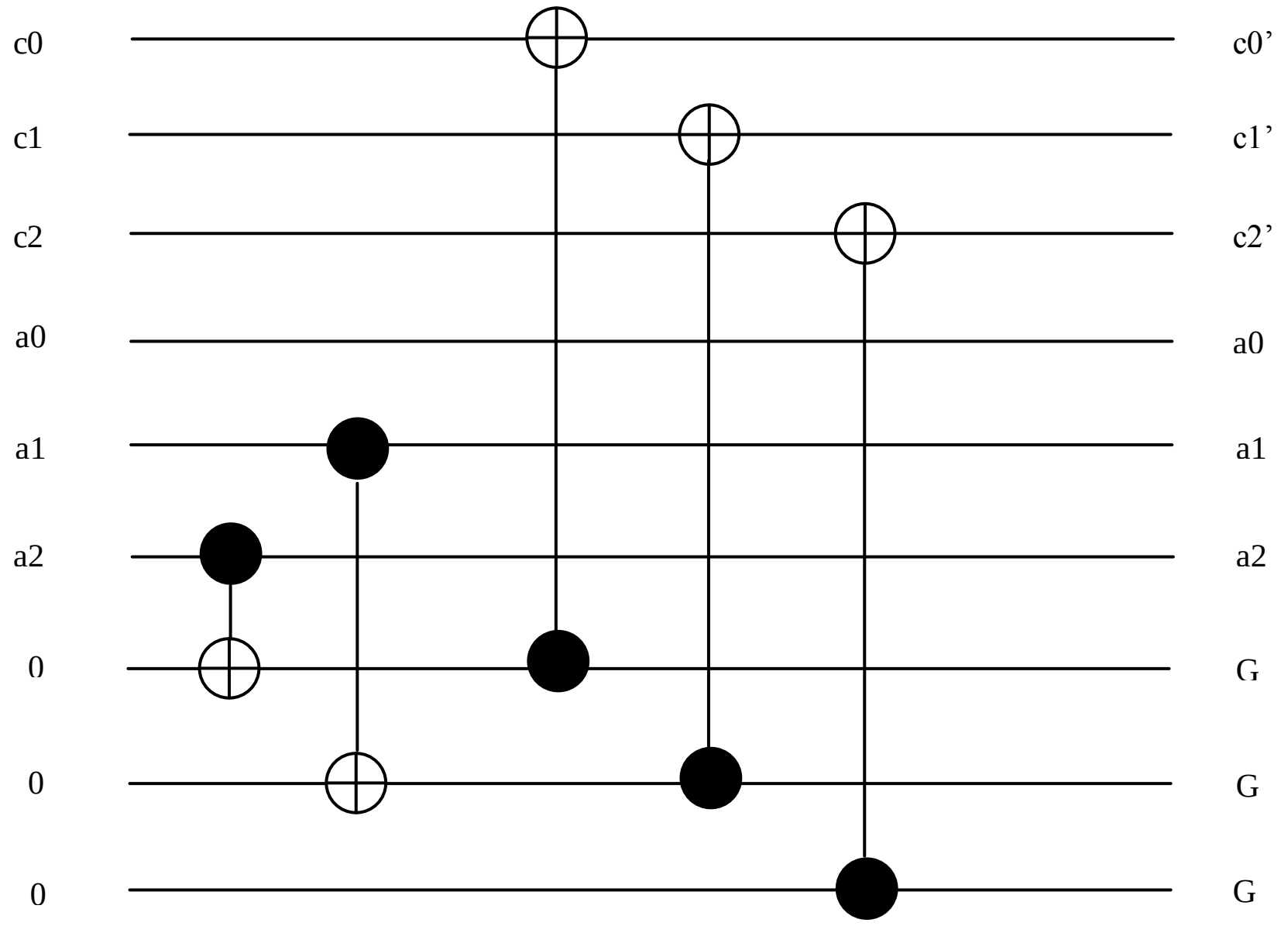

**Figure 3.25: Realization of a** >> 1

### 3.2.5 Arithmetic Operations

Distinct feature of the operations mentioned in this section is that the result of operation is stored in a third operand. The circuits described below correspond to operations between 2-bit operands.

**i) Addition:** Denoted as c^ = (a + b), implying addition of a and b and storing result in c. Table 3.18 shows truth table and figure 3.26 shows reversible circuit for this operation. We need some constant inputs too for reversibility, and the number of constant input lines depends on bit width of operands, here two. Accordingly, two extra output lines will be obtained, here treated as garbage (g0 and g1).Note that the lower most circuit line in figure 3.26 indicates how the circuit can be expanded when number of bits in the operands increased.

**Table 3.18: Truth table of c^ = a + b**

| **Inputs** | | | | | | | | **Outputs** | | | | | | | |
|---|---|---|---|---|---|---|---|---|---|---|---|---|---|---|---|
| **c1** | **c0** | **b1** | **b0** | **a1** | **a0** | **0** | **0** | **c1’** | **c0’** | **b1** | **b0** | **a1** | **a0** | **g0** | **g1** |
| 0 | 0 | 0 | 0 | 0 | 0 | 0 | 0 | 0 | 0 | 0 | 0 | 0 | 0 | G | G |
| 0 | 0 | 0 | 0 | 0 | 1 | 0 | 0 | 0 | 1 | 0 | 0 | 0 | 1 | G | G |
| 0 | 0 | 0 | 0 | 1 | 0 | 0 | 0 | 1 | 0 | 0 | 0 | 1 | 0 | G | G |
| 0 | 0 | 0 | 0 | 1 | 1 | 0 | 0 | 1 | 1 | 0 | 0 | 1 | 1 | G | G |
| 0 | 0 | 0 | 1 | 0 | 0 | 0 | 0 | 0 | 1 | 0 | 1 | 0 | 0 | G | G |
| 0 | 0 | 0 | 1 | 0 | 1 | 0 | 0 | 1 | 0 | 0 | 1 | 0 | 1 | G | G |
| 0 | 0 | 0 | 1 | 1 | 0 | 0 | 0 | 1 | 1 | 0 | 1 | 1 | 0 | G | G |
| 0 | 0 | 0 | 1 | 1 | 1 | 0 | 0 | 0 | 0 | 0 | 1 | 1 | 1 | G | G |
| 0 | 0 | 1 | 0 | 0 | 0 | 0 | 0 | 1 | 0 | 1 | 0 | 0 | 0 | G | G |
| 0 | 0 | 1 | 0 | 0 | 1 | 0 | 0 | 1 | 1 | 1 | 0 | 0 | 1 | G | G |
| 0 | 0 | 1 | 0 | 1 | 0 | 0 | 0 | 0 | 0 | 1 | 0 | 1 | 0 | G | G |
| 0 | 0 | 1 | 0 | 1 | 1 | 0 | 0 | 0 | 1 | 1 | 0 | 1 | 1 | G | G |
| 0 | 0 | 1 | 1 | 0 | 0 | 0 | 0 | 1 | 1 | 1 | 1 | 0 | 0 | G | G |
| 0 | 0 | 1 | 1 | 0 | 1 | 0 | 0 | 0 | 0 | 1 | 1 | 0 | 1 | G | G |
| 0 | 0 | 1 | 1 | 1 | 0 | 0 | 0 | 0 | 1 | 1 | 1 | 1 | 0 | G | G |
| 0 | 0 | 1 | 1 | 1 | 1 | 0 | 0 | 1 | 0 | 1 | 1 | 1 | 1 | G | G |

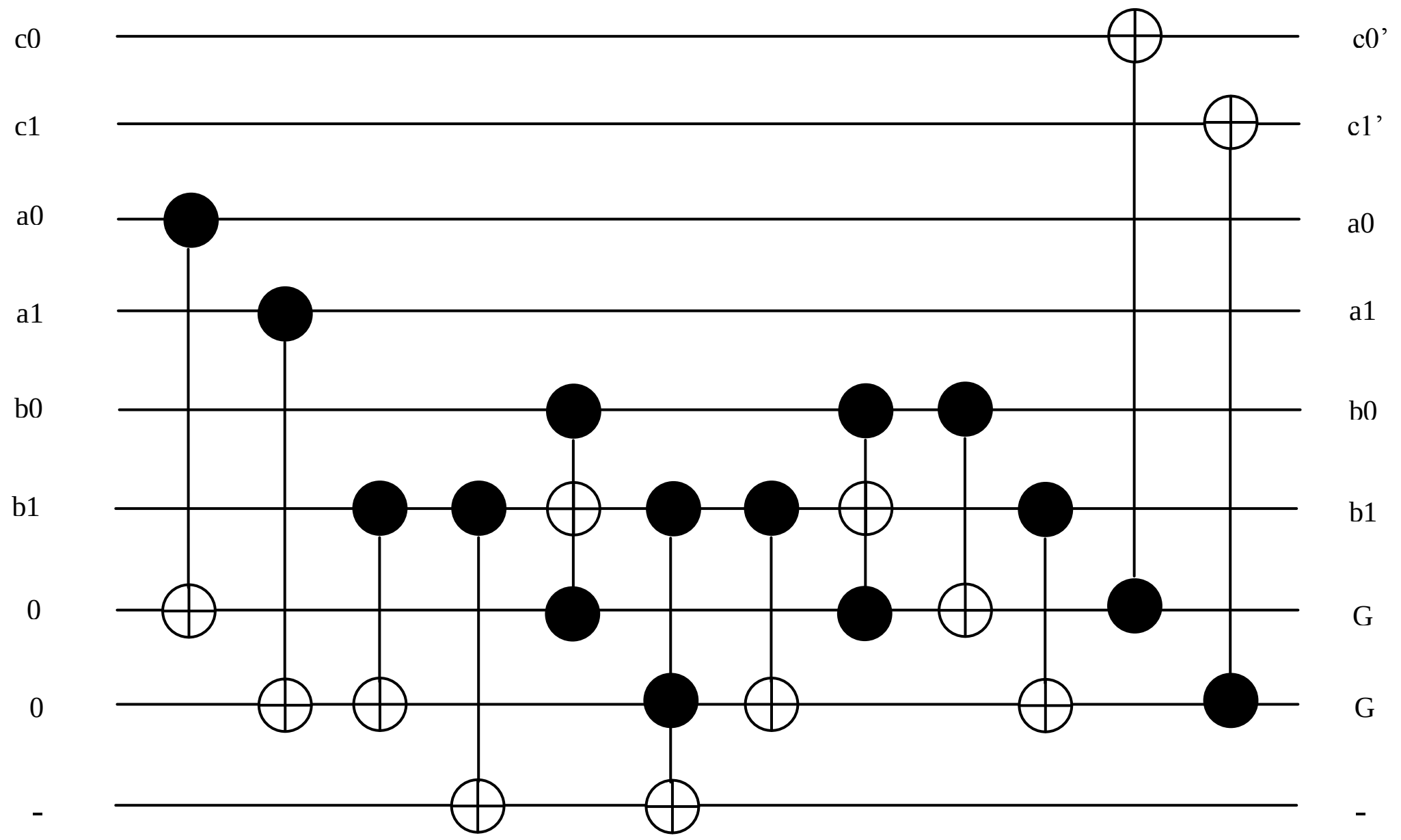

**Figure 3.26: Realization of c^ = (a + b)**

**ii) Subtraction:** Denoted as c^ = (a − b), implying subtraction of a and b and storing result in c. Table 3.19 shows truth table and figure 3.27 shows reversible circuit for this operation. We need some constant inputs too for reversibility, and the number of constant input lines depends on bit width of operands, here two. Accordingly, two extra output lines will be obtained, here treated as garbage (g0 and g1).Note that the lower most circuit line in figure 3.27 indicates how the circuit can be expanded when number of bits in the operands increased.

**Table 3.19: Truth table of c^ = a − b**

| Inputs | | | | | | | | Outputs | | | | | | | |
|---|---|---|---|---|---|---|---|---|---|---|---|---|---|---|---|
| **c0** | **c1** | **b1** | **b0** | **a1** | **a0** | **0** | **0** | **c1'** | **c0'** | **b1** | **b0** | **a1** | **a0** | **g0** | **g1** |
| 0 | 0 | 0 | 0 | 0 | 0 | 0 | 0 | 0 | 0 | 0 | 0 | 0 | 0 | G | G |
| 0 | 0 | 0 | 0 | 0 | 1 | 0 | 0 | 0 | 1 | 0 | 0 | 0 | 1 | G | G |
| 0 | 0 | 0 | 0 | 1 | 0 | 0 | 0 | 1 | 0 | 0 | 0 | 1 | 0 | G | G |
| 0 | 0 | 0 | 0 | 1 | 1 | 0 | 0 | 1 | 1 | 0 | 0 | 1 | 1 | G | G |
| 0 | 0 | 0 | 1 | 0 | 0 | 0 | 0 | 1 | 1 | 0 | 1 | 0 | 0 | G | G |
| 0 | 0 | 0 | 1 | 0 | 1 | 0 | 0 | 0 | 0 | 0 | 1 | 0 | 1 | G | G |
| 0 | 0 | 0 | 1 | 1 | 0 | 0 | 0 | 0 | 1 | 0 | 1 | 1 | 0 | G | G |
| 0 | 0 | 0 | 1 | 1 | 1 | 0 | 0 | 1 | 0 | 0 | 1 | 1 | 1 | G | G |
| 0 | 0 | 1 | 0 | 0 | 0 | 0 | 0 | 1 | 0 | 1 | 0 | 0 | 0 | G | G |
| 0 | 0 | 1 | 0 | 0 | 1 | 0 | 0 | 1 | 1 | 1 | 0 | 0 | 1 | G | G |
| 0 | 0 | 1 | 0 | 1 | 0 | 0 | 0 | 0 | 0 | 1 | 0 | 1 | 0 | G | G |
| 0 | 0 | 1 | 0 | 1 | 1 | 0 | 0 | 0 | 1 | 1 | 0 | 1 | 1 | G | G |
| 0 | 0 | 1 | 1 | 0 | 0 | 0 | 0 | 0 | 1 | 1 | 1 | 0 | 0 | G | G |
| 0 | 0 | 1 | 1 | 0 | 1 | 0 | 0 | 1 | 0 | 1 | 1 | 0 | 1 | G | G |
| 0 | 0 | 1 | 1 | 1 | 0 | 0 | 0 | 1 | 1 | 1 | 1 | 1 | 0 | G | G |
| 0 | 0 | 1 | 1 | 1 | 1 | 0 | 0 | 0 | 0 | 1 | 1 | 1 | 1 | G | G |

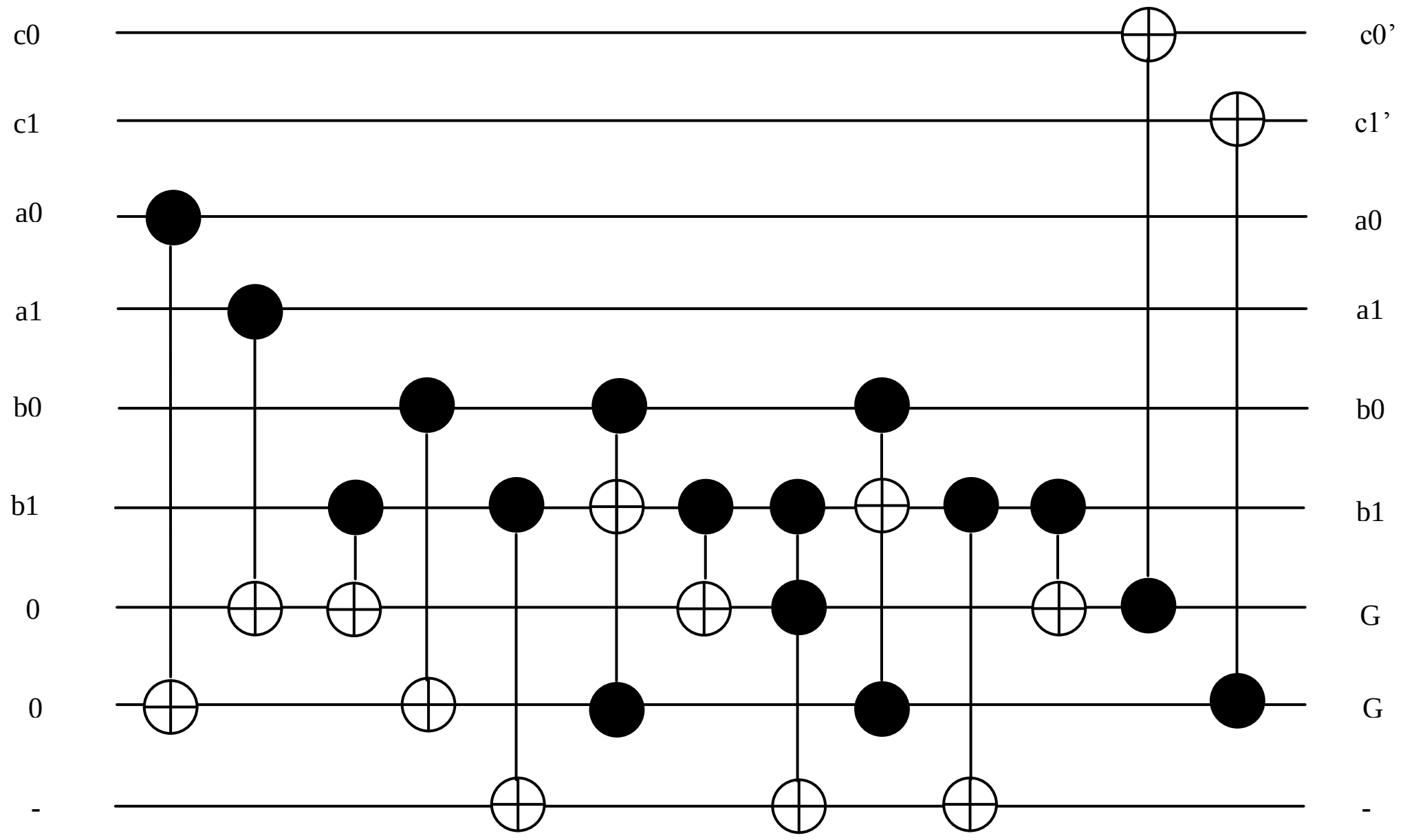

**Figure 3.27: Realization of $c\hat{} = (a - b)$**

**iii) Multiplication:** Denoted as $c\hat{} = (a * b)$, implying multiplication of a and b and storing result in c. Table 3.20 shows truth table and figure 3.28 shows reversible circuit for this operation. We need some constant inputs too for reversibility, and the number of constant input lines depends on bit width of operands, here two. Accordingly, two extra output lines will be obtained, here treated as garbage (g0 and g1).Note that the lower most circuit line in figure 3.28 indicates how the circuit can be expanded when number of bits in the operands increased.

**Table 3.20: Truth table of c^ = a ∗ b**

| Inputs | | | | | | | | Outputs | | | | | | | |
|---|---|---|---|---|---|---|---|---|---|---|---|---|---|---|---|
| **c0** | **c1** | **b1** | **b0** | **a1** | **a0** | **0** | **0** | **c1'** | **c0'** | **b1** | **b0** | **a1** | **a0** | **g0** | **g1** |
| 0 | 0 | 0 | 0 | 0 | 0 | 0 | 0 | 0 | 0 | 0 | 0 | 0 | 0 | G | G |
| 0 | 0 | 0 | 0 | 0 | 1 | 0 | 0 | 0 | 1 | 0 | 0 | 0 | 1 | G | G |
| 0 | 0 | 0 | 0 | 1 | 0 | 0 | 0 | 1 | 0 | 0 | 0 | 1 | 0 | G | G |
| 0 | 0 | 0 | 0 | 1 | 1 | 0 | 0 | 1 | 1 | 0 | 0 | 1 | 1 | G | G |
| 0 | 0 | 0 | 1 | 0 | 0 | 0 | 0 | 1 | 1 | 0 | 1 | 0 | 0 | G | G |
| 0 | 0 | 0 | 1 | 0 | 1 | 0 | 0 | 0 | 0 | 0 | 1 | 0 | 1 | G | G |
| 0 | 0 | 0 | 1 | 1 | 0 | 0 | 0 | 0 | 1 | 0 | 1 | 1 | 0 | G | G |
| 0 | 0 | 0 | 1 | 1 | 1 | 0 | 0 | 1 | 0 | 0 | 1 | 1 | 1 | G | G |
| 0 | 0 | 1 | 0 | 0 | 0 | 0 | 0 | 1 | 0 | 1 | 0 | 0 | 0 | G | G |
| 0 | 0 | 1 | 0 | 0 | 1 | 0 | 0 | 1 | 1 | 1 | 0 | 0 | 1 | G | G |
| 0 | 0 | 1 | 0 | 1 | 0 | 0 | 0 | 0 | 0 | 1 | 0 | 1 | 0 | G | G |
| 0 | 0 | 1 | 0 | 1 | 1 | 0 | 0 | 0 | 1 | 1 | 0 | 1 | 1 | G | G |
| 0 | 0 | 1 | 1 | 0 | 0 | 0 | 0 | 0 | 1 | 1 | 1 | 0 | 0 | G | G |
| 0 | 0 | 1 | 1 | 0 | 1 | 0 | 0 | 1 | 0 | 1 | 1 | 0 | 1 | G | G |
| 0 | 0 | 1 | 1 | 1 | 0 | 0 | 0 | 1 | 1 | 1 | 1 | 1 | 0 | G | G |
| 0 | 0 | 1 | 1 | 1 | 1 | 0 | 0 | 0 | 0 | 1 | 1 | 1 | 1 | G | G |

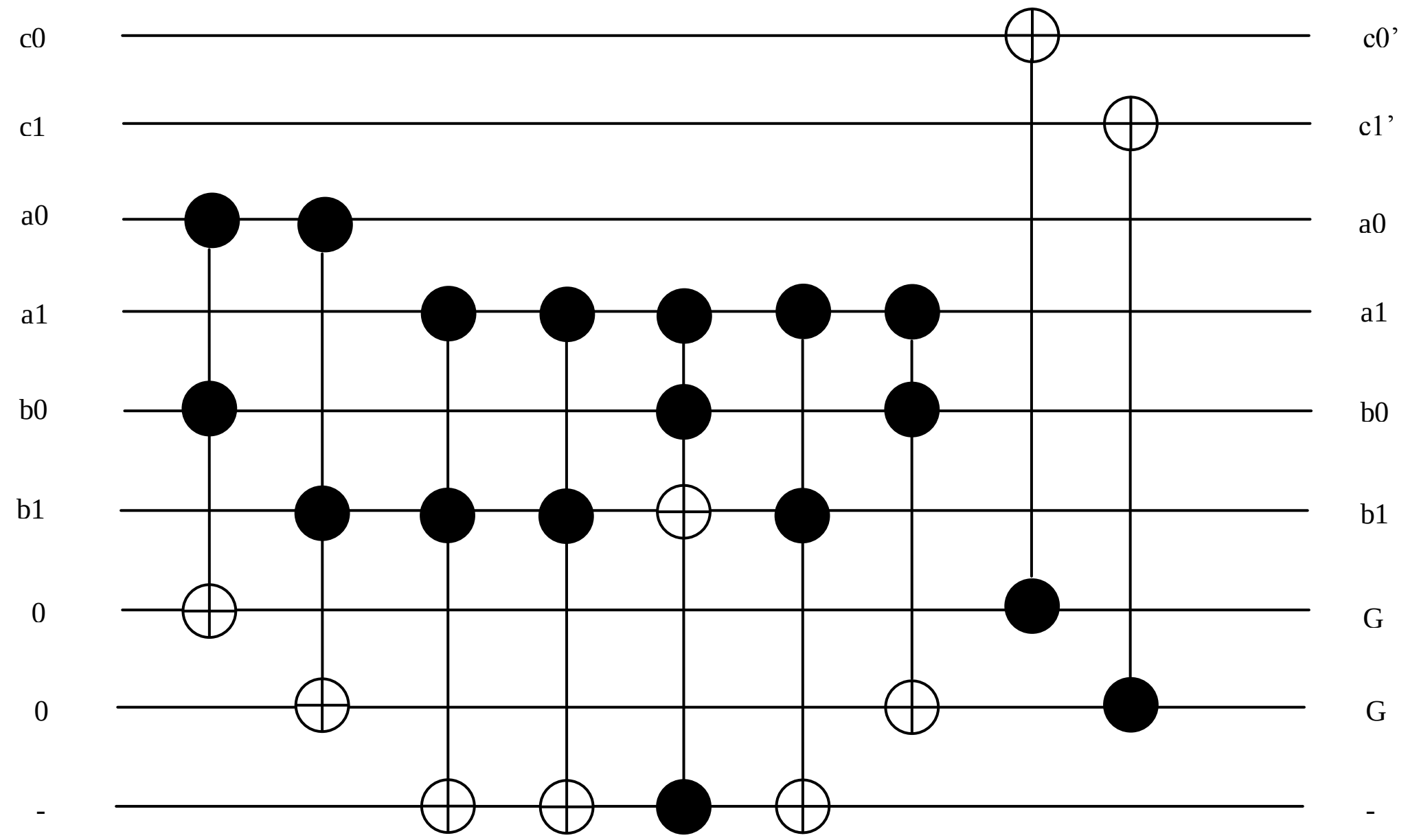

**Figure 3.28: Realization of c^ = (a ∗ b)**

## 3.3 Unary operations

These operations apply on a single operand. The unary operations provided in SyReC and their corresponding circuits are given below.

**i) Bitwise negation of x:** Denoted as ~ = x. The reversible circuit is shown in figure 3.29.

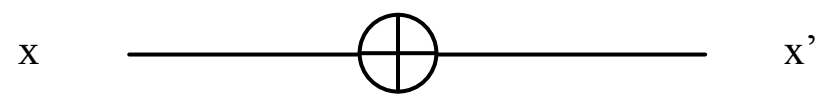

**Figure 3.29: Realization of ~ = x**

**ii) Increment of x:** Denoted as x+= 1 or ++= x

Figure 3.30 shows reversible circuit for the increment operation when x is 4 bit-wide.

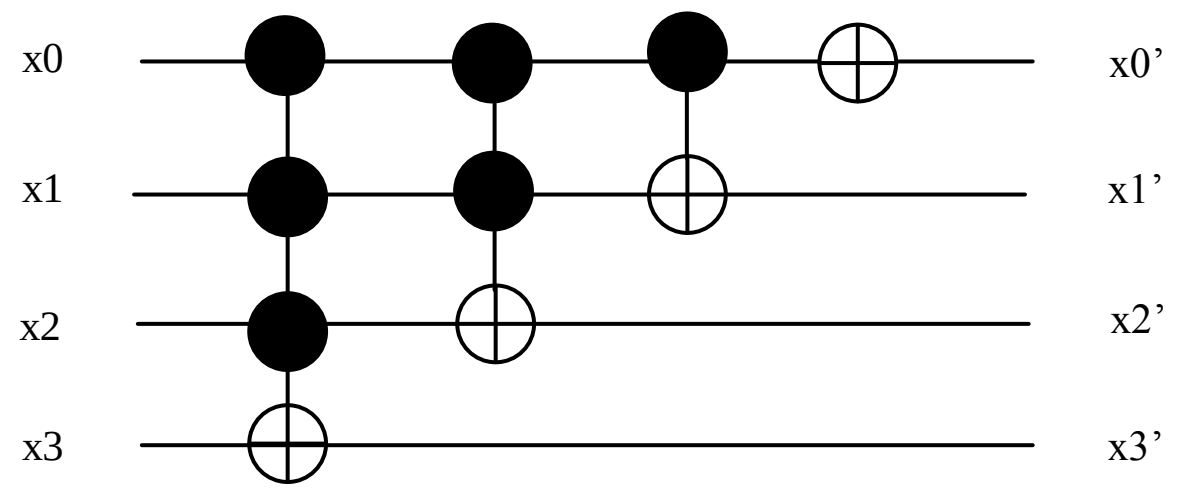

**Figure 3.30: Realization of ++= x**

**iii) Decrement of x:** Denoted as x−= 1 or −−= x

Figure 3.31shows reversible circuit for the increment operation when x is 4 bit-wide.

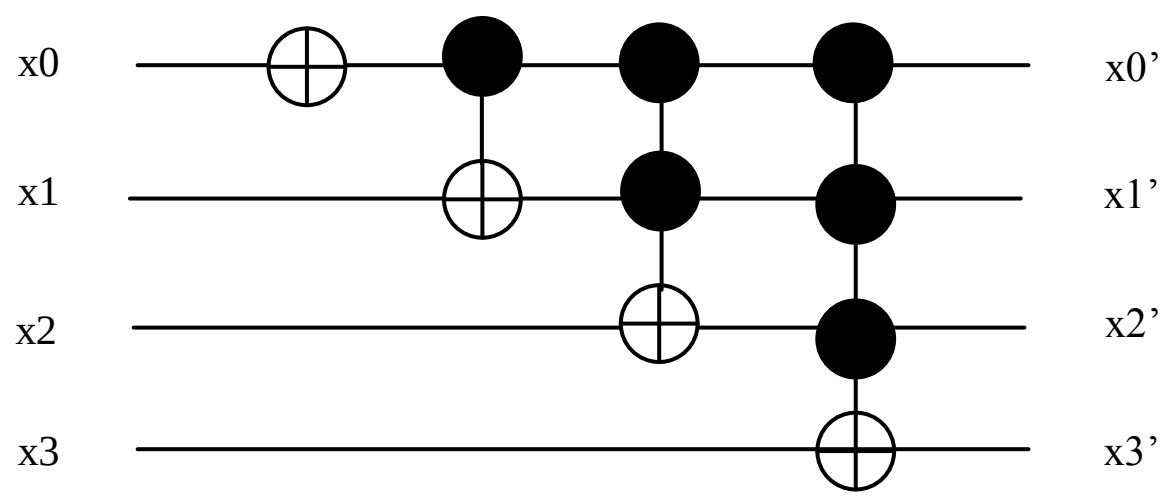

**Figure 3.31: Realization of −−= x**

## 3.4 SWAP operation

Swap gate is reversible gate which SWAP simply exchanges the bit values it is handed. Figure 3.32 shows a realization of swap operation.

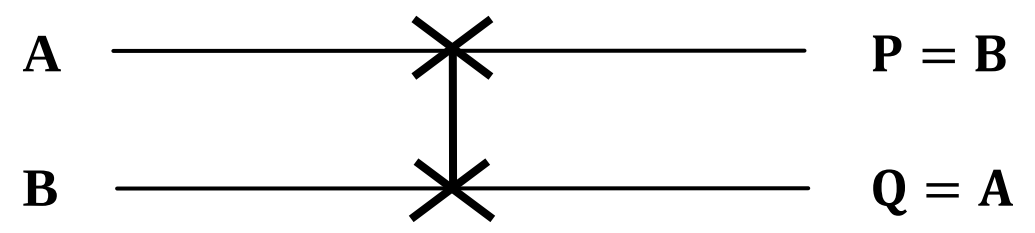

**Figure 3.32: Realization of Swap operation**

## 3.5 Conditional Statements

A code in SyReC contains constructs like loops, modules and conditional statements with the bodies consisting of statements as described in above sections. Loops and procedure calls/uncalls can be realized in a straightforward way by simple cascading (i.e. unrolling) the respective statements within a loop block for each iteration. Since then number of iterations must be available before synthesis, this results in a finite number of statements which is subsequently processed. Call and uncall of procedures are handled similarly. Here, the respective statements in the procedures are cascaded together. The realization of control operations as reversible cascades is not possible in such straight forward manner. To realize conditional statements, two variants are proposed, discussed below with the help of an example (figure 3.32).

1. The first approach is of duplication. The values of all signals that will be affected in an if- or else-block are copied through an extra input line with a constant value (shown by Signal a and Signal c in Figure 3.32(a)). The sub-circuits realizing the respective if/else block are added, denoted by the boxes in fig. the block actually to be executed depends on the result of the conditional statement (Signal e in Figure 3.32(a)), leading either to duplication of the values from the original lines or being swapped to give the desired result.

2. The second realization (shown in Figure 3.32(b)) makes intensive use of control connections. The outcome of the conditional statement behaves as control line for both if and else blocks. Hence, control lines are added to all gates in the realization of the respective then- and else-block. The gates in these blocks are triggered if and only if the result of the conditional statement (i.e. signal e) is assigned to 1 or 0, respectively. A NOT gate (i.e. a Toffoli gate t($\phi$, {e}) without control lines) is thereby applied to flip the value of e so that the gates of the else block can be “controlled” as well.

**eg.** **if e then**
**a+= b**
**else**
**c+= d**
**fi e**

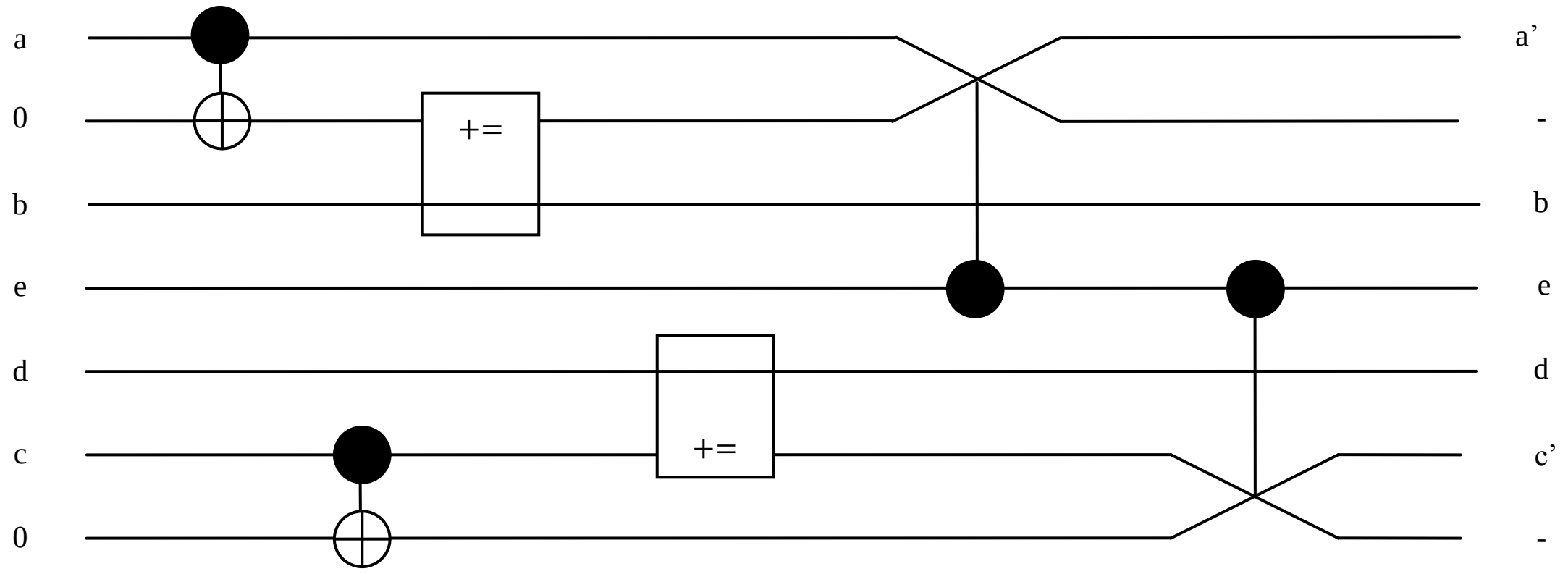

**(a) Using duplication**

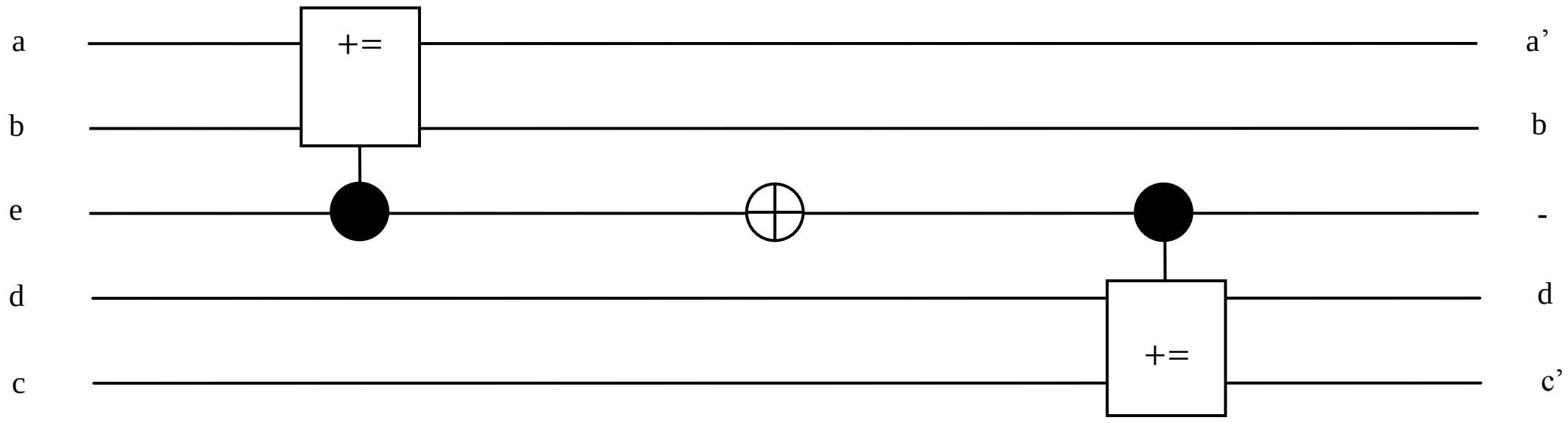

**(b) Without duplication**

**Figure 3.33: Realization of an if-else statement**

Deciding which approach to use during synthesis is a cost trade-off. Using the approach of duplication leads to additional circuit lines, which is already a restricted resource in quantum logic. While second approach uses many control lines which makes the quantum costs significantly larger in this solution.

**Chapter 4**

# EXPANDABLE CIRCUIT DESIGN

The purpose of this chapter is to show that complex reversible circuit can be constructed from SyReC specification using a hierarchical approach similar to the conventional approach of circuit synthesis for irreversible circuits. This approach has dual advantage of making the conversion from SyReC code to reversible circuit easy along with making such circuits expandable. Expanding a circuit is important from the point of view of real life applications.

We begin by giving circuits for some complex circuits of low bit width, and proceed towards a circuit of elevator controller of varying bit widths, thus showing the expandability of a circuit.

## 4.1 General Approach for Circuit Realization

We adopt hierarchical approach towards realization of complex circuit. Hierarchical synthesis involves identifying basic blocks of a complex circuit and using existing realization of them.

As discussed in chapter 3, individual operations of any SyReC specification can be realized into reversible circuits. These operations can be treated as components, which can further be combined into the desired circuits. Thus, SyReC specification of any circuit can be analyzed into sequence of individual operations and the respective circuit realizations be combined into a single circuit. Putting it formally, the approach is:

1. Traverse whole program written in SyReC
2. Identify individual operations and their sequence
3. Map statements, expressions and operations to circuit blocks
4. Cascade circuit components as per sequence.

Figure 4.1 shows a flow of generation of schematic from SyReC code.

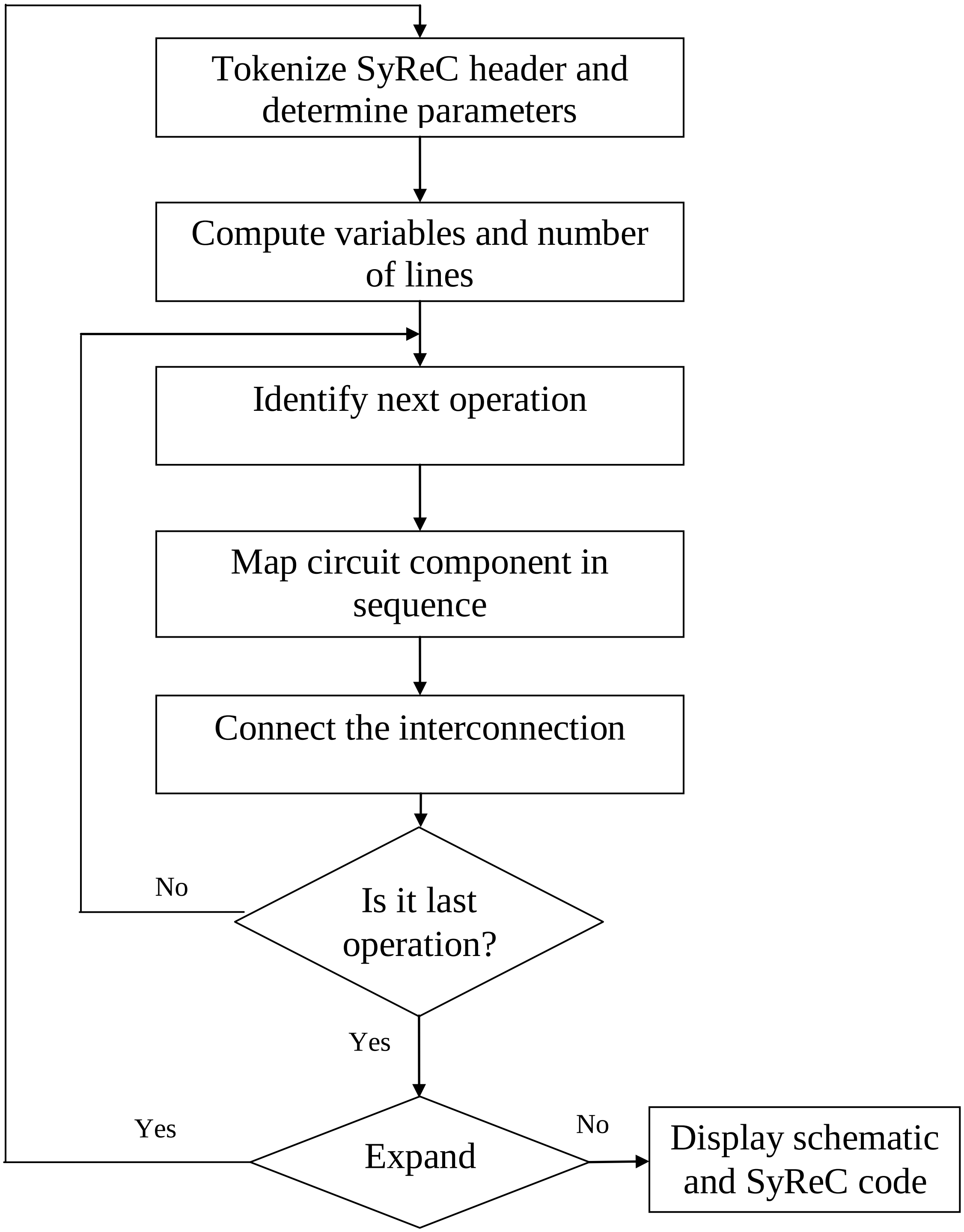

**Figure 4.1: Schematic generation steps**

## 4.2 Few Complex Reversible Circuits

In this section we present a few popular combinational circuits along with their SyReC specification. Further, we translate the specification into a circuit realization using the hierarchical approach. The sub-circuits have been realized using the control-intensive approach discussed in section 3.

### 4.2.1 Program Counter

Program counter is an essential component in any processor architecture. A program counter circuit can be viewed as a combinational circuit which modifies the value of a variable in one of the following manners:

i. resets to 0 (or any set initial value)

ii. increments the value

iii. sets the value (to a branch target address)

Taking 2-bit wide program counter and initial value as 0, SyReC specification of the circuit is given in figure 4.2(a).

```
module program_counter (in reset(1), in inc(1), in jump(2), inout pc(2)) wire zero(2)
if (reset = 1) then
        pc <=> zero
else
        if (inc = 1) then
                pc += 1
        else
                pc <=> jump
        fi (inc = 1)
fi (reset = 1)
```

**Figure 4.2(a): SyReC code of Program Counter**

SyReC code in figure 4.2(a) treats the program counter variable '*pc'* both as an input and output. Here '*jump*' denotes a new value to be assigned to '*pc*', indicating a new branch addressor unconditional branch label. Control line '*inc*' decides incrementing '*pc*' by 1, or setting value to '*jump*'. Control line '*reset*' decides resetting of '*pc*' to '*zero*' (initial value).

As shown in figure 4.2(b), the circuit is built stage by stage, where each successive stage is in fact circuit realization of successive instructions in SyReC specification. The condition of the outermost if, i.e. value of '*reset*' pin, is copied to control line of the '*then*' portion of circuit. The statements of '*then*' portion are realized as simple assignment operations (as explained in chapter 3). The inverted '*reset*' pin value is controlling the '*else*' portion of circuit. Nested if-else is realized similarly by copying value from control pin '*inc*'.

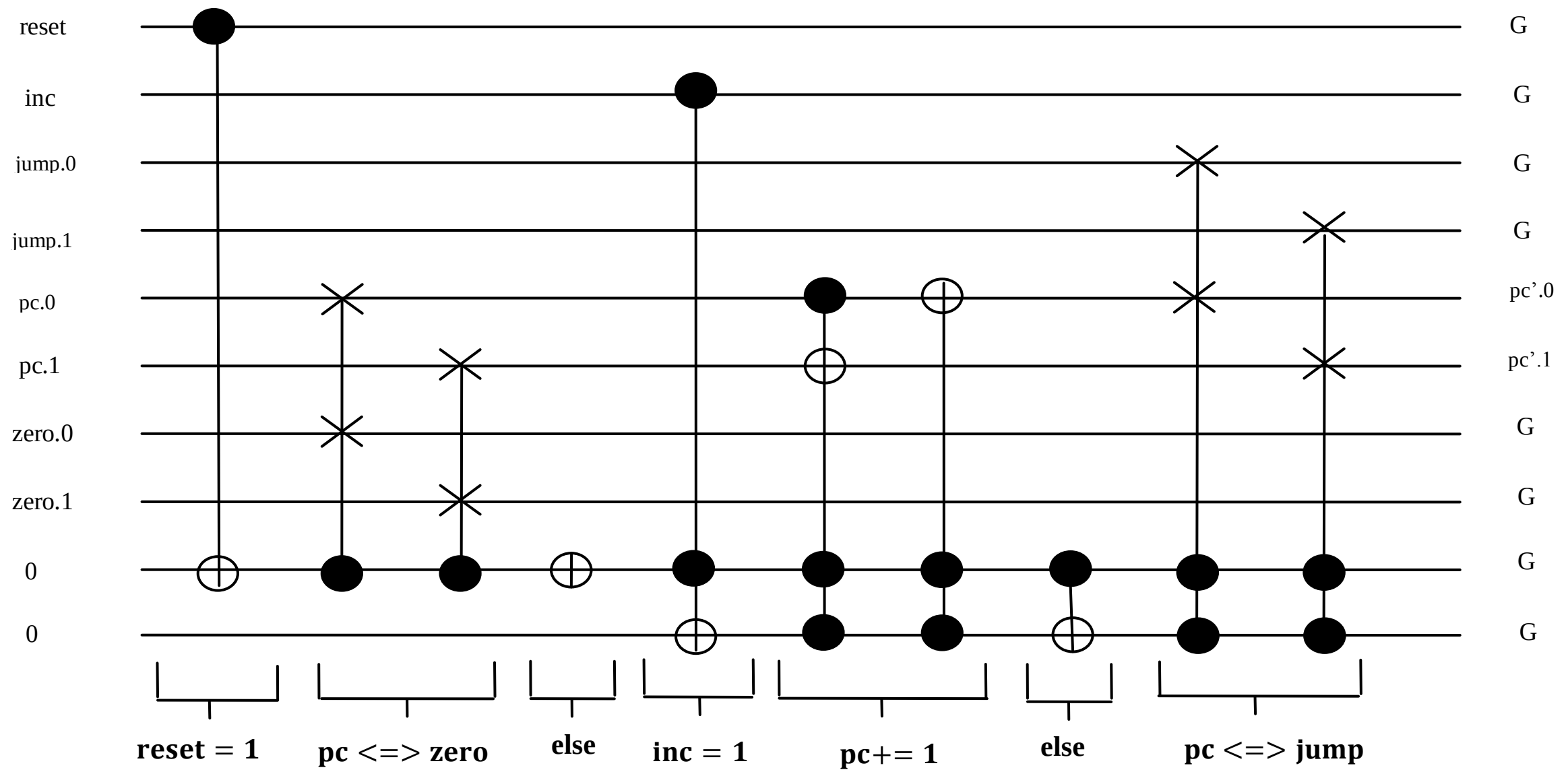

**Figure 4.2(b): Realization of Program Counter**

### 4.2.2 Arithmetic Unit (au)

In its simplest form, we can think of an arithmetic unit to perform certain mathematical operations between provided operands. For the purpose of illustration we have taken a unit with four operations and two operands (provided as an input to the circuit), here of 2-bit width.

Operations to be performed are decided by the control line '*op*', As follows:

- op=0 corresponds to addition of two operands *a* and *b*, result is stored in *c*.
- op=1 corresponds to subtraction of two operands *a* and *b*, result is stored in *c*.
- op=2 corresponds to multiplication of two operands *a* and *b*, result is stored in *c*.
- op=3 corresponds to copying value of operand *a* into output *c*.

The SyReC specification of au is given in figure 4.3(a). The circuit is specified as nested if-else structures. Individual operations are implemented directly as assignment-type statements as discussed in chapter 3.

```
module au (in op(2), in a(2), in b(2), out c(2))
if (op = 0) then
        c ^= (a + b)
else
        if (op = 1) then
                c ^= (a − b)
        else
                if (op = 2) then
                        c ^= (a ∗ b)
                else
                        c ^= a
                fi (op = 2)
        fi (op = 1)
fi (op = 0)
```

**Figure 4.3(a): SyReC code of Arithmetic Unit**

Realization of au is given in figure 4.3(b). Individual operations are the four basic components of the circuit which are connected to each other within a larger circuit of the if-else structures. The two bits of '*op*' are treated as two different control lines *op.0*, *op.1*, whose combination activates any one basic component (one mathematical operation), at a time.

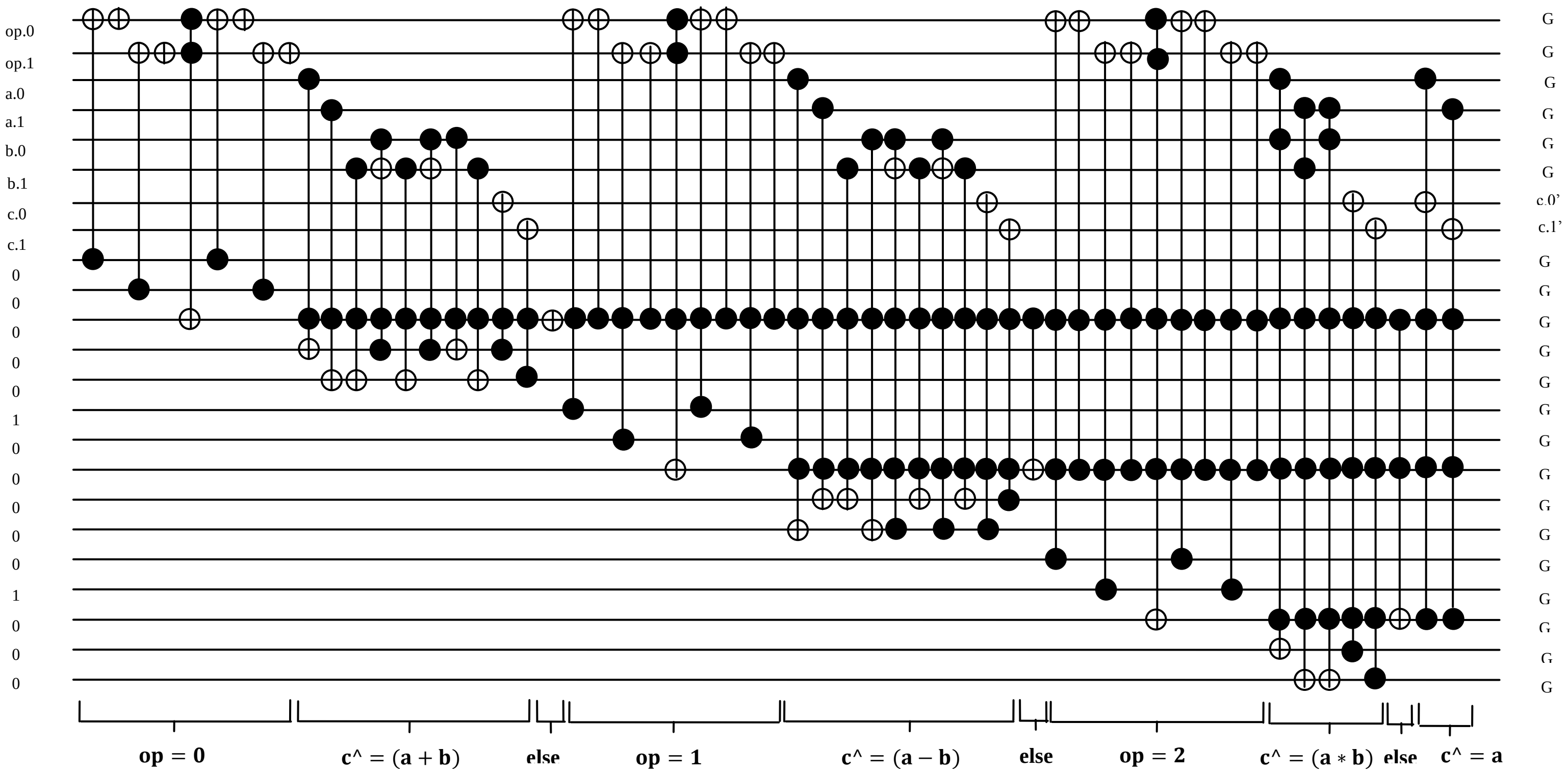

**Figure 4.2(b): Realization of Arithmetic Unit**

### 4.2.3 Logic Unit (lu)

In its simplest form, we can think of a logic unit to perform logical operations of AND, OR, XOR and NOT. As an example we have taken two operands of 2-bit width (provided as an input to the circuit).

Operations to be performed are decided by the control line '*op*', As follows:

- op=0 corresponds to AND of two operands $X_1$ and$X_2$, result is stored in $X_0$.
- op=1 corresponds to OR of two operands $X_1$ and$X_2$, result is stored in $X_0$.
- op=2 corresponds to XOR of two operands $X_1$ and $X_2$, result is stored in $X_0$.
- op=3 corresponds to copying value of operand $X_1$ into output $X_0$ and then negation of $X_0$.

The SyReC specification of lu is given in figure 4.4(a) and realization of lu given in figure 4.4(b).

```
module lu (in op(2), out x0(2), inout x1(2), inout x2(2))
if (op = 0) then
        x0 ^= (x1&x2)
else
        if (op = 1) then
                x0 ^= (x1 | x2)
        else
                if (op = 2) then
                        x0 ^= (x1 ^ x2)
                else
                        x0 ^= x1 ; ~= x0
                fi (op = 2)
        fi (op = 1)
fi (op = 0)
```

**Figure 4.4(a): SyReC code of Logic Unit**

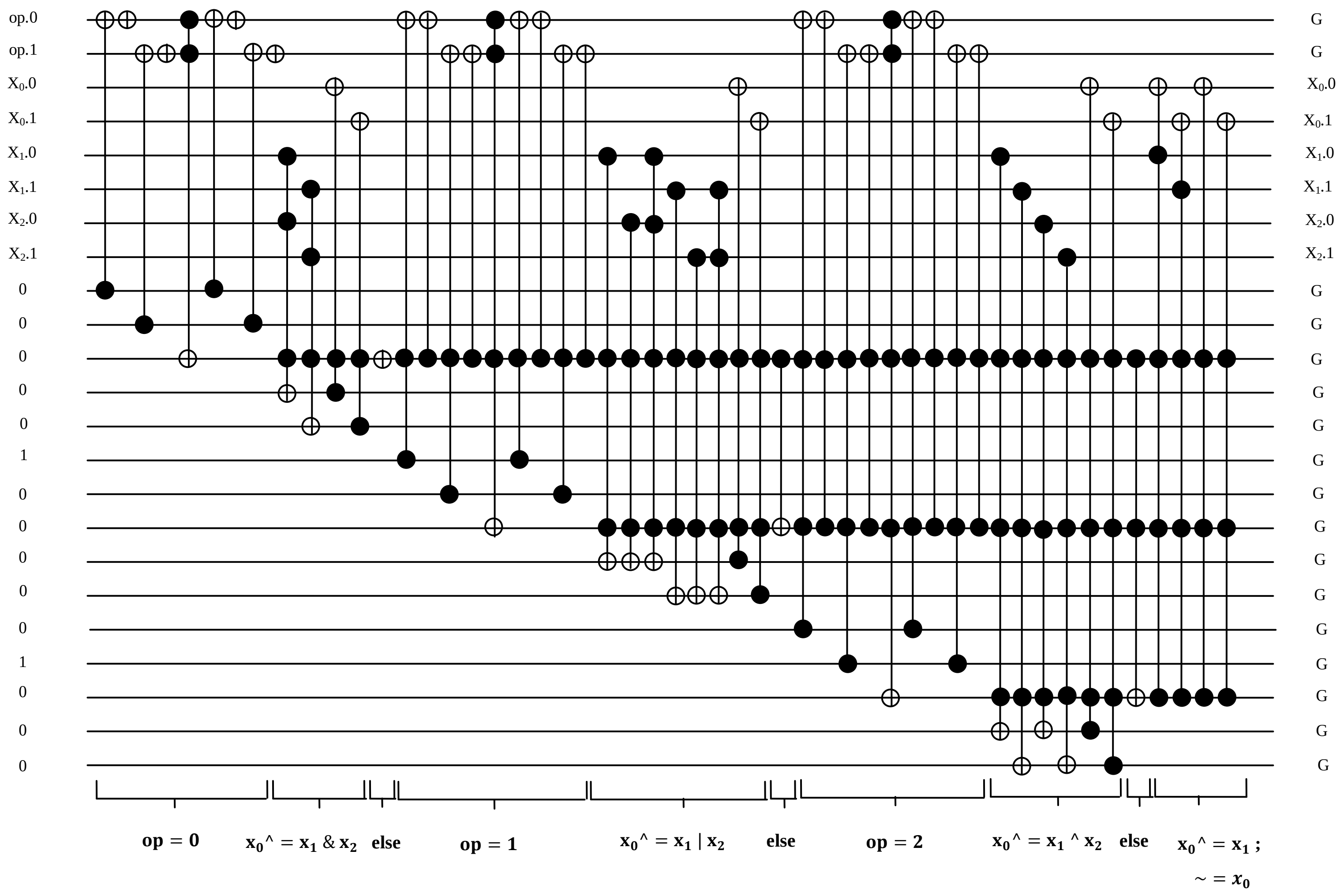

**Figure 4.3(b): Realization of Logic Unit**

## 4.3 Expanding a Complex Reversible Circuit

Expanding a circuit refers to increasing the number of bits of its input and/or output signals. Thus, we increase the total number of input and output lines of the circuit without changing its behavior. To retain the behavior of a circuit, number, nature and purpose of control lines does not change during expansion.

In order to save the effort required in building a circuit from scratch, we modify an existing circuit with lower bit width parameters and realize the expand circuit. Hence, we can say a circuit is expandable if a circuit of larger bit width can be obtained from a circuit of smaller bit width.

General approach is to identify the sections which are affected by increase in bit width, and then modifying them accordingly. The modifications are usually replication of steps (gate stages) for each added bit according to operation performed. Taking the circuit of program counter of section 4.2.1 as an example we illustrate its expansion to 3-bit and 4-bit.The circuit of figure 4.2(b) is for 2-bit wide program counter variable. Figure 4.5(a) is the SyReC code for program counter 3-bit wide and figure 4.5(b) is its circuit realization. When width of '$pc$' is increased to 3-bits, the sections affected by it are only those where value of '$pc$' is modified. The operations "$pc <=> zero$", "$pc <=> jump$" require replication of a gate for the 3$^{rd}$ bit whereas the operation $pc += 1$ requires extra stage for the 3$^{rd}$ bit. This extra stage to be added depends upon the operation of increment.

In a similar fashion we can obtain circuit for program counter 4-bit wide from the circuit for 3-bit '$pc$'. Figure 4.6(a) is the SyReC code for program counter 4-bit wide and figure 4.6(b) is its circuit realization. Again we can observe that operations "$pc <=> zero$" and "$pc <=> jump$" require replication of a gate for the 4$^{th}$ bit, while the operation $pc += 1$ requires extra stage for the 4$^{th}$bit.

It can easily be observed upon comparison that certain sections, like if and else condition checking, are same in all circuits (figure 4.2(b), 4.5(b) and 4.6(b)). Also the control lines – '$inc$' and '$reset$' do not change.

```
module program_counter (in reset(1), in inc(1), in jump(3), inout pc(3)) wire zero(3)
if (reset = 1) then
        pc <=> zero
else
        if (inc = 1) then
                pc += 1
        else
                pc <=> jump
        fi (inc = 1)
fi (reset = 1)
```

**Figure 4.5(a): SyReC code of Program Counter (3-bit)**

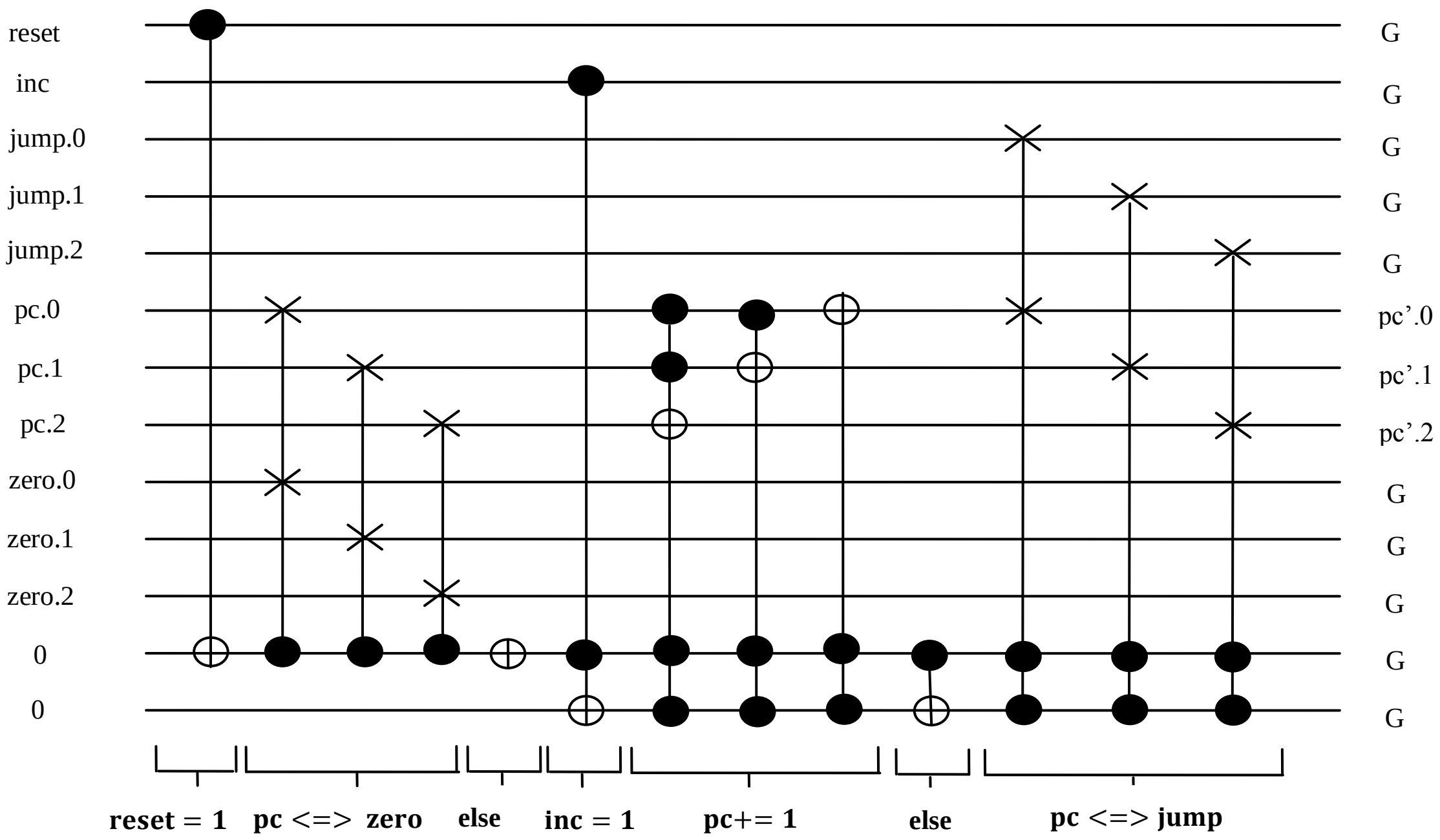

**Figure 4.5(b): Realization of Program Counter (3-bit)**

```
module program_counter (in reset(1), in inc(4), in jump(2), inout pc(4)) wire zero(4)
if (reset = 1) then
        pc <=> zero
else
        if (inc = 1) then
                pc += 1
        else
                pc <=> jump
        fi (inc = 1)
fi (reset = 1)
```

**Figure 4.6(a): SyReC code of Program Counter (4-bit)**

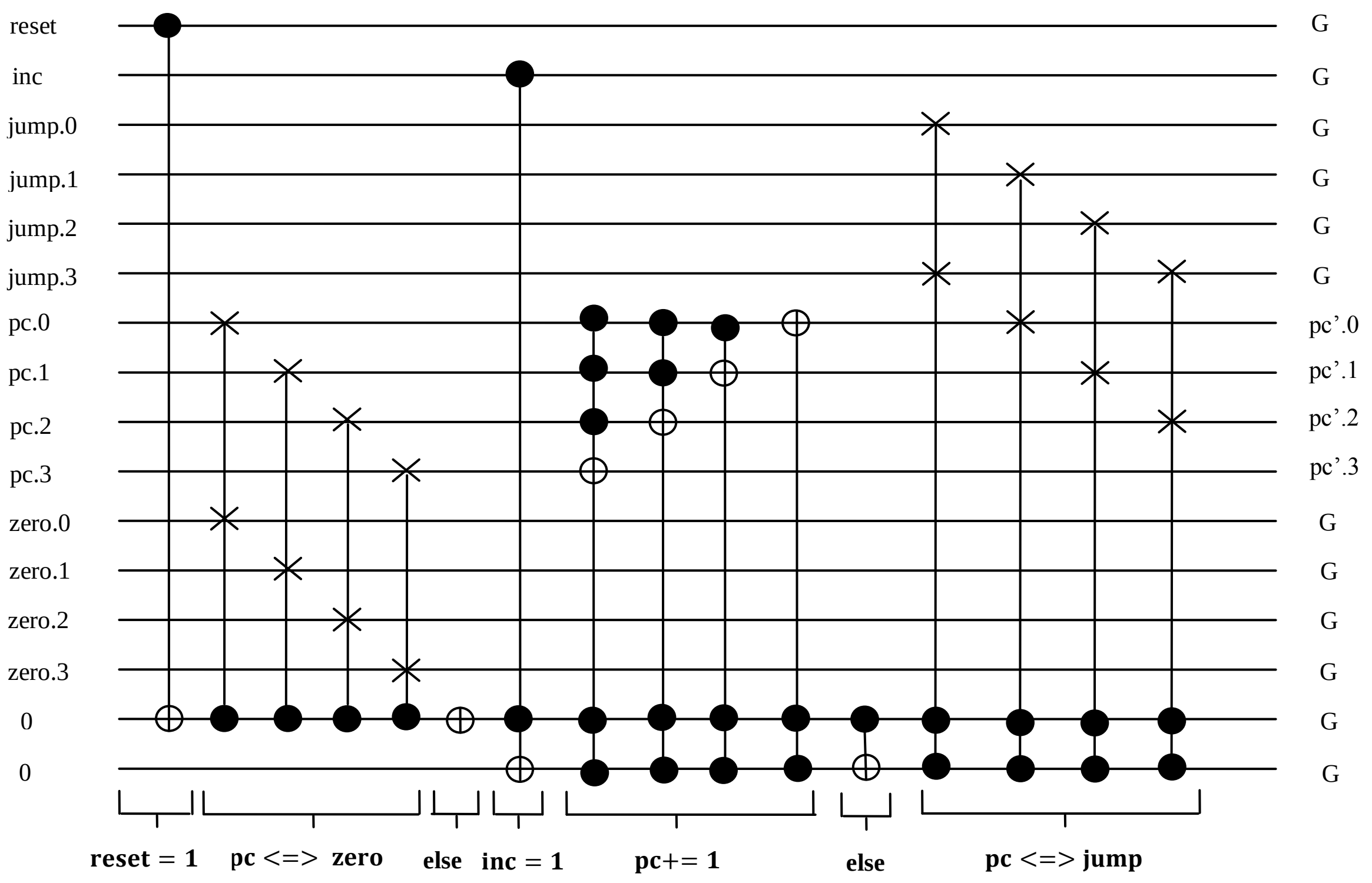

**Figure 4.6(b): Realization of Program Counter (4-bit)**

### 4.3.1 Elevator Controller

Elevator controller, as the name suggests, is a circuit for controlling movement and other functions of an elevator system. Through, the system requires a sequential reversible circuit; we present the combinational portion of the circuit since current version of SyReC does not include specification method for reversible sequential components. Hence, this circuit needs to be elaborated under certain system and its specifications.

**System description:** The elevator serves n floors (floor 0 to floor n-1) and each floor has its request/call button at the external door of elevator well. Each floor has only one call button (not two, as is common in modern elevators). A sensor is located at every floor. We can use this sensor to locate the current position of the elevator. The elevator system itself consists of several parts: an elevator to carry persons, a door, which can be opened and closed by a motor, floor buttons which can be pushed by passengers for their destination floor and a controller circuit to control all these actions. A sensor informs the control system about the door position. The elevator engine moves it up or down. Elevator can also remain idle on any floor when it is not moving.

Informally, the elevator behavior is defined as follows. The sensor which detects current floor of the elevator is connected to a variable '*c_f*'. The value of this variable is set by the floor sensor and can be read by the controller circuit. Requests for an elevator to reach a certain floor can come from two sources: call buttons and floor buttons. For this purpose we maintain arrays of Boolean variables, which when set indicate request for stopping elevator at that floor. If we press the call button at a floor, or a passenger inside the elevator presses a floor button, the request is stored by setting the corresponding Boolean variable. Whenever the elevator moves to this floor it will stop and reset its Boolean variable. When the floor is reached, the door opens. The door stays open for some time to allow passengers to enter or exit the elevator. After this time, the door closes again. A Boolean variable 'door' is used by the controller to command the door to be opened or closed. If the elevator should stop at a certain floor, the motor is stopped immediately after the reception of a signal from the corresponding floor sensor. The controller uses another variable to signal whether motor should move the elevator or stop it. Also, the direction of movement needs to be signaled through a variable ‘dir’. These signals associated with the controller circuit are shown in Figure 4.7, for an elevator serving four floors.

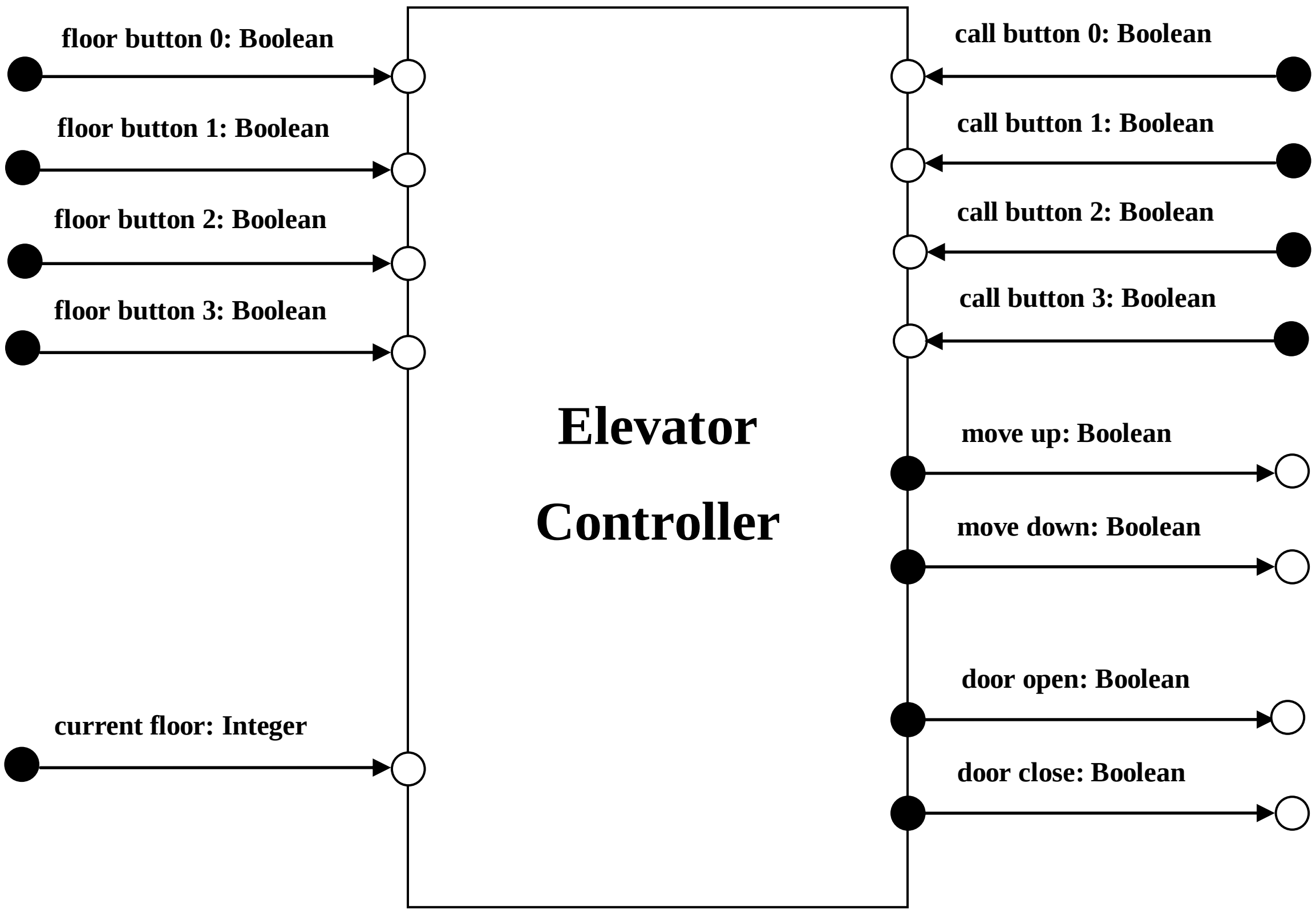

**Figure 4.7: System model: Elevator Controller and its signals**

All the actions of elevator controller are decided according to the request serving strategy. Different strategies exist for handling the requests from the individual floors. Here we adopt simplest strategy that does follows:

- Continue travelling in same direction until there are pending requests in that direction
- If there are no further request, stop and become idle or change direction if there are requests in the opposite direction

**The elevator controller reversible circuit:** We present here a combinational circuit for elevator controller, which deals with following signals:

- **Door open/close:** 1-bit signal with 0 indicating close and 1 for open
- **Move:** 1-bit signal with value 1 indicating that elevator is moving and 0 indicating stopping of elevator
- **Direction:** 1-bit signal for direction of movement, 0 is for down and 1 is for up
- **Current floor:** used to hold the current floor position of the elevator. Its bit width depends on number of floors
- **Floor buttons:** array of variables each 1-bit wide. Size of array is equal to number of floors. This corresponds to the destination floor buttons inside the elevator, with $i^{th}$ variable for $i^{th}$ floor. An event of a passenger pressing the floor button leads to setting of corresponding variable. Serving a request leads to resetting of the corresponding variable.
- **Call buttons:** array of variables each 1-bit wide. Size of array is equal to number of floors. This corresponds to the call buttons outside the elevator, one on each floor, with $i^{th}$ variable for $i^{th}$ floor. An event of pressing the call button leads to setting of corresponding variable. Serving a request leads to resetting of the corresponding variable.

Since many variables depend on the total number of floors, n, we can use n as a parameter to decide the scale of the circuit. Beginning with the smallest, that is n=2, circuit needs to check only the current floor and pending request. Depending on it, the elevator would either move up or down or stop. The SyReC code for the controller of an elevator serving two floors is given in figure 4.8(a). Realizing this circuit by hierarchical approach we obtain circuit as shown in figure 4.8(b).

Expanding this circuit in fact means to increase the number of floors. Hence, taking n=4, the SyReC code is as shown in figure 4.9(a). Line 1-5 check whether the elevator was requested to stop at current floor. If so it stops, opens the door and the corresponding request variables '*fb[c_f]*' and '*cb[c_f]*' are reset. Line 7-13 are for checking whether there are any pending requests, from any floor between current floor and third floor. Accordingly elevator will move up. If there are no requests elevator will remain idle on current floor. Line 15-21 check whether there are any pending request from any floor between current floor and ground floor. Accordingly elevator will move down. If there are no requests elevator will remain idle on current floor. We expand the circuit of fig 4.8(b), by the expansion method described above to obtain circuit shown in figure 4.9(b).

Since, expansion approach is a general approach; we can produce a generalized SyReC code for elevator controller program, shown in figure 4.10. Line 1-5 check whether the elevator was requested to stop at current floor. If so it stops, opens the door and the corresponding request variables '*fb[c_f]*' and '*cb[c_f]*' are reset. Line 7-13 are for checking whether there are any pending requests, from any floor between current floor and topmost floor. Accordingly elevator will move up. If there are no requests elevator will remain idle on current floor. Line 15-21 check whether there are any pending request from any floor between current floor and ground floor. Accordingly elevator will move down. If there are no requests elevator will remain idle on current floor.

Thus, this circuit adopts a strategy to give priority of serving requests from higher floors than to the requests from lower floors.

```
module elevator (inout c_f(1), inout fb[2](1), inout cb[2](1), out door(1), out move(1), out dir(1))
if ((fb[c_f] = 1) || (cb[c_f] = 1)) then
    door^ = 1; move^ = 0; fb[c_f] = 0; cb[c_f] = 0
else
    door^ = 0; move^ = 1
fi ((fb[c_f] = 1) || (cb[c_f] = 1))

if (c_f < 1) then
    for$i = (c_f + 1) to 1 do
        if ((fb[$i] = 1) || (cb[$i] = 1)) then
            door^ = 0; move^ = 1; dir^ = 1
        else
            move^ = 0
        fi((fb[$i] = 1) || (cb[$i] = 1))
    rof

else if (c_f > 0) then
    for$i = (c_f − 1) to 0 step -1 do
        if ((fb[$i] = 1) || (cb[$i] = 1)) then
            door^ = 0; move^ = 1; dir^ = 1
        else
            move^ = 0
        fi((fb[$i] = 1) || (cb[$i] = 1))
    rof

else
        move^ = 0
    fi(c_f > 0)
fi (c_f < 1)
```

**Figure 4.8(a): SyReC code of Elevator Controller (Floors=2)**

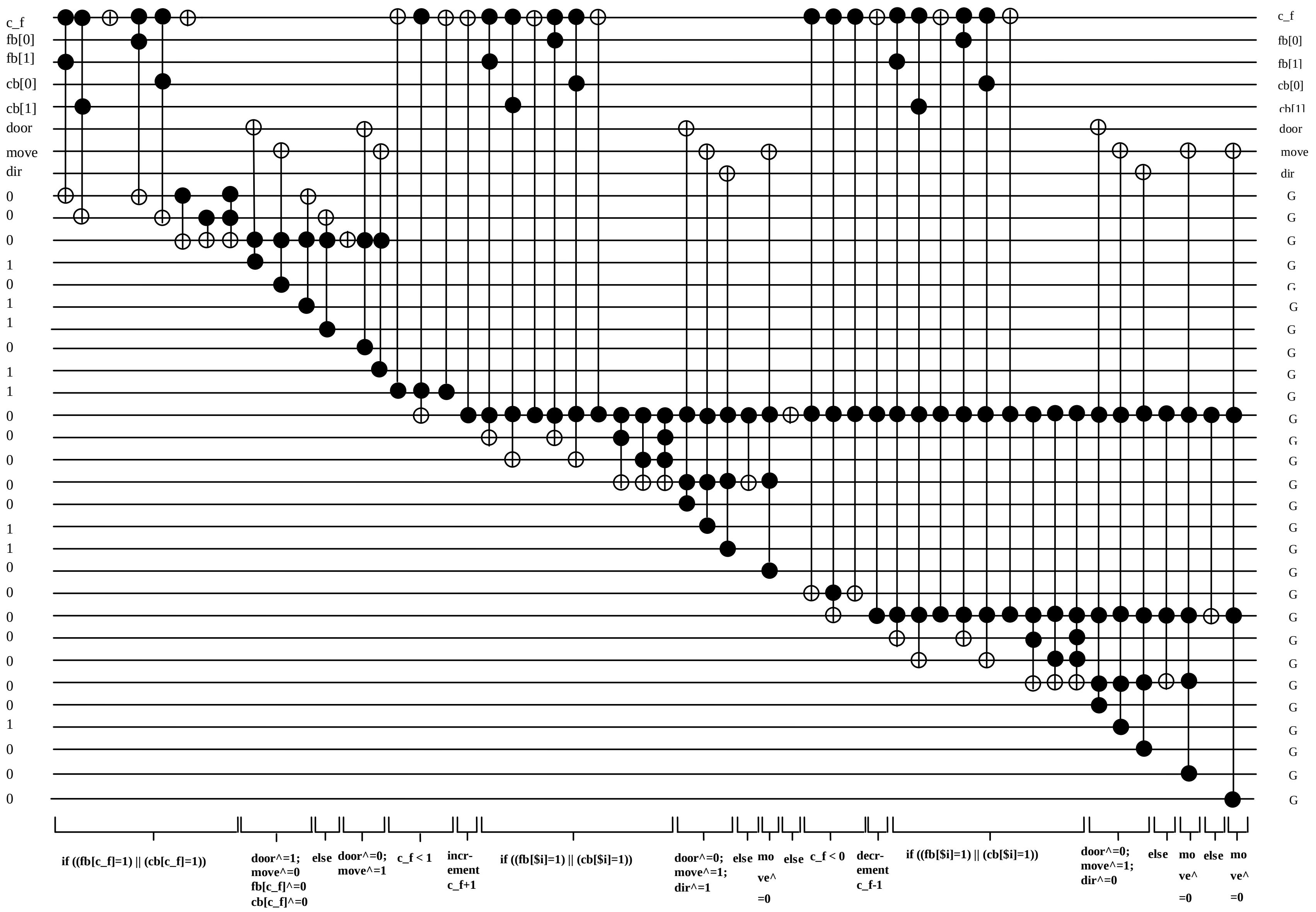

**Figure 4.8(b): Realization of Elevator Controller (Floors=2)**

```
module elevator (inout c_f(2), inout fb[4](1), inout cb[4](1), out door(1), out move(1), out dir(1))
    if ((fb[c_f] = 1) || (cb[c_f] = 1)) then
        door^ = 1; move^ = 0; fb[c_f] = 0; cb[c_f] = 0
    else
        door^ = 0; move^ = 1
    fi ((fb[c_f] = 1) || (cb[c_f] = 1))

    if (c_f < 3) then
        for$i = (c_f + 1) to 3 do
            if ((fb[$i] = 1) || (cb[$i] = 1)) then
                door^ = 0; move^ = 1; dir^ = 1
            else
                move^ = 0
            fi((fb[$i] = 1) || (cb[$i] = 1))
        rof

    else if (c_f > 0) then
        for$i = (c_f − 1) to 0 step -1 do
            if ((fb[$i] = 1) || (cb[$i] = 1)) then
                door^ = 0; move^ = 1; dir^ = 1
            else
                move^ = 0
            fi((fb[$i] = 1) || (cb[$i] = 1))
        rof

        else
            move^ = 0
        fi(c_f > 0)
    fi (c_f < 3)
```

**Figure 4.9(a): SyReC code of Elevator Controller (Floors=4)**

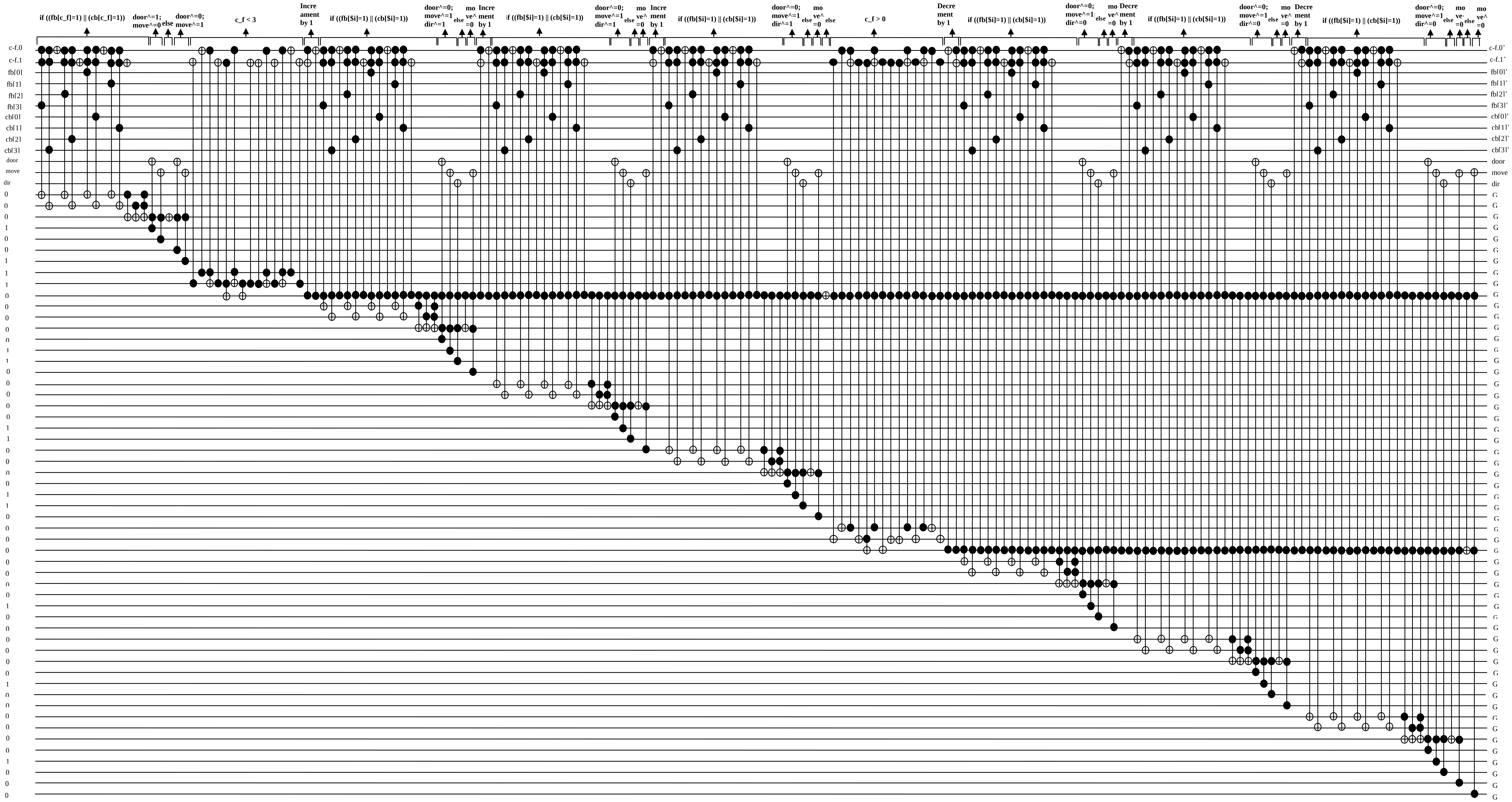

**Figure 4.9(b): Realization of Elevator Controller (n=4)**

```
module elevator (inout c_f(log2 n), inout fb[n](1), inout cb[n](1), out door(1), out
move(1), out dir(1))
    if ((fb[c_f] = 1) || (cb[c_f] = 1)) then
        door^ = 1; move^ = 0
    else
        door^ = 0; move^ = 1
    fi ((fb[c_f] = 1) || (cb[c_f] = 1))

    if (c_f < n − 1) then
        for$i = (c_f + 1) to n − 1 do
            if ((fb[$i] = 1) || (cb[$i] = 1)) then
                door^ = 0; move^ = 1; dir^ = 1
            else
                move^ = 0
        fi((fb[$i] = 1) || (cb[$i] = 1))
    rof

    else if (c_f > 0) then
        for$i = (c_f − 1) to 0 step -1 do
            if ((fb[$i] = 1) || (cb[$i] = 1)) then
                door^ = 0; move^ = 1; dir^ = 1
            else
                move^ = 0
            fi((fb[$i] = 1) || (cb[$i] = 1))
        rof

        else
            move^ = 0
        fi(c_f > 0)
    fi (c_f < n − 1)
```

**Figure 4.10: SyReC code of Elevator Controller (Floors=n)**

**Chapter 5**

# IMPLEMENTATION AND RESULTS

In this chapter we discuss about our tool **"RCHDL Realizer"**, its features and use. Our results consist mainly of effect on cost of a circuit due to expansion.

## 5.1 Tool Description

The tool has been developed using Netbeans on java platform.

**Purpose:** The basic purpose of this tool is to facilitate expansion of a reversible circuit realization provided a SyReC code specification of the circuit. Also this tool is helpful for analyzing the growth of number of lines, gates and quantum cost of a reversible circuit as the reversible circuit expands.

**Inputs and output:** The generalized SyReC specifications of certain circuit serve as static data for the program while the bit width of these circuits is obtained as user inputs. Accordingly a proper SyReC code of the desired circuit is generated and then transform into circuit realization.

**Approach:** The approach used for circuit realization is the hierarchical approach as discussed in chapter 4. Circuit for higher bits is obtained from the respective circuit of lowest bit-width using the expansion technique in chapter 4.

## 5.2 GUI for Expand Reversible Circuit

Figure 5.1 shows the GUI for the tool. It contains following components:

**A:** Drop down list box to select any one reversible circuit from the list.

**B:** Display text field area to display SyReC code of selected reversible circuit.

**C:** Display area for the generated reversible circuit.

**D:** Menu bar for selecting different actions.

**E:** User input area where user can edit bit width of selected reversible circuit.

**F:** View area for viewing certain parameters of the reversible circuit.

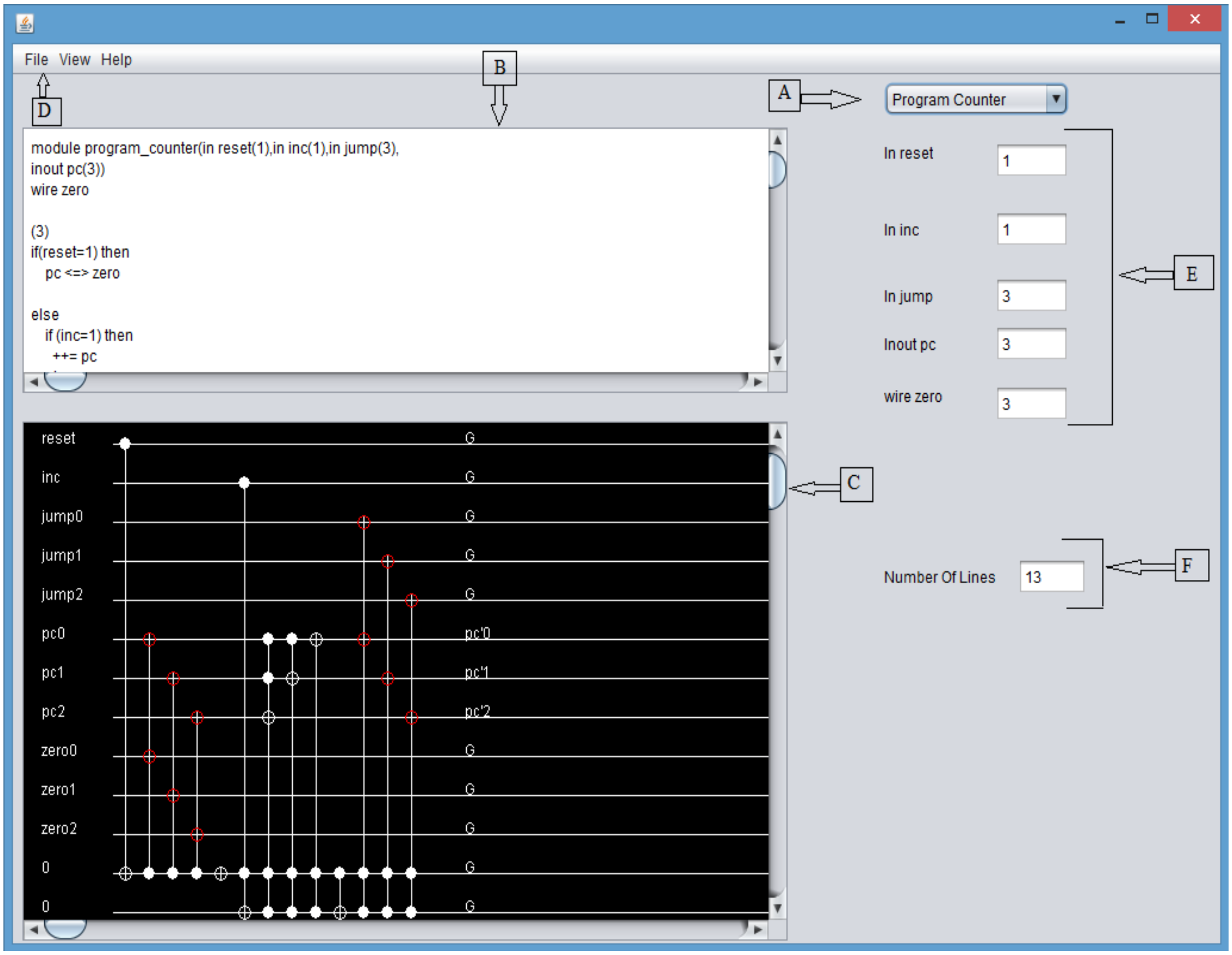

**Figure 5.1: GUI of tool**

## 5.3 Description of Menu Bar

Tool contains several actions which can be performed by choosing appropriate menu command.

**1. File menu:** This contains commands related to reversible circuit generation. The commands are:

- **a. Set of parameters:** To change parameters of the selected reversible circuit, such as bit width.

b. **Run:** To generate reversible circuit according to set parameters.

c. **Exit:** To exit the tool.

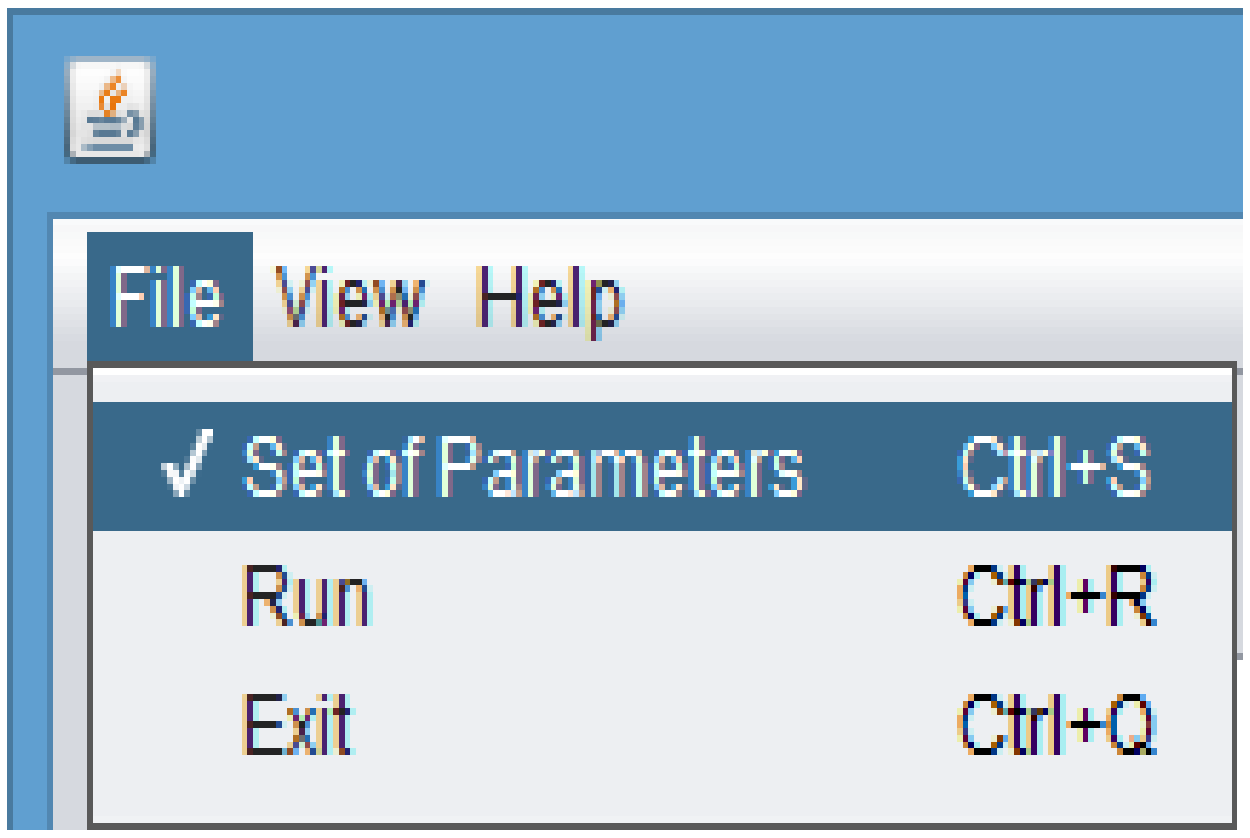

**Figure 5.2: File menu**

**2. View menu:** This contains commands to view certain auxiliary information about the current circuit. The commands correspond to the information required, which are:

a. Number of lines

b. Number of gates

c. Quantum cost

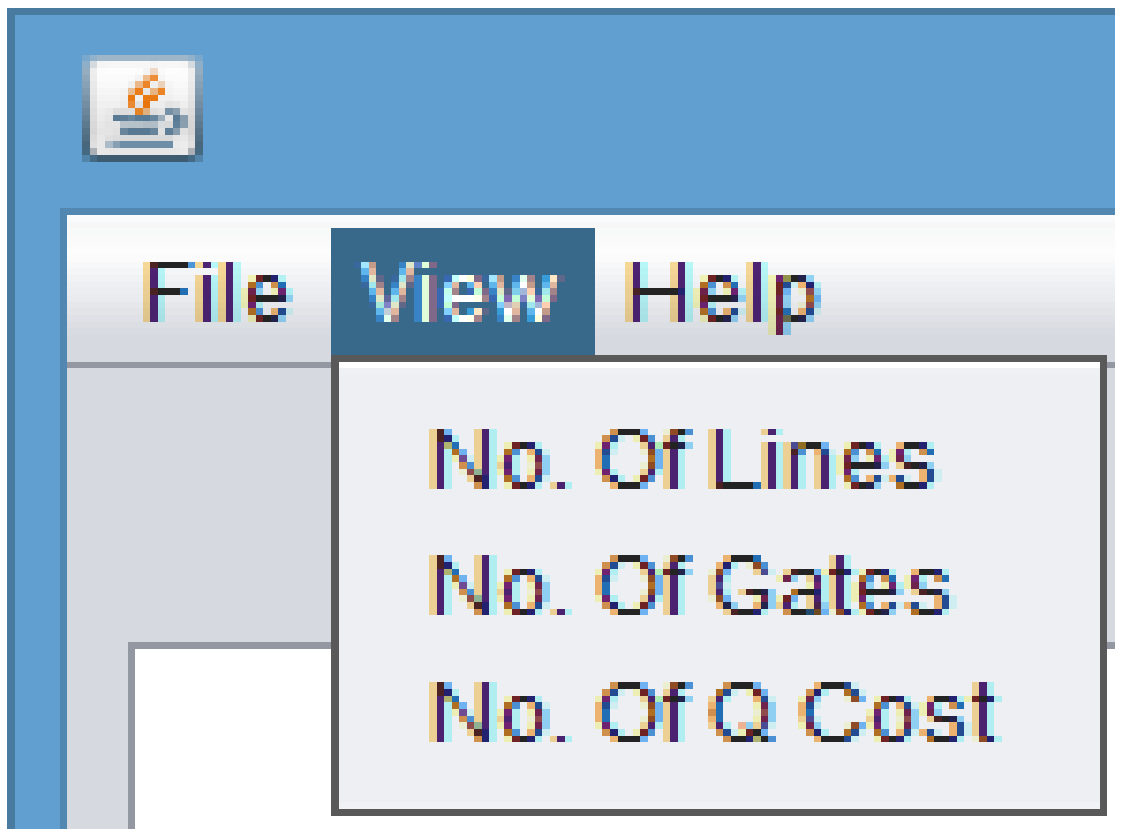

**Figure 5.3: View menu**

**3. Help:** Gives information about tool.

## 5.4 User Actions

In this section we describe the step through which a user may interact appropriately with the tool.

Sequence of actions by user to expand a reversible circuit:

**Step 1:** Select a circuit from drop down list of control ‘A’.

**Step 2:** Choose option ‘Set of parameters’ from file menu.

**Step 3:** Change the parameters in user input area (component E).

**Step 4:** Choose option ‘Run’ from File menu.

This would lead to display of SyReC code in area ‘B’ and reversible circuit in area ‘C’. These actions show in figure 5.4 to figure 5.10.

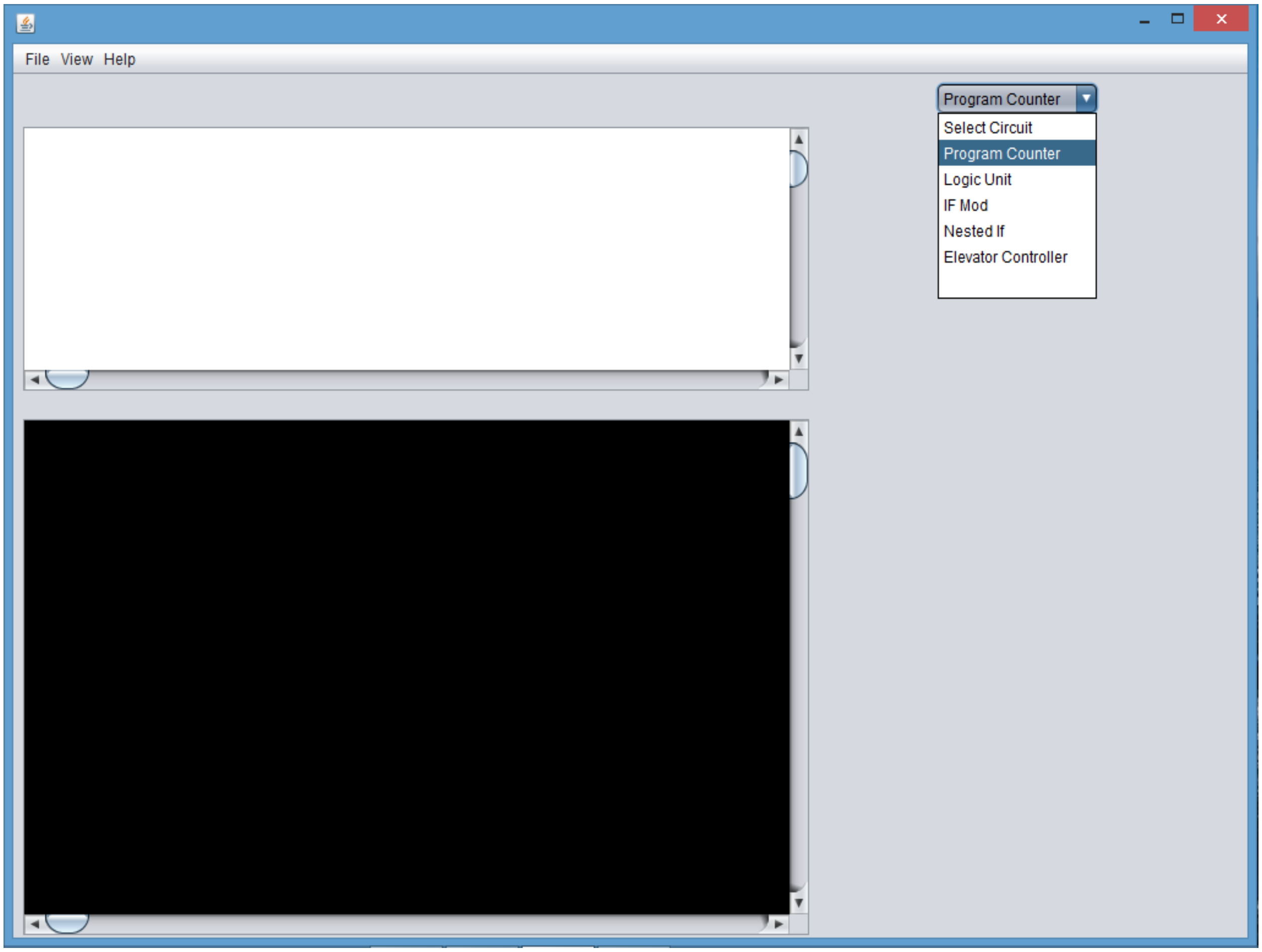

**Figure 5.4: Select circuit from drop down list**

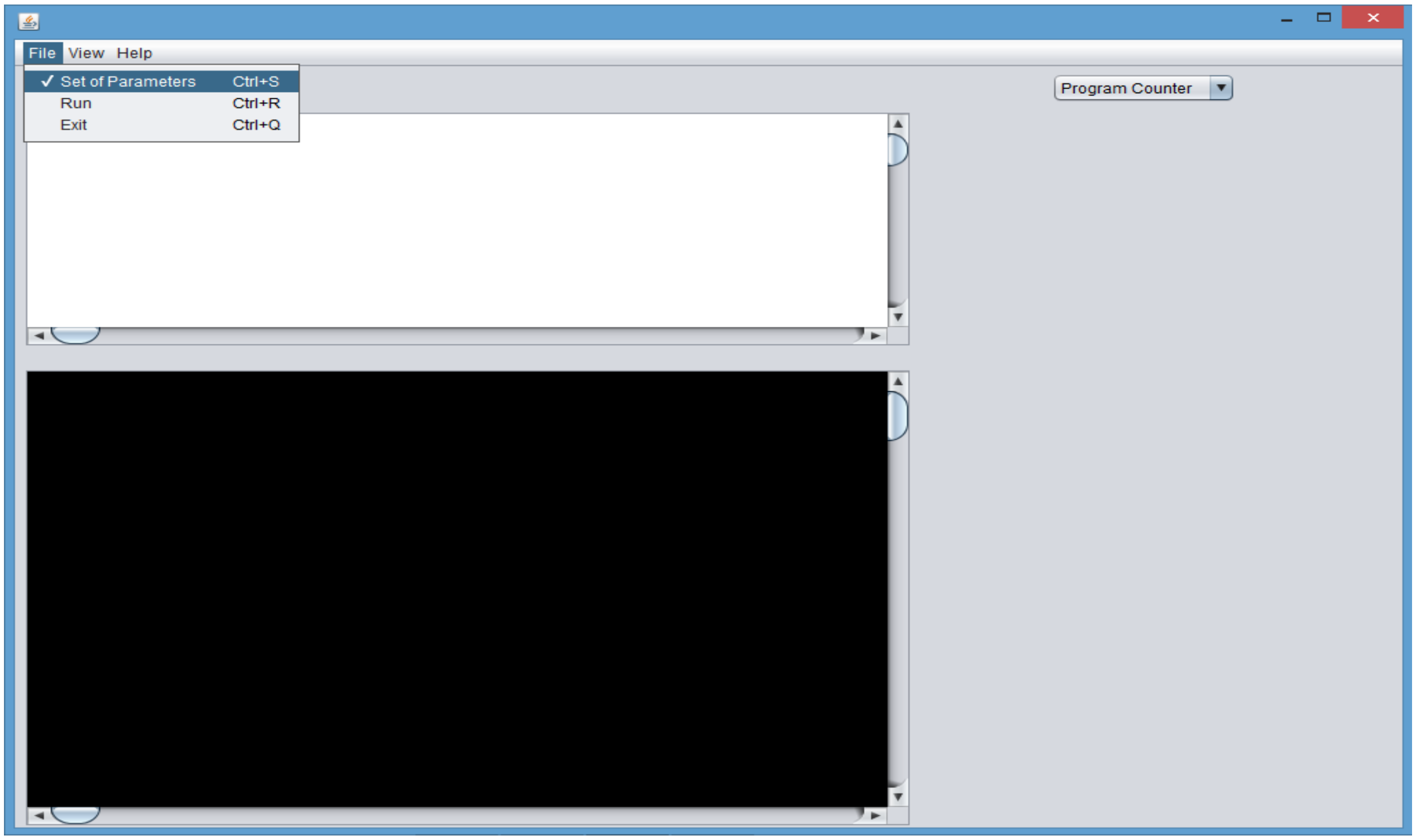

**Figure 5.5: Choose option set of parameters from file menu**

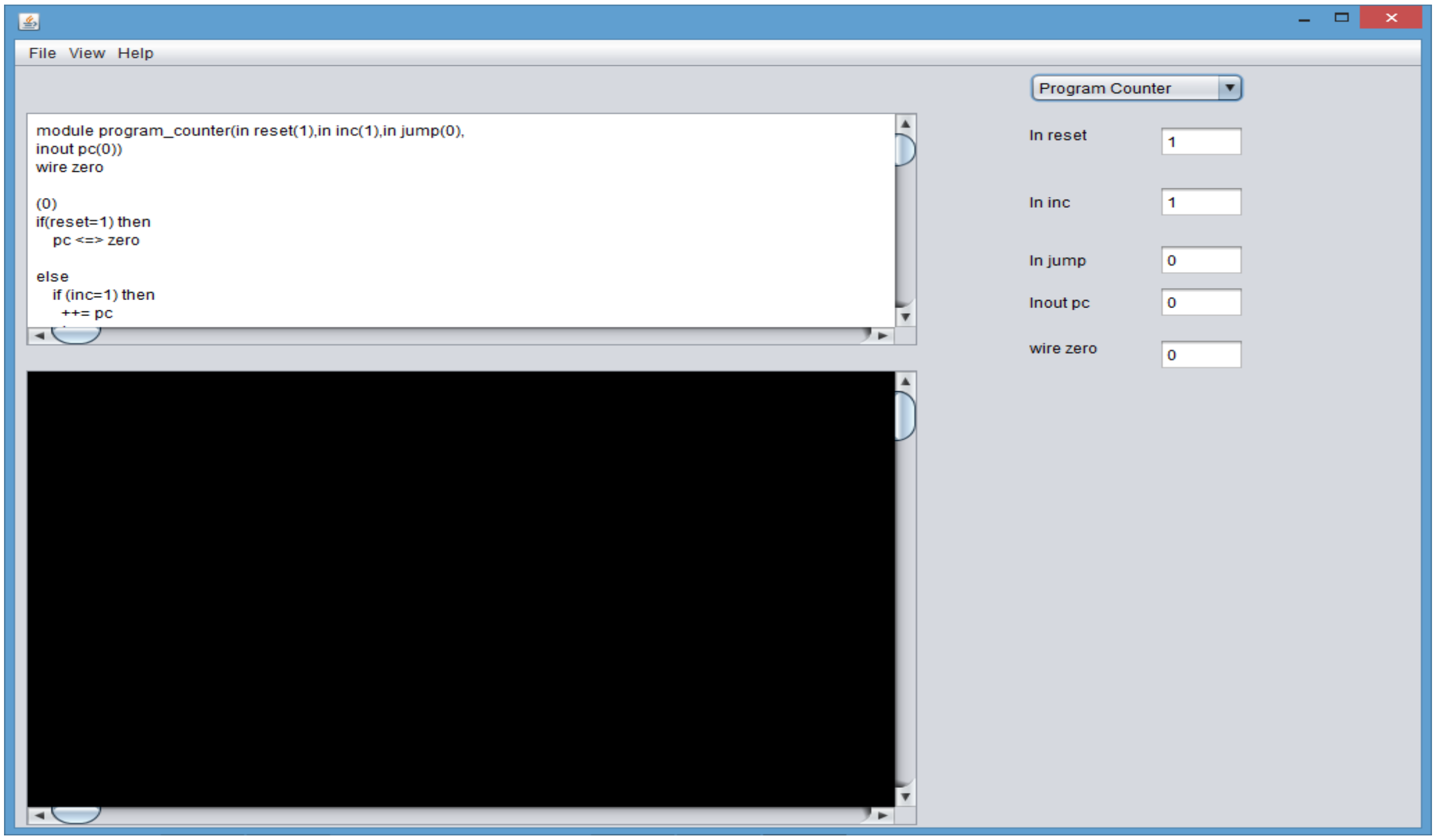

**Figure 5.6: Display SyReC code and variables of selected circuit**

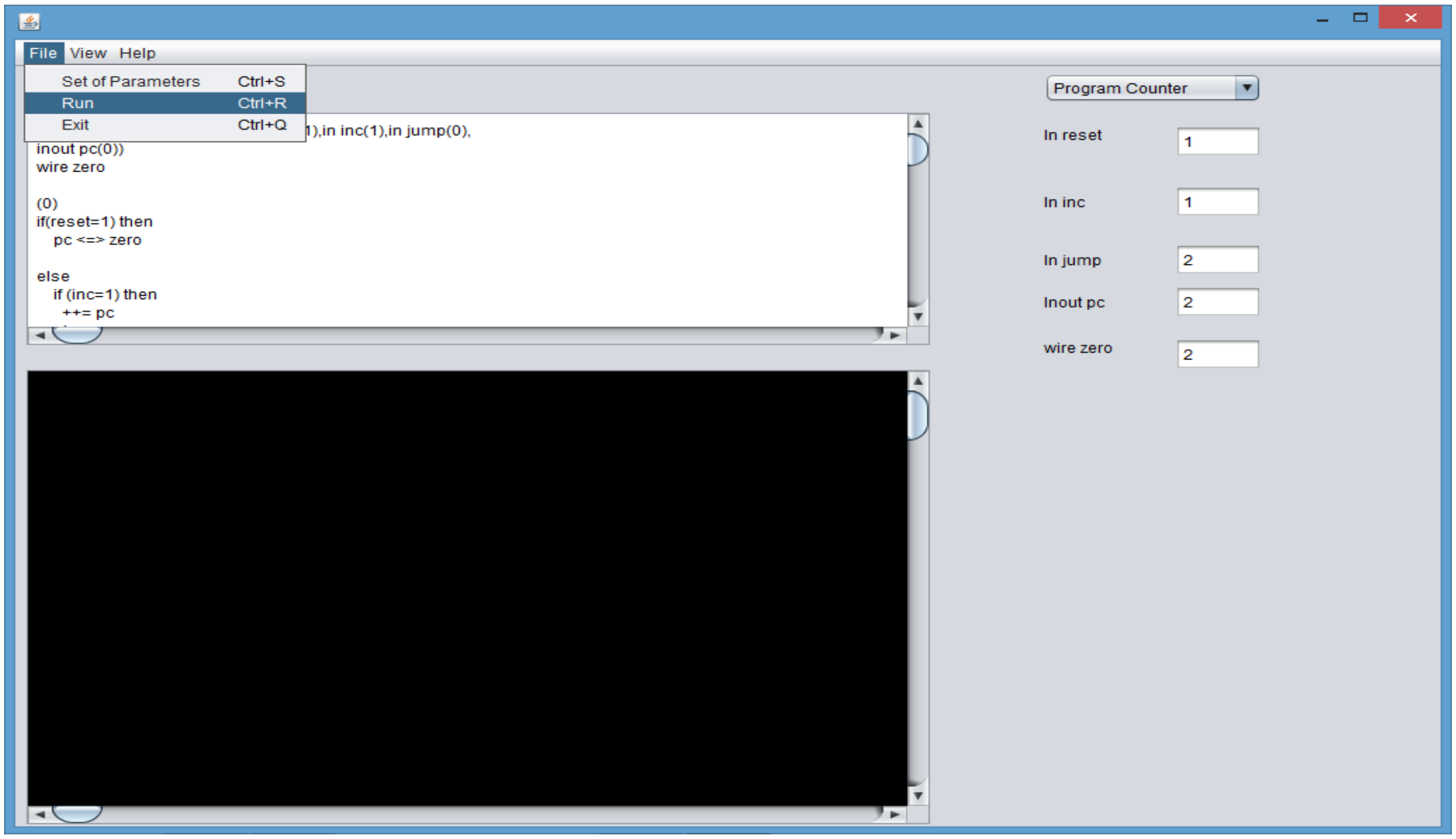

**Figure 5.7: Put bit width of the selected circuit and run**

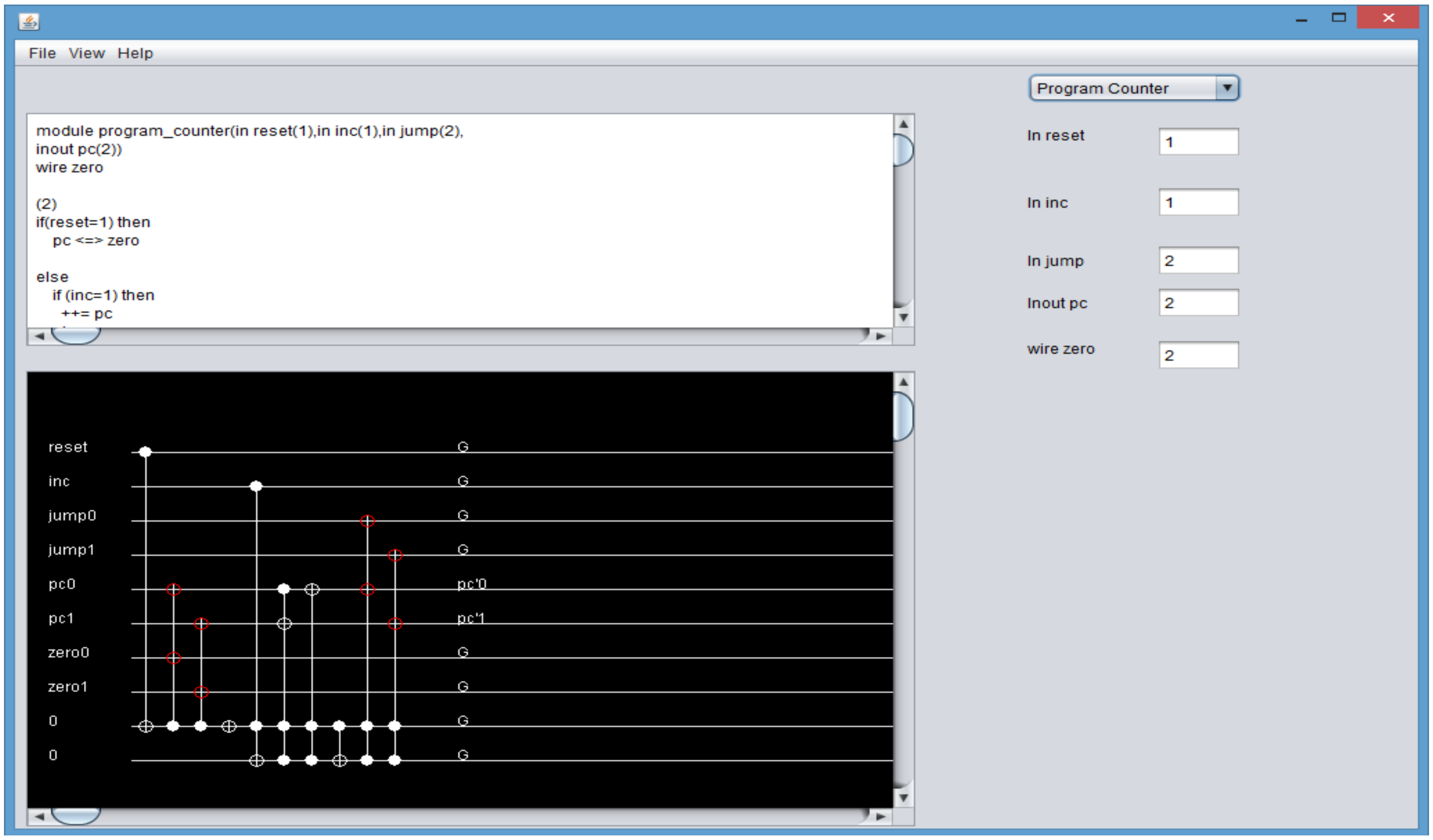

**Figure 5.8: Display schematic and generated SyReC code**

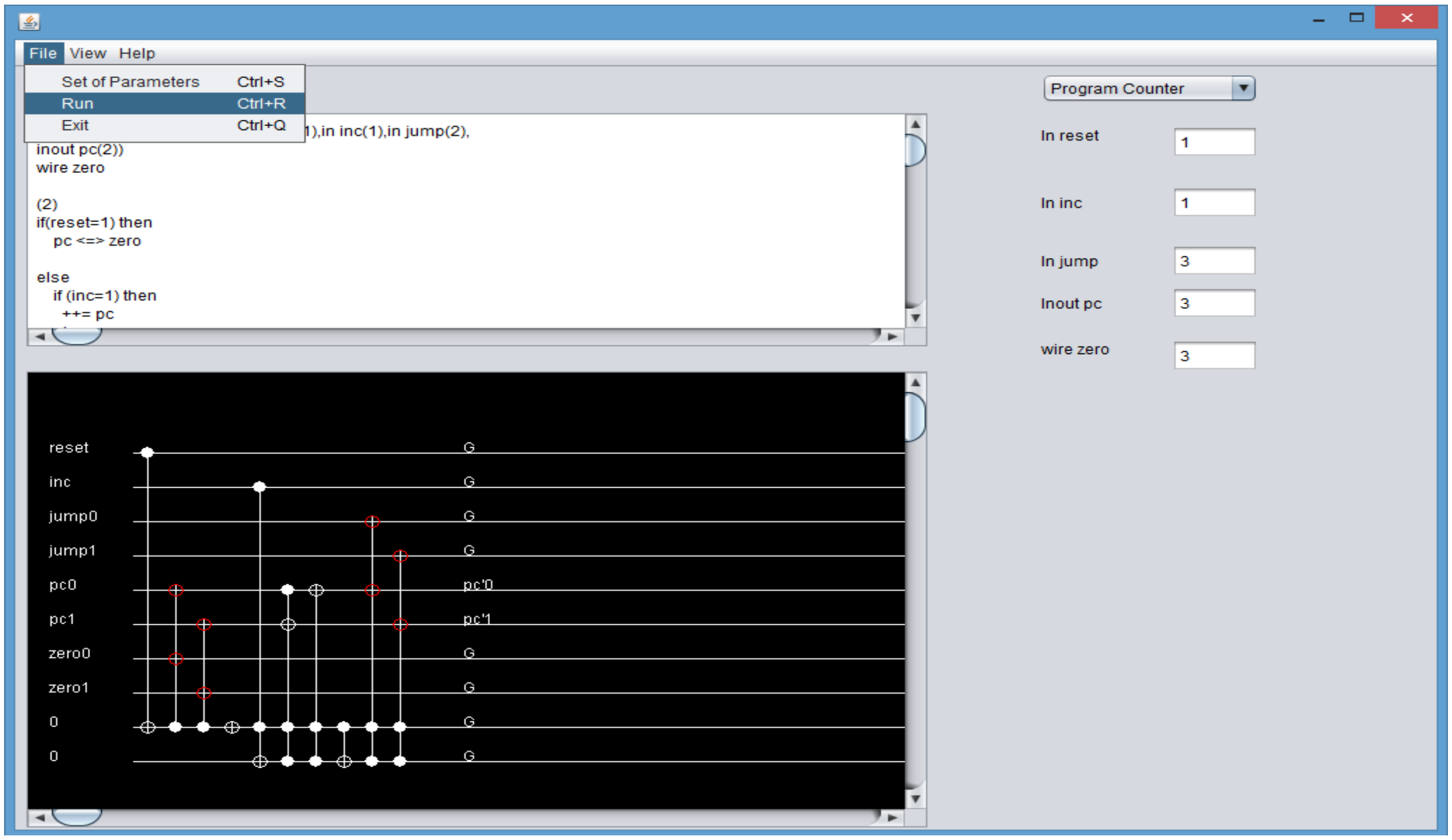

**Figure 5.9: Expand selected circuit**

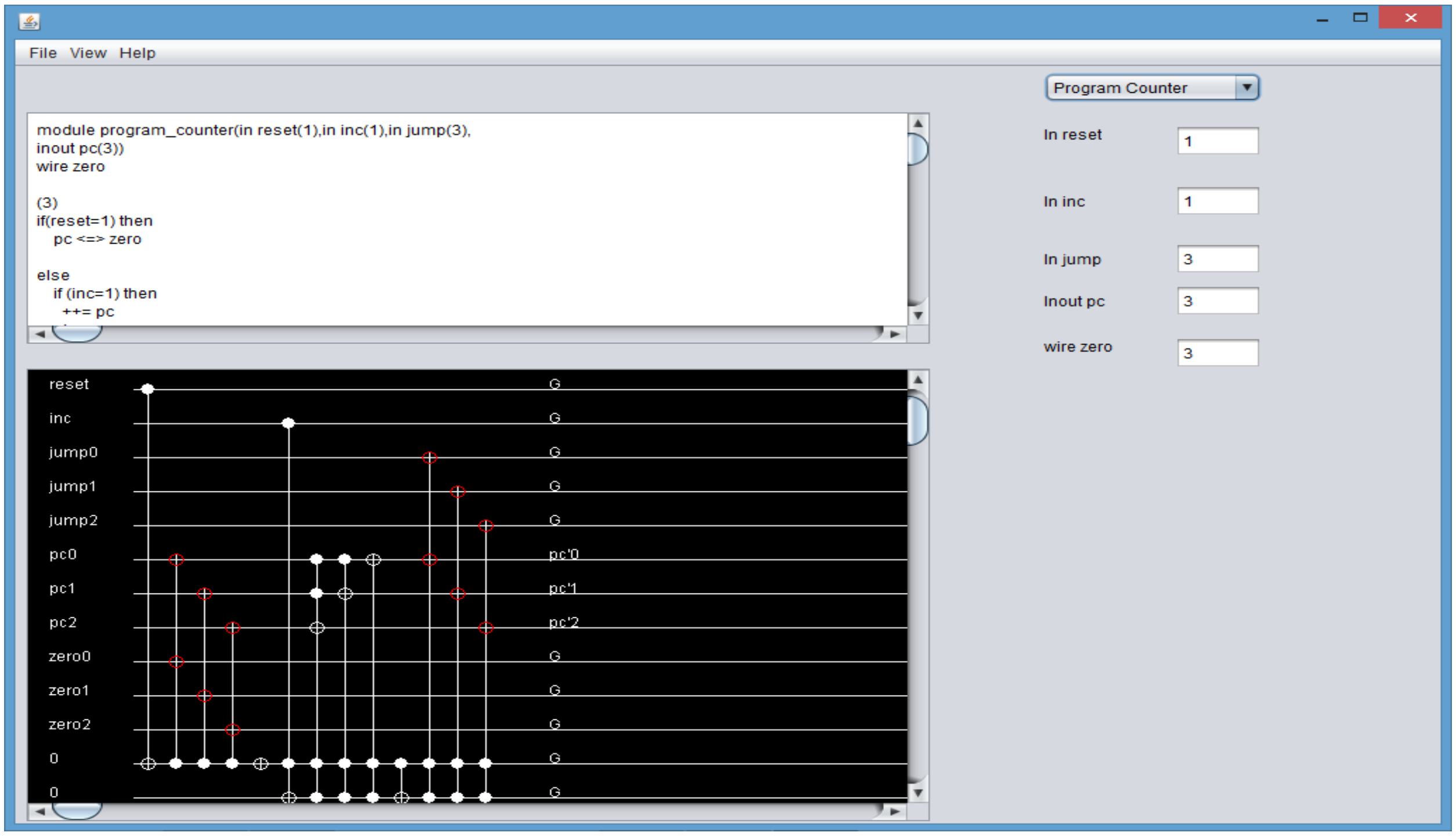

**Figure 5.10: Display schematic and generated SyReC code**

Sequence of actions to view auxiliary information (it assumes that a reversible circuit has already been selected and run):

**Step 1:** Select view from Menu bar.

**Step 2:** Select the respective information tab out of three.

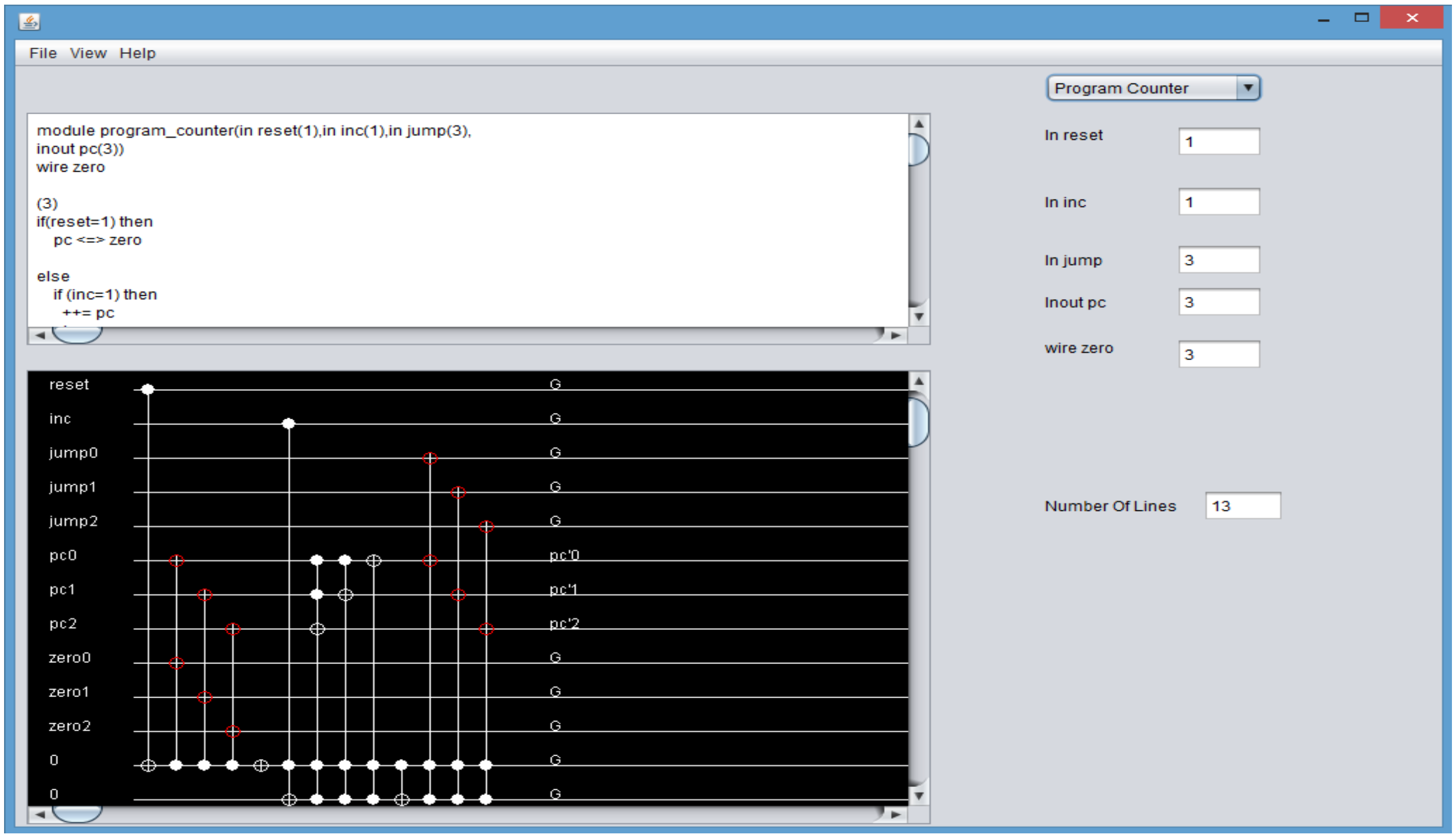

**Figure 5.11: View auxiliary information of selected circuit from view menu**

## 5.5 Analyze effects of expansion

We have focused on three parameters concerning reversible circuits, namely:

- **Number of lines:** Total number of lines in the reversible circuit (input or output). Lines are considered an important resource in any reversible circuit. Optimizing a reversible circuit often refers to reducing number of lines.

- **Number of gates:** Total number of reversible gates (multiple Toffoli or Fredkin) used in the reversible circuit. These are helpful for deriving quantum cost of reversible circuit.

Subsequently, during synthesis number of gates given idea about hardware cost, and area of circuit.

- **Quantum cost:** Quantum cost denotes the effort needed to transform a reversible circuit to a quantum circuit. The sum of the quantum cost for all gates defines the quantum cost of the whole circuit. Quantum cost associated with a Multiple Toffoli gate is computed as $2^n - 3$ as the size n and quantum cost associated with a Multiple Fredkin gate is computed as $2^n - 1$ as the size n (without considering garbage lines).

Table 5.1 shows the effect of expansion on these above mentioned parameters for reversible circuit program counter. As we observe that with every expansion (expanding one bit) the number of lines and number of gates increase by constant value of 3. Also the quantum cost is polynomial in bit width.

**Table 5.1: Effect of expansion on Program counter**

| Bit-width | Number of lines | Number of gates | Quantum cost |
|---|---|---|---|
| 2-bit | 10 | 10 | 70 |
| 3-bit | 13 | 13 | 121 |
| 4-bit | 16 | 16 | 204 |
| 5-bit | 19 | 19 | 351 |

Table 5.2 shows parameters of Logic unit reversible circuit for various bit-widths. As the bit-width increases, number of lines increase by constant 6. Similarly numbers of gates get increased by 11 per bit increase. Here, even quantum cost is also a linear function. We observe an increase of 215 units per bit.

**Table 5.2: Effect of expansion on Logic unit**

| Bit-width | Number of lines | Number of gates | Quantum cost |
|---|---|---|---|
| 2-bit | 23 | 52 | 588 |
| 3-bit | 29 | 63 | 803 |
| 4-bit | 35 | 74 | 1018 |
| 5-bit | 41 | 85 | 1233 |

Table 5.3 shows effect of expansion on Elevator controller reversible circuit for various bit-widths.

**Table 5.3: Effect of expansion on Elevator controller**

| Floor (n) | Number of lines | Number of gates | Quantum cost |
|---|---|---|---|
| n = 2 | 36 | 55 | 543 |
| n = 4 | 71 | 185 | 3365 |

Chapter 6

# CONCLUSION AND FUTURE WORK

In this dissertation, we proposed an approach for expansion reversible circuits. Also, we have designed a tool to assist in reversible circuit expansion and study the impact on related costs like number of lines, number of gates and quantum costs. The tool generates a SyReC specification of a selected circuit and realizes it into a reversible circuit which can be expanded by increasing number of bits of operands involved. Moreover, we have presented a reversible circuit for elevator controller.

The approach used for circuit realization at basic component level is control-intensive and has scope of optimization. This has already been researched in [31], yet there is further scope of improvement.

## 6.1 Future Scope

- Elevator controller is better implemented as a sequential circuit, which further needs a SyReC specification for sequential circuits. This being an open problem till date.
- SyReC processors are not available.
- Our tool puts a limit on the bit-width of a reversible circuit to be expanded. Consequently, elevator controller reversible circuit cannot be further expanded using the tool (through it can be done manually using our expansion approach). Thus, enhancing the capabilities of the tool could be a designable improvement.

**Vandana Maheshwari**

**Address:** "Shree Bhawan",

B-43, Talwandi, Kota, Rajasthan.

**Email:** vandana.mhshwri@gmail.com

**Mobile No.:** 9462966086

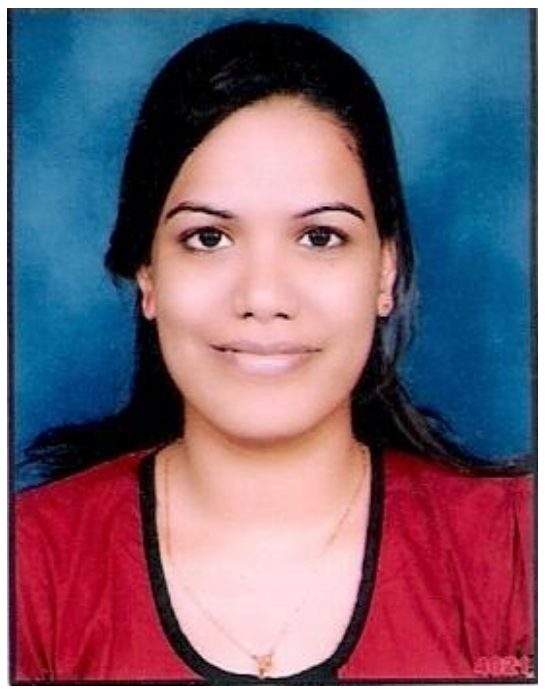

**OBJECTIVE:** To work in highly challenging and competitive atmosphere which offers growth with opportunities to enrich my knowledge and skills.

**EDUCATION QUALIFICATION:**

- Successfully completed **M.Tech** in **Computer Science & Engg.** Branch from **University College of Engineering**, Rajasthan Technical University, Kota, Rajasthan.
- Successfully completed **B.Tech** in **Computer Science & Engg.** Branch from **Arya College of Engg. & I.T.**, Jaipur, Rajasthan.

**PERSONAL PROFILE:**

- **Date of Birth** 21st April,1987
- **Father's Name** Sh. Mahesh Chand Ajmera
- **Language Known** Hindi & English

**Place: Kota**

**Date: 11-March-2014** **(Vandana Maheshwari)**